\documentclass[useAMS]{mn2e}
\usepackage{epsfig,natbib_vw,times}

\usepackage{float}

\newfloat{code}{h}{lop}

\usepackage{amsmath,amssymb,upgreek}

\bibpunct[, ]{(}{)}{;}{a}{}{,}

\def \aj {AJ}
\def \mnras {MNRAS}
\def \apj {ApJ}
\def \apjs {ApJS}
\def \apjl {ApJL}
\def \aap {A\&A}

\def \araa {ARAA}
\def \pasp {PASP}

\def\gsim{\mathrel{\lower0.6ex\hbox{$\buildrel {\textstyle >}
 \over {\scriptstyle \sim}$}}}
\def\lsim{\mathrel{\lower0.6ex\hbox{$\buildrel {\textstyle <}
 \over {\scriptstyle \sim}$}}}

\def \ha {H$\alpha$}
\def \nii {[N~{\sc ii}]}
\def \oiii {[O~{\sc iii}]}

\def \hb {H$\beta$}

\def \tauvline {$\uptau_{\mathsf{V,line}}$}
\def \tauvcont {$\uptau_{\mathsf{V,cont}}$}
\def \Dtauvline {$\Delta\uptau_{\mathsf{V,line}}$}
\def \Dtauvcont {$\Delta\uptau_{\mathsf{V,cont}}$}

\def \EBV {\textit{E(B$-$V)}}
\def \ebv {\textit{E(B$-$V)}}
\def \mustar {$\mu^*$}
\def \Mstar {M$^*$}
\def \mum {$\mu{\rm m}$}
\def \ssfr {$\psi_S$}
\def \logssfr {$\log_{10}\psi_S$}
\def \ab {$b/a$}
\def \ba {$b/a$}
\def \Rpet {$R_{\rm pet}$}
\def \rpet {$R_{\rm pet}$}
\def \ql {$Q_\lambda$}
\def \sopt {$s_{\mathsf{opt}}$}
\def \snir {$s_{\mathsf{nir}}$}
\def \sfuvnuv {$s'_{\rm FUV-NUV}$}
\def \sfuvu {$s'_{\rm FUV-u}$}

\def \dtk {$\Delta\uptau_{\mathsf{K}}$}
\def \dtv {$\Delta\uptau_{\mathsf{V}}$}

\def \tauv {$\uptau_{\mathsf{V}}$}
\def \dtl {$\Delta\uptau_{\lambda}$}

\title[Dust attenuation curves]{Empirical determination of the
  shape of dust attenuation curves in star-forming galaxies}
\author[V. Wild et al.]{
\parbox[t]{\textwidth}{\raggedright 
Vivienne Wild$^{1,2}$\thanks{vw@roe.ac.uk},  St\'{e}phane Charlot$^1$,
Jarle Brinchmann$^{3,4}$,  Timothy Heckman$^5$, Oliver Vince$^{1,6}$, Camilla Pacifici$^1$,
Jacopo Chevallard$^1$
}
\vspace*{6pt}\\
$^1$Institut d'Astrophysique de Paris, CNRS, Universit\'e Pierre \& Marie Curie, UMR 7095, 98bis bd Arago, 75014 Paris, France\\
$^2$Institute for Astronomy, University of Edinburgh, Royal Observatory, Blackford Hill, Edinburgh, EH9 3HJ, U.K. (SUPA) \\
$^3$Leiden Observatory, Leiden University, 2300RA, Leiden, The
Netherlands\\
$^4$Centro de Astrof\'{\i}sica, Universidade do Porto, Rua das Estrelas, 4150-762 Porto, Portugal\\
$^5$Department of Physics and Astronomy, Johns Hopkins University, Baltimore, MD 21218, USA\\
$^6$Astronomical Observatory of Belgrade, Volgina 7, 11060 Belgrade, Serbia
}

\begin{document}
\maketitle
\begin{abstract}

  We present a systematic study of the shape of the dust attenuation
  curve in star-forming galaxies from the far-ultraviolet to the near
  infrared ($\sim0.15-2$\mum), as a function of specific star
  formation rate (\ssfr) and axis ratio (\ba), for galaxies with and
  without a significant bulge. Our sample comprises 23,000 (15,000)
  galaxies with a median redshift of 0.07, with photometric entries in
  the SDSS, UKIDSS-LAS (and GALEX-AIS) survey catalogues and emission
  line measurements from the SDSS spectroscopic survey. We develop a
  new pair-matching technique to isolate the dust attenuation curves
  from the stellar continuum emission. The main results are: (i) the
  slope of the attenuation curve in the optical varies weakly with
  \ssfr, strongly with \ba, and is significantly steeper than the
  Milky Way extinction law in bulge-dominated galaxies; (ii) the NIR
  slope is constant, and matches the slope of the Milky Way extinction
  law; (iii) the UV has a slope change consistent with a dust bump at
  2175\AA\ which is evident in all samples and varies strongly in
  strength with \ba\ in the bulge-dominated sample; (iv) there is a
  strong increase in emission line-to-continuum dust attenuation
  (\tauvline/\tauvcont) with both decreasing \ssfr\ and increasing
  \ba; (v) radial gradients in dust attenuation increase strongly with
  increasing \ssfr, and the presence of a bulge does not alter the
  strength of the gradients. These results are consistent with the
  picture in which young stars are surrounded by dense `birth clouds'
  with low covering factor which disperse on timescales of $\sim10^7$
  years and the diffuse interstellar dust is distributed in a
  centrally concentrated disk with a smaller scaleheight than the
  older stars that contribute the majority of the red and NIR
  light. Within this model, the path length of diffuse dust, but not
  of birth-cloud dust, increases with increasing inclination and the
  apparent optical attenuation curve is steepened by the differential
  effect of larger dust opacity towards younger stars than towards
  older stars.  Additionally, our findings suggest that: (i) galaxies
  with higher star formation rates per unit stellar mass have a higher
  fraction of diffuse dust, which is more centrally concentrated; (ii)
  the observed strength of the 2175\AA\ dust feature is affected
  predominantly by global geometry; (iii) only highly inclined disks
  are optically thick.  We provide new empirically-derived attenuation
  curves for correcting the light from star-forming galaxies for dust attenuation.

\end{abstract}


\section{Introduction}\label{sec:intro}

Dust is an ubiquitous component of the baryonic Universe, potentially
condensing out of the interstellar medium as soon as the first
generation of stars has expired and provided the requisite metals 
  \citep{Meurer:1997p6688, Hughes:1998p6753, Bertoldi:2003p6722,
    Draine:2009p5441}. The presence of dust in galaxies hampers our
measurements of intrinsic galaxy properties, including stellar masses
and star formation rate (SFR), when observations are limited in
  wavelength coverage or availability of spectroscopic data
  \citep{Calzetti:2007p4471}. It also impacts on our measurements of
global population statistics, such as luminosity functions and
colour-magnitude diagrams. The strong systematic variation of dust
content with stellar mass, SFR and morphology can lead to significant
biases in results if not correctly accounted for
\citep{2004MNRAS.351.1151B, Driver:2007p6274,
  daCunha:2010p6453,Gilbank:2010p5643, Maller:2009p6273}.  Dust
expelled from galaxies and residing in galaxy halos
\citep{2005MNRAS.361L..30W, ca_ukirt} can even have
implications for the estimation of cosmological parameters
\citep{Menard:2010p6225}.

For several decades research into dust in galaxies focused around the
question of whether spiral disks are optically thick or thin. By
  studying large samples of galaxies with varying inclinations to the
  line of sight, in combination with models for the dust distribution
  in galaxies, NIR photometry, or rotation curves, the amount of
  attenuation suffered by starlight before it reaches the observer can
  be constrained \citep[e.g.][]{Giovanelli:2002p6625,
    Driver:2007p6274, Maller:2009p6273, Yip:2010p4536}.  The consensus
  is that the stellar continua in galaxy disks suffer 0 - 1 magnitude
  of attenuation in B (or SDSS $g$-band), dependent on
  inclination. Galaxy bulges may well suffer greater attenuation due
  to radial gradients in dust disks, \citet{Driver:2007p6274} find
  inclination dependent attenuations in bulges of 0.8-2.6 magnitudes in the
  B-band. 

While the extinction\footnote{We follow the standard terminology
  \citep[e.g.][]{2001PASP..113.1449C}: extinction = absorption +
  scattering out of the line-of-sight; attenuation = absorption +
  scattering in to and out of the line-of-sight caused by local and
  global geometric effects.} of starlight by dust can be measured with
relative ease along lines of sight to stars in the Milky Way and
closest galaxies (SMC and LMC), the integration of light over the
entire stellar population of a distant galaxy introduces considerable
complications. Firstly, the dust is not distributed uniformly
  throughout the galaxy: there are dense birthclouds around the
  youngest stars, there is a diffuse component exhibiting strong
  radial gradients \citep[e.g.][]{LonsdalePersson:1987p6754,
    Peletier:1995p6267, Boissier:2004p6288, MunozMateos:2009p5435},
  and there is some evidence that diffuse interstellar dust is
  distributed throughout the disk with a smaller scaleheight than that
  of the stars \citep{Xilouris:1999p6157}. Secondly, the dust does
not approximate a uniform screen in front of the stars: dust and stars
are mixed, potentially causing an additional scattering component
which can affect the shape of the attenuation curve relative to the
underlying extinction curve. Indeed, \citet{Boquien:2009p4652} suggest
that variation in extinction laws and dust-star geometry may cause the
scatter in the dust attenuation vs. reddening (e.g. IRX-$\beta$)
relation observed in star-forming galaxies, and \citet{Buat:2011p6721}
find tentative evidence for a steeper dust attenuation curve in
ordinary star-forming galaxies than found in starburst galaxies by
\citet{Calzetti:2000p4473}. 

On the other hand, the sensitivity of the shape of the dust
attenuation curve to dust grain properties and dust-star geometry
allows us to extract information on the properties and distribution of
dust in galaxies. For example, a flatter blue/UV attenuation curve is
generally expected in the case of spatially mixed dust and stars, as
the least extincted stars contribute a greater fraction of total
emergent blue light than red light \citep[e.g.][]{Gordon:1997p6428,
  2000ApJ...539..718C}. Dust features, the strongest one being at
2175\AA, can provide further constraints. Models suggest that the
2175\AA\ feature should vary in strength in galaxy spectral energy
distributions (SEDs) depending upon the properties and geometry of the
dust \citep{Granato:2000p6294, Witt:2000p6300}. Observationally, there
is some evidence for a 2175\AA\ absorption feature in ordinary
star-forming galaxies \citep{Conroy:2010p6451} and high-redshift
galaxies \citep{Noll:2009p5915}, but clear absence in local starburst
galaxies \citep{1994ApJ...429..582C} and the Small Magellanic Cloud
\citep{Pei:1992p4812,2003ApJ...594..279G}.

In this paper we measure empirically the optical-NIR and UV-to-NIR
attenuation curves, using a sample of $\sim$23,000 and $\sim$15,000
star-forming galaxies respectively, as a function of radius, specific
star formation rate (\ssfr$\equiv {\rm SFR/M^*}$), inclination angle
(\ba), for galaxies both with and without massive bulges. We achieve
this by carefully pair-matching galaxies to remove unwanted stellar
population signal.  Spectroscopic information provides a measure of
relative dust content via the Balmer emission lines, allowing us to
combine galaxy pairs in such a way as to reveal an average attenuation
curve with very high SNR.

The development of this new, robust, statistical method to measure
dust attenuation curves in galaxies with a range of different
underlying stellar populations, allows us for the first time to
observe trends in the shape of the attenuation curve as a function of
galaxy properties.  Deriving attenuation curves using different
  sized apertures, enables us to measure the strong radial gradients
  in dust opacity with \ssfr. We compare dust opacity suffered by
  emission lines to that suffered by the optical continuum, finding
  strong trends with \ssfr\ and axis ratio. Finally, we study the
  strength of the 2175\AA\ dust bump which we find to depend on
  inclination and the presence of a bulge. All these empirical results
  relate to the spatial distribution of dust in galaxies, relative to
  stars of different ages.

In Section \ref{sec:method} we present our methodology, along with a
brief summary of the formalism used to describe dust attenuation in
galaxies. In Section \ref{sec:datasets} we describe the datasets used
and provide details of the photometric information extracted from each
one. In Section \ref{sec:sample} we present the attenuation curves,
the shape of which we characterise by measuring the slope in small
wavelength ranges in Section \ref{sec:paramfits}. In this section we
also present the radial gradients, a comparison of attenuation in the
emission lines and stellar continua, and the strength of the 2175\AA\
dust bump. In Section \ref{sec:empfits} we provide an empirically
derived fitting formula to correct the spectral energy distributions
of a wide range of star-forming galaxies for dust attenuation. Finally
in Section \ref{sec:disc} we present a qualitative model for the local
and global geometry of dust in galaxies, which is consistent with the
majority of our results. 

 Throughout the paper we compare our results to the Milky Way
  (MW), large and small Magellanic cloud extinction curves (LMC and
  SMC), and the local starburst attenuation curve of
  \citet[][C00]{Calzetti:2000p4473}. We adopt the MW extinction curve
  of \citet{1989ApJ...345..245C}, which is defined over the wavelength
  range $0.125<\lambda/$\mum$<3.5$, with the correction given by
  \citet{1994ApJ...422..158O} in the optical region
  $0.3<\lambda/$\mum$<0.9$. We adopt the tabulated LMC and SMC-bar
  curves of \citet{2003ApJ...594..279G}, spline interpolating onto our
  wavelength points.

  Absolute quantities such as stellar mass and star formation rate are
  required to match pairs of galaxies, and for these we assume the
  standard $\Lambda$CDM cosmology with $\Omega_{\rm M}=0.3$,
  $\Omega_\Lambda=0.7$ and $h=0.7$. The attenuation curves are based
  on flux ratios, and therefore independent of cosmology, magnitude
  types and filter zero-points.

\section{Methodology} \label{sec:method}

The basic principle of our method is the same as that used to measure
the dust extinction along lines-of-sight through the Milky Way and
Magellanic Clouds, where the spectral energy distributions (SED) of
pairs of stars of similar spectral type are compared. Here, we
calculate the ratio of pairs of galaxy SEDs, where one galaxy is more
dusty than the other, combining many pairs to obtain a high SNR
measurement of the attenuation curve. Just as with sight-lines to
stars in the Milky Way or local galaxies, the method is only valid
where no significant bias exists between the spectral types of the
more dusty and less dusty galaxies. To ensure this, we match galaxy
pairs in gas-phase metallicity, specific star formation rate (\ssfr,
corrected for dust attenuation using the ratio of \ha\ to \hb\ luminosity), axis ratio
and redshift.

To ensure that we always divide the SED of the dustier member of the
pair with that of the less dusty member, we use the ratio of \ha\
to \hb\ luminosity as a measure of
\emph{relative} dust content. Our method is independent of, and
insensitive to, the \emph{absolute} dust content of the galaxies. In
the first part of the paper we calculate the average attenuation curve
for samples split by specific star formation rate (\ssfr), axis ratio
(\ab) and stellar surface mass density (\mustar), in circular
apertures of increasing physical size. We use the resulting
attenuation curves to study the changing shape and amplitude of the
attenuation curve with each of these properties. In the second part of
the paper we provide an empirically derived prescription for the
attenuation of starlight by dust in external galaxies as a function of
their \ssfr, \ba\ and \mustar.

\subsection{Background to dust attenuation laws}\label{sec:bg}
The many different formalisms used to describe dust curves
necessitate a brief overview of the subject to allow orientation of
our work relative to other papers in the literature.

 Light emitted by stars in external galaxies is affected by dust
  in three different ways: (i) absorption; (ii) scattering out of the
  line-of-sight to the observer; (iii) scattering into the
  line-of-sight to the observer. The first two combined are usually
termed ``extinction''. In this paper we use the term ``attenuation''
to mean the average loss of light after integration over all
lines-of-sight to light sources throughout the entire galaxy,
encompassing all effects arising from the local and global geometric
configuration of the dust and stars\footnote{ Some other papers refer to
  ``single-line-of-sight'' attenuation, which includes effects
  (i)-(iii) but excludes global geometric effects.}. We use the term
``effective optical depth'' to describe the final observed optical
depth, which is related to the attenuation through $A_\lambda =
1.086\uptau_\lambda$. Because blue light is scattered and absorbed
more than red light by dust grains, an overall ``reddening'' of the
light is observed. In this section, we introduce the basic formalism
used to describe the effects of dust on the integrated light from
galaxies.  \citet{2001PASP..113.1449C} gives an extensive review of
the literature up until a decade ago, although note that the
terminology differs in places from that adopted here.

The attenuation $A_\lambda$, in magnitudes, at a
given wavelength $\lambda$ is given by:
\begin{eqnarray}
A_\lambda &=& -2.5\log_{10} \left(\frac{I_{\lambda}^o}{I_{\lambda}^e}\right)\\
&=& -2.5\log_{10} \left[\exp(-\uptau_\lambda)\right]\\
&=& 1.086\uptau_\lambda
\end{eqnarray}
where $I_{\lambda}^o$ is the observed luminosity, $I_{\lambda}^e$
is the intrinsic (emitted) luminosity of the source, and $\uptau_\lambda$ is the
effective optical depth of the dust.  Ultimately it is this quantity
that we wish to know in order to correct the SEDs of galaxies for dust
attenuation.

Observationally, the measurement of the overall amplitude of
extinction and reddening of the light are usually separate tasks. It
is therefore practical to separate the attenuation into two
components, the amplitude (usually specified in the $V$-band,
$\uptau_V$) and the shape of the attenuation curve
($Q_\lambda$)\footnote{Conversions between different parameterisations
  of attenuation curve shapes are: $\frac{A_\lambda}{E(B-V)}
  \frac{E(B-V)}{A_V} = \frac{k_\lambda}{k_V} =
  \frac{\uptau_\lambda}{\uptau_V} = Q_\lambda$}:
\begin{equation}
\uptau_\lambda  =  \uptau_V Q_\lambda
\end{equation}
where $Q_V=1$. In the case of extinction, \ql\ is a function of the
distribution of grain composition, size and shape.  When studying
  the integrated light from external galaxies \ql\ is additionally a
  function of the \emph{local geometry} of the dust and stars, and the
  \emph{global geometry} of the galaxy (bulge and disk, and
  inclination).

 Extinction curves have been measured along sight lines toward
  stars in the MW, LMC and SMC. While the average curves of the three
  galaxies are apparently entirely distinct, the individual
  sight-lines in all three galaxies show a continuum of properties
  from steep slopes with weak 2175\AA\ dust bumps like the average
  SMC-bar curve, to shallow slopes with strong bumps like the MW curve
  \citep{2003ApJ...594..279G}. Unfortunately, current data is not good
  enough to determine robustly the ``hidden parameter(s)'' responsible
  for this variation, although grain destruction by local star
  formation is often assumed.

The  attenuation suffered by light from stars in nearby
starburst galaxies, where the unattenuated stellar continuum varies
relatively little between galaxies, has been measured empirically from
the ultra-violet (UV) to near infra-red (NIR) in a series of papers by
D. Calzetti and collaborators \citep{Kinney:1994p5666,
  1994ApJ...429..582C, Calzetti:1997p5089, Boker:1999p4704,
  Calzetti:2000p4473}. The attenuation curve was found to be
``greyer'' (i.e. flatter) in the optical than the Milky Way extinction
curve, qualitatively as expected from the geometrical effects of mixed
dust and stars. Additionally, no evidence was found for a strong
``2175\AA'' dust feature which is observed in the majority of extinction
curves measured along lines-of-sight in the MW \citep[but see ][ and
Section \ref{sec:uv}]{Noll:2009p5915,Conroy:2010p6451}.

\subsection{Measuring dust attenuation curves from
  pair-matched galaxy samples}\label{sec:formalism}

An observed galaxy SED can be described as an intrinsic SED of unit
stellar mass ($I_{\lambda}^e$), multiplied by factors to account for
the total stellar mass, cosmological distance of the galaxy and
internal dust attenuation. Ideally the two factors of normalisation
should be estimated from a region of the SED that is not attenuated by
dust.  In this paper, the longest wavelength point available to us is
the K-band at 2.2\mum. We therefore normalise each galaxy by its
K-band flux and retain a small correction factor throughout this work
which allows for the small amount of dust attenuation in the K-band
(\dtk) during the fitting procedures. The normalised SED ($F_\lambda$)
can be written as:
\begin{equation}\label{eqn:fluxnorm}
F_\lambda = \frac{I_\lambda^e A}{f_{\mathsf{K}}} \exp(-\uptau_\lambda)
\end{equation}
where $A$ accounts for stellar mass and cosmological distance,
$\exp(-\uptau_\lambda)$ for internal dust attenuation, and
$f_{\mathsf{K}}$ is the observed K-band flux. 

Directly from Eqn. \ref{eqn:fluxnorm}, the ratio of a pair of
normalised galaxy SEDs (labelled 1 and 2), where both members of the
pair have the same intrinsic SED but different dust contents and
absolute normalisations, can be written after some algebra as:
\begin{equation}
\left(\frac{F_{1}}{F_{2}}\right)_\lambda = \exp(-\Delta\uptau_\lambda + \Delta\uptau_{\mathsf{K}})
\end{equation}
where $\Delta\uptau_\lambda \equiv \uptau_{\lambda,1} -
\uptau_{\lambda,2}$ and $\Delta\uptau_{\mathsf{K}}$ accounts for the
small, unknown extinction in the K-band. The geometric
mean\footnote{The geometric mean of flux ratios is preferred for dust
  curves, as this is equivalent to the arithmetic mean of the log of
  flux ratios, which is directly related to optical depth. } of a
sample of flux ratios, where in each pair galaxy $F_{1}$ is more dusty
than galaxy $F_{2}$, results in a dust attenuation curve 
  normalised to zero at K which we denote $\mathcal{T}_\lambda$ to
  distinguish it from the true attenuation:
\begin{eqnarray}
  \mathcal{T}_\lambda &=& \frac{1}{n} \sum_{i=1}^n\ln\left(\frac{F_{1,i}}{F_{2,i}}\right)_\lambda \label{eqn:Fobs}\\
  &=& -\left<\Delta\uptau_{\mathsf{V}}\right>\, Q_\lambda + \left<\Delta\uptau_{\mathsf{K}}\right>\label{eqn:Ftheory}
\end{eqnarray}
where $n$ is the number of pairs of galaxies in the sum,
$\left<\Delta\uptau_\lambda\right>$ is the arithmetic mean of the
difference in dust attenuation between galaxies 1 and 2, and we have
expanded \dtl\ into the \emph{amplitude} of attenuation at 5500\AA\ (\dtv)
and the \emph{shape} of the attenuation curve (\ql), as
described above.  From now on we will drop the brackets and write
simply $\Delta\uptau$ as the effective optical depth averaged over
multiple galaxy pairs. 

Some points to note about the method are:
\begin{itemize}
\item Because we are working with broadband photometry, strictly
  Eqn. \ref{eqn:fluxnorm} should include an integration over the
  filter response function ($R_\lambda$). In the case that \ql\ varies
  with $\lambda$ over the wavelength range of the filter, it can be
  shown that we measure $\int Q_\lambda R_\lambda d\lambda$ rather
  than \ql, even in the case that $I_\lambda^e$ varies with
  $\lambda$. Due to the generally weak variation of attenuation curves
  with $\lambda$, this distinction is only important in the UV
  (Section \ref{sec:uv}) and we have therefore omitted the integrand
  for clarity.
\item Eqn. \ref{eqn:Ftheory} holds only in the case where \ql\ is the
  same for both members of the pair i.e. when there is no trend
  between \ql\ and dust content. See Appendix \ref{app:method} and
  Figure \ref{fig:app1} for the result of tests showing that this is
  the case in our samples.
\item \ql\ may be different for different pairs in the sum, in this
  case the measured \ql\ will be the \dtv\ weighted mean \ql\ of the
  sample.
\item The method does not require that the intrinsic stellar
  populations are exactly the same in each galaxy pair, any residual
  stellar population remaining after division of the two SEDs will
  only add noise to the final measurement. However, it does require
  that there is no systematic correlation between the residual and the
  dust content of the galaxy. See Appendix \ref{app:method} and Figure
  \ref{fig:app2} for an example of how such a correlation can occur
  using photometry to estimate \ssfr\ rather than the \ha\ emission
  line.
\item The method is almost, but not quite, the same as that employed
  by Calzetti et al. for starburst galaxies:
  \begin{itemize}
  \item They use a single ``dust-free'' composite spectrum with which
    to compare dustier composite spectra. Because we use a whole range
    of types of galaxies, with many different shapes of SEDs, we first
    define strict galaxy pairs, the SEDs of which we divide before
    combining galaxies into samples. This approach makes our method
    more robust, as each pair is treated individually meaning outliers
    can be easily identified and removed. Additionally, errors can be
    estimated on the resulting attenuation curves.
  \item Each of our binned samples of galaxies contains 14-40 times
    more galaxies than the original starburst sample of
    \citet{1994ApJ...429..582C} which contained 39 galaxies.
  \item We define the attenuation curve in terms of the shape of the
    curve (\ql), and the amplitude of attenuation in the \emph{stellar
      continuum} at $V$ (\tauv). This differs in formalism to that of
    Calzetti et al. who directly incorporate the amplitude of dust
    attenuation in the emission lines into the attenuation curve for
    the stellar continuum.
  \end{itemize}
\end{itemize}

\subsection{Pair-matching procedure} \label{sec:procedure}

In order to achieve the requirement for a close match in intrinsic
stellar population, we identify pairs of galaxies that match in
metallicity, \ssfr, \ba\ and redshift. The
restrictions on metallicity and \ssfr\ are to ensure that both
galaxies have similar intrinsic SEDs.  The restriction on \ba\ also
minimises the effect of seeing different stellar populations in
inclined vs. face-on galaxies simply due to changes in
transparency. 

In order to identify galaxies which are more- or less-dusty we use the
ratio of \ha\ to \hb\ luminosity measured from the SDSS spectra. At
this stage, the sole purpose of using this line ratio as a
measure of dust content is to ensure that, in as many pairs as
possible, the denominator SED is more dusty than the numerator in
Eqn. \ref{eqn:Fobs}. Only in this case will we measure an overall
attenuation.

Because the \ha\ and \hb\ emission lines are measured within a small
fixed angular aperture (see Section \ref{sec:sdss}), and radial
gradients in dust exist, both galaxies in the pair must have the same
redshift to minimise the numbers of pairs in which the denominator SED
is in fact less dusty than the numerator simply because the lines are
measured from different regions of each member of the pair. The
restriction on redshift also means that galaxy pairs do not need to be
corrected for any relative shifting of band-passes with redshift, as
the photometric bands probe the same rest-frame wavelength of the SED
in each member of the pair.

The procedure is as follows. The galaxies are ordered by increasing
ratio of \ha\ to \hb\ luminosity, and starting from the least dusty
galaxy in the sample, the dustiest suitable partner is found using the
criteria mentioned above.  We require that there is a significant
difference between the dust content of galaxy-pairs, as measured by
the difference in emission line optical depth. This maximises the
signal-to-noise of the final attenuation curves, by excluding pairs
which contribute only noise and no signal.

To summarise the pair-matching criteria:
\begin{itemize}
\item $|\Delta \log {\rm O/H}| < 0.05$
\item $|\Delta \log \psi_s| <0.1$
\item $|\Delta (b/a) | <0.1$
\item $|\Delta z| <0.01$
\item \Dtauvline$>0.2$
\end{itemize}
Details of how the derived quantities are measured are given in
Section \ref{sec:datasets}. If a suitable pair is not found, the
initial galaxy is removed from the sample, and the algorithm moves
onto the next galaxy. Once a pair has been identified, both members
are removed from the sample, thus each galaxy can only contribute once
to the attenuation curve.

Once all possible pairs have been identified within a particular
sample, the galaxy SEDs are normalised by their K-band flux and the
flux ratios calculated in each band for each pair. These flux ratios
are combined using a geometric mean (Eqn. \ref{eqn:Fobs}). A preliminary
stack is made, outlying pairs with flux ratios which are in the upper
or lower 1\% of the distribution in any band are removed from the
sample, and the attenuation curves recalculated using the cleaned
sample. 

We verified that the precise details of the pair-matching criteria
given above are unimportant to the final results by varying the widths
of the bins and repeating our analysis.  Tests showed that the
  best SNR was achieved when the difference between the dustiness of
  each pair of galaxies is maximised. We verified that selecting pairs
  to have approximately the same \Dtauvline\ instead does not
  significantly alter the results (see Figure \ref{fig:app1}). 
Additionally, we repeated our analysis replacing the pair-match on
metallicity with a pair-match on stellar mass or stellar surface mass
density. These latter quantities are global measurements from the
total SDSS photometry, whereas metallicity is measured from the
central 3\arcsec\ of the galaxies. Once again, our results were
generally found not to be sensitive to exactly which properties we
used in the pair-matching, presumably by virtue of the observed strong
relations between these parameters. Where differences were found they
were consistent with our expectation that dustier galaxies have higher
metallicities and therefore redder intrinsic SEDs, therefore a failure
to match on metallicity results in a slight bias in the attenuation
curves as discussed in the previous subsection.

We note that no correction is applied for the shifting of the
photometric bands with redshift (K-correction). We attempted to
correct each SED ratio for this shifting, using an iterative approach
in which the previously derived attenuation curve was used to
determine the small corrections to apply to each flux ratio. However,
this was found to make no difference to the results. The smoothness of
the attenuation curves compared to individual galaxy SEDs, the similar
redshift of both members of the pairs, and the relatively small
redshift range covered by the majority of the galaxies (70\% with
$0.03<z<0.09$) likely all help to minimise the effects. For the final
derivation of the attenuation curve in Section \ref{sec:empfits} the
K-correction is irrelevant as all pairs are combined at their correct
rest-frame wavelength.

\section{The datasets}\label{sec:datasets}

\begin{figure*}
\includegraphics[scale=0.9]{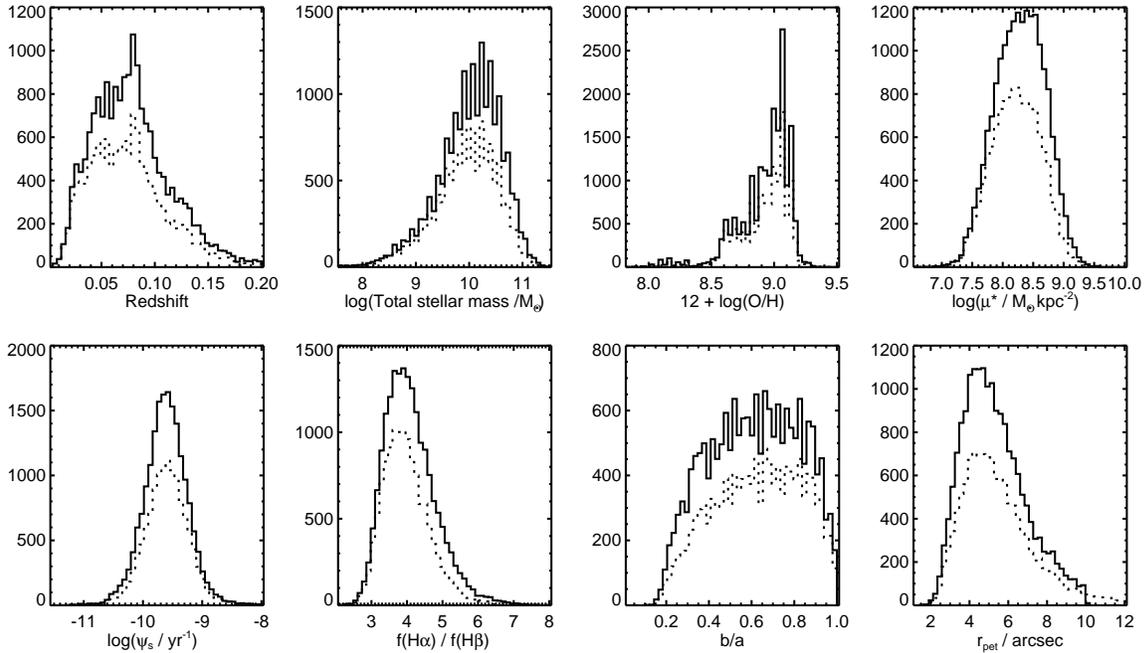}
\caption{Distribution of basic galaxy properties for the full sample
  (full line) and UV sample (dotted line). Individual panels show (a)
  redshift; (b) total stellar mass measured from SED fitting to the
  5-band SDSS photometry; (c) gas-phase metallicity in the central
  3\arcsec; (d) stellar surface mass density, calculated from the
  physical size measured in the $z$-band and total stellar mass; (e)
  specific star formation rate (\ssfr) measured within the central
  3\arcsec; (f) ratio of \ha\ to \hb\ luminosity; (g) axis ratio,
  measured in the $r$-band; (h) $r$-band Petrosian radius in arcsec.
}\label{fig:distbn}
\end{figure*}

In order to measure the dust attenuation curves from the UV to the
NIR, in a wide range of star-forming galaxies, we combine three of the
largest modern day astronomical datasets: the Sloan Digital Sky Survey
(SDSS) in the optical; the UKIRT Infrared Deep Sky Survey (UKIDSS) in
the near-infrared (NIR); the Galaxy Evolution Explorer (GALEX) in the
ultra-violet (UV).

All fluxes are corrected for Galactic extinction using the values
provided in the respective catalogs which are based upon the
\citet{1998ApJ...500..525S} dust emission maps and the MW extinction
curve of \citet{1989ApJ...345..245C}, with the small correction in the
optical wavelength regime of \citet{1994ApJ...422..158O}. The GALEX
catalog provides E(B-V) which we convert into extinction for the UV
fluxes using the MW extinction curve.

Figure \ref{fig:distbn} shows the distributions of redshift and
physical properties for the galaxies used in this paper.

\subsection{The SDSS spectroscopic survey}\label{sec:sdss}

The SDSS is an optical photometric and spectroscopic survey of local
galaxies. The final data release \citep[DR7,][]{Abazajian:2009p5430}
includes 9380 sq. degrees with spectroscopic coverage, targeting
nearly $10^6$ galaxies with Petrosian $r$-band magnitudes
$<$17.77. The spectra have good signal-to-noise and moderate
resolution which allows the deconvolution of stellar continuum from
nebular emission that is crucial to accurately measure the \ha\
and \hb\ line strengths. Such a deconvolution has been carried out by
\citet{2004MNRAS.351.1151B} and \citet{2004ApJ...613..898T} and the
emission-line measurements are available
online\footnote{http://www.mpa-garching.mpg.de/SDSS}. The SDSS provides the $u,g,r,i,z$
aperture photometry used in this paper, with central effective
wavelengths ($\lambda_{\rm eff}$) of 3546, 4670, 6156, 7471 and
8918\AA\ respectively
\citep{Schneider:1983p5662,1996AJ....111.1748F}. The SQL query used to
extract the parameters from the SDSS Catalog Archive Server is
provided in Appendix \ref{app:sql}. A total of 797026 unique galaxies
are extracted from the catalogue, with 552577 that have spectral
per-pixel signal-to-noise ratio SNR$>6$ in the $g$-band, are in the
SDSS-MPA database and have ``best'' magnitudes within the formal
survey requirements ($14.5<m_r<17.77$), where $m_r$ is the $r$-band
Petrosian magnitude corrected for Galactic extinction.

Because we are interested in the ISM properties of star-forming
galaxies, we use the \nii/\ha\ and \oiii/\hb\ emission line ratios
\citep[BPT ratios]{1981PASP...93....5B} to remove galaxies which show
signs of an Active Galactic Nucleus (AGN), selecting only those
galaxies that lie below the demarcation line of
\citet{2003MNRAS.346.1055K}. We also select only those galaxies with
all 4 lines measured at $>$$3\sigma$ confidence, after scaling the
formal errors by the scaling factors suggested on the SDSS-MPA
webpages. This results in 140737 galaxies.
 
The parameters that we extract from the SDSS catalogues are:
\begin{description}
\item[{\bf Emission line dust (\tauvline):}] the \ha\ and \hb\ line
  fluxes, measured after accurate subtraction of the underlying
  stellar continuum, give the total dust attenuation suffered by the
  emission lines at 5500\AA. We assume the dust attenuation curve for
  emission lines of \citet{Wild:2011p6538}:
  $\uptau_{\mathsf{V,line}}=3.226\ln(f_{\rm H\alpha}/f_{\rm
    H\beta})_{\rm obs} - 3.226\ln(f_{\rm H\alpha}/f_{\rm H\beta})_{\rm
    int}$. This was measured through a comparison of the luminosity of
  mid-IR and optical emission lines. In the case that the dust free
  intrinsic line ratio ($(f_{\rm H\alpha}/f_{\rm H\beta})_{\rm int}$)
  is the same for both galaxies in a pair, this term cancels when the
  ratio of two galaxy SEDs is calculated\footnote{The intrinsic line
    ratio depends upon $T_e, n_e$ and $Z$ and therefore varies
    slightly with galaxy type and dust content. Over most of parameter
    space relevant to galaxies, this causes a variation in true
    optical depth for a given observed line ratio at a level of less
    than 10\%. However, for massive, high metallicity galaxies with
    low dust contents, optical depths can be up to a factor of 2 lower
    than predicted from the observed line ratio assuming a standard
    intrinsic ratio of 2.87. This effect can not bias our attenuation curves, but
    may act as an additional source of noise in certain bins. }.
\item[{\bf Stellar mass (\Mstar):}] measured from the 5-band SDSS
  Petrosian photometry in units of solar mass using a Bayesian fit to
  a library of stochastic star formation histories similar to those
  used in \citet{Gallazzi:2005p6450} are publicly
  available\footnote{http://www.mpa-garching.mpg.de/SDSS/DR7/Data/stellarmass.html}. Stellar
  masses are calculated from both total and fibre photometry.
\item[{\bf Star formation rate:}] calculated from the dust corrected
  \ha\ luminosities, using the conversion factor of
  \citet{1998ARA&A..36..189K} and \tauvline\ given above. We do not
  use the SFR of \citet{2004MNRAS.351.1151B} as this assumes a
  different dust attenuation curve. The precise conversions assumed to
  calculate SFR from \ha\ luminosity do not affect our final
  results. Specific star formation rate (\ssfr) is calculated as
  SFR/\Mstar$_{\rm fib}$. It was found that \ssfr\ calculated from the
  5-band SDSS photometry alone was not a good enough estimate of the
  intrinsic spectral shape of the galaxies (see Appendix
  \ref{app:method} and Figure \ref{fig:app2}).
\item[{\bf Gas phase metallicity:}] measured from a combination of
  the strongest optical emission line fluxes by
  \citet{2004ApJ...613..898T}, using the method of Charlot \&
  Longhetti (2001).
\item[{\bf Stellar surface mass density (\mustar):}] calculated as
  \Mstar$/(2 \pi r^2)$ where $r$ is the physical size in kpc of the
  radius which contains 50\% of the $z$-band Petrosian flux.  The
  $z$-band probes the older stellar population and thus gives greater
  weight to galaxy bulges than a bluer band would do.
\item[{\bf Size:}] we adopt the $r$-band Petrosian radius as a measure
  of the size of the galaxies. This is the radius at which the ratio
  of local surface brightness to mean surface brightness is 0.2. The
  $r$-band probes both young and old populations with almost equal
  weight in star forming galaxies, and therefore provides a good
  average measure of galaxy size. 
\item[{\bf Inclination (\ba):}] the exponential profile axis ratio,
  measured in the $r$-band which has the highest SNR of all red SDSS
  bands. The majority of our star-forming sample are expected to be
  disk galaxies, and therefore the exponential profile fit is the most
  appropriate. A detailed study of the link between \ba\ and the
  physical shape of galaxies in the SDSS is given by
  \citet{Maller:2009p6273}.
\item[{\bf Recent star formation history:}] we remove all galaxies
  with post-starburst spectral features using the method of
  \citet{wild_psb}. These galaxies have unusual SED shapes which may
  correlate with dust, thus causing problems during our pair-matching
  procedure.
\end{description}

A feature of SDSS is that the spectra are observed through 3\arcsec\
diameter optical fibres, therefore probing a small central fraction of
the total light of the galaxies. For the main focus of this work, this
``aperture bias'' is irrelevant as we use the spectroscopic
information primarily to identify relative dust content.  Where
necessary, i.e. when we compare fibre and photometric quantities, we
use photometry extracted in $\sim3$\arcsec\ apertures.

\subsection{The UKIDSS Large Area Survey}

We match the SDSS catalogue resulting from the previous subsection
with the seventh data release of the ongoing UKIDSS Large Area Survey
\citep[LAS,][]{2007MNRAS.379.1599L} to obtain NIR Y, J, H, K photometry
for the galaxies ($\lambda_{\rm eff} = 1.0305, 1.2483, 1.6313,
2.2010\mu$m). UKIDSS uses the UKIRT Wide Field Camera
\citep[WFCAM,][]{2007A&A...467..777C}, the photometric system is
described in \citet{2006MNRAS.367..454H}, the calibration in
\citet{Hodgkin:2009p4949} and the science archive in
\citet{2008MNRAS.384..637H}. We locate all objects with SDSS-UKIDSS
cross-matches within 2\arcsec\ and select the nearest neighbour if
more than one match exists.  We exclude all galaxies for which
photometry has been ``de-blended''. Problems with the UKIDSS
pipeline mean that the flux in such objects is overestimated
\citep{Hill:2010p4946}. The SQL query used to extract the parameters
from the UKIDSS online database, and cross-match the SDSS and UKIDSS
catalogues, is provided in Appendix \ref{app:sql}.  The UKIDSS-SDSS
cross match results in our primary optical-NIR sample of 22902
star-forming galaxies.

\subsection{The GALEX All-Sky Imaging Survey}

There is considerable interest in the shape of the attenuation curve
in the rest-frame UV wavelength range, particularly for high-redshift
galaxy studies. Therefore, we build a second, smaller galaxy sample in
which we match our optical-NIR galaxy sample to the GALEX All-Sky
Imaging Survey catalogue \citep[AIS,][]{2005ApJ...619L...1M}. The SQL
query used to query the GALEX-MAST
database\footnote{http://galex.stsci.edu/casjobs/} is shown in
Appendix \ref{app:sql}. We use the table xSDSSDR7 to identify the
closest (rank 1) galaxies within 2''
\citep{Budavari:2009p5346}\footnote{http://galex.stsci.edu/doc/CASJobsXTutorial.htm},
and retain only those galaxies with both NUV ($\lambda_{\rm
  eff}=$2267\AA) and FUV ($\lambda_{\rm eff}=$1516\AA) fluxes measured
at greater than 3$\sigma$ confidence. This second UKIDSS-SDSS-GALEX
sample contains 15305 galaxies.

\subsection{Aperture photometry}

The precision to which extremely large modern surveys allow us to
measure stacked quantities demands the use of extreme care in the
handling of the data if tiny systematics are not to dominate the final
result. While our pair-matching methodology eliminates absolute
offsets in photometric scales between different bands and datasets,
relative offsets can cause noticeable errors. For example, radial
gradients in the stellar populations and dust contents of galaxies
mean that the use of different photometric apertures at different
wavebands would result in incorrect results. Fortunately both the SDSS
and UKIDSS catalogues provide sets of aperture photometry measured
with various circular aperture sizes.  For each galaxy we interpolate
these annular flux densities onto a grid fixed relative to the
galaxy's $r$-band Petrosian radius (\Rpet). This allows us to
calculate dust attenuation curves at fixed cumulative fractions of
\Rpet, namely 25\%, 35\%, 50\%, 70\%, 90\% and 100\%. The large PSF of
GALEX  ($\sim$5\arcsec\ FWHM) prevents a similar such analysis of the
UV data, for which we are restricted to a single ``total'' magnitude.

Annular aperture photometry is provided in the SDSS catalog at radii
of 0.23, 0.68, 1.03, 1.76, 3, 4.63, 7.43 and 11.42 arcsec, which we
convert into cumulative fluxes. Similarly, cumulative aperture
photometry is provided in the UKIDSS catalog at radii of 0.5, 0.71, 1,
1.41, 2, 2.83, 4, 5, 6, 7, 8 and 10 arcsec. For each galaxy, we
linearly interpolate both sets of cumulative fluxes onto the grid of
fixed fractions of \Rpet\ given above. Although linear interpolation
between measured apertures is not sufficiently accurate to obtain
absolute magnitudes, it is entirely sufficient for our purpose of
obtaining accurate dust curves.

The use of fixed-aperture photometry does not account for a changing
point-spread-function (PSF) as a function of band or survey. However,
the atmospheric seeing conditions under which the SDSS and UKIDSS
photometry were obtained were generally excellent, and the PSF is
typically smaller than 0.35\Rpet\ (the smallest aperture size that we
analyse in Section \ref{sec:paramfits}) which is $\gsim$1.5\arcsec\
for galaxies in our sample. The small number of very extended galaxies
with their Petrosian radius larger than the maximum apertures
extracted from the catalogs were removed from the sample (see Section
\ref{sec:sdss} for a definition of Petrosian radius).

\section{The attenuation curves}\label{sec:sample}

\begin{table*}
  \centering
  \caption{\label{tab:sample} Number of pairs, median redshift, median
    \ssfr, median \ba, and mean \Dtauvline\ for each SDSS+UKIDSS
    sample. }
  \vspace{0.2cm}
  \begin{tabular}{cccccc}\hline\hline
    & ${\rm n_{pair}}$ & $\bar{z}$ & $\overline{\psi_s}$ &
    $\overline{(b/a)}$ & $\Delta\uptau_{\mathsf{V,line}}$\\ \hline 

     $\mu^*<3\times10^8$ & & & & & \\ \hline
All galaxies & $3431$ & $0.067$ & $-9.62$ & $0.62$ & $0.60\pm_{0.00}^{0.00}$\\
$-11.00<\log(\psi_S/{\rm yr^{-1}})<-9.90$ & $541$ & $0.060$ & $-10.03$ & $0.55$ & $0.65\pm_{0.01}^{0.02}$\\
$-9.90<\log(\psi_S/{\rm yr^{-1}})<-9.75$ & $547$ & $0.069$ & $-9.82$ & $0.61$ & $0.62\pm_{0.01}^{0.01}$\\
$-9.75<\log(\psi_S/{\rm yr^{-1}})<-9.62$ & $578$ & $0.072$ & $-9.68$ & $0.66$ & $0.59\pm_{0.01}^{0.01}$\\
$-9.62<\log(\psi_S/{\rm yr^{-1}})<-9.46$ & $689$ & $0.072$ & $-9.54$ & $0.64$ & $0.60\pm_{0.01}^{0.01}$\\
$-9.46<\log(\psi_S/{\rm yr^{-1}})<-9.25$ & $613$ & $0.069$ & $-9.37$ & $0.65$ & $0.56\pm_{0.01}^{0.01}$\\
$-9.25<\log(\psi_S/{\rm yr^{-1}})<-7.90$ & $366$ & $0.057$ & $-9.12$ & $0.60$ & $0.53\pm_{0.02}^{0.02}$\\
 
$0.0<b/a<0.4$ & $896$ & $0.056$ & $-9.68$ & $0.34$ & $0.69\pm_{0.01}^{0.01}$\\
$0.4<b/a<0.6$ & $911$ & $0.065$ & $-9.60$ & $0.55$ & $0.60\pm_{0.01}^{0.01}$\\
$0.6<b/a<0.8$ & $697$ & $0.074$ & $-9.60$ & $0.72$ & $0.55\pm_{0.01}^{0.01}$\\
$0.8<b/a<1.0$ & $721$ & $0.077$ & $-9.61$ & $0.87$ & $0.55\pm_{0.01}^{0.01}$\\
 
\hline $\mu^*>3\times10^8$ & & & & &  \\ \hline
All galaxies & $1877$ & $0.083$ & $-9.75$ & $0.62$ & $0.72\pm_{0.01}^{0.01}$\\
$-11.00<\log(\psi_S/{\rm yr^{-1}})<-9.90$ & $503$ & $0.079$ & $-10.09$ & $0.58$ & $0.70\pm_{0.02}^{0.01}$\\
$-9.90<\log(\psi_S/{\rm yr^{-1}})<-9.75$ & $334$ & $0.086$ & $-9.82$ & $0.59$ & $0.70\pm_{0.02}^{0.02}$\\
$-9.75<\log(\psi_S/{\rm yr^{-1}})<-9.62$ & $290$ & $0.092$ & $-9.69$ & $0.62$ & $0.72\pm_{0.02}^{0.03}$\\
$-9.62<\log(\psi_S/{\rm yr^{-1}})<-9.46$ & $314$ & $0.086$ & $-9.55$ & $0.65$ & $0.69\pm_{0.02}^{0.02}$\\
$-9.46<\log(\psi_S/{\rm yr^{-1}})<-7.90$ & $359$ & $0.085$ & $-9.32$ & $0.72$ & $0.73\pm_{0.02}^{0.03}$\\
 
$0.0<b/a<0.4$ & $485$ & $0.071$ & $-9.88$ & $0.35$ & $0.75\pm_{0.02}^{0.02}$\\
$0.4<b/a<0.6$ & $555$ & $0.084$ & $-9.72$ & $0.55$ & $0.73\pm_{0.02}^{0.02}$\\
$0.6<b/a<0.8$ & $399$ & $0.090$ & $-9.69$ & $0.72$ & $0.70\pm_{0.02}^{0.02}$\\
$0.8<b/a<1.0$ & $349$ & $0.089$ & $-9.70$ & $0.88$ & $0.63\pm_{0.02}^{0.02}$\\
\hline 

  \end{tabular}
  
\end{table*}

\begin{table*}
  \centering
  \caption{\label{tab:data} Averaged attenuation curves, normalised to
    zero at $K$ ($\mathcal{T}_\lambda$, Eqn. \ref{eqn:Fobs}), for each
    sample studied in this paper [full table available online]. } 
  \vspace{0.2cm}
  \begin{tabular}{cccccccccc}\hline\hline
    Sample & Aperture & $u$ & $g$ & $r$ & $i$ & $z$ & Y & J & H \\
\hline
All galaxies&0.25\Rpet&$-0.493$&$-0.362$&$-0.261$&$-0.209$&$-0.161$&$-0.123$&$-0.074$&$-0.042$\\
&0.35\Rpet&$-0.473$&$-0.349$&$-0.250$&$-0.200$&$-0.153$&$-0.116$&$-0.068$&$-0.038$\\
&0.50\Rpet&$-0.455$&$-0.336$&$-0.239$&$-0.192$&$-0.146$&$-0.110$&$-0.062$&$-0.035$\\
&0.70\Rpet&$-0.439$&$-0.325$&$-0.231$&$-0.186$&$-0.141$&$-0.105$&$-0.056$&$-0.033$\\
&0.90\Rpet&$-0.431$&$-0.319$&$-0.227$&$-0.183$&$-0.139$&$-0.101$&$-0.054$&$-0.032$\\
&1.00\Rpet&$-0.428$&$-0.316$&$-0.225$&$-0.182$&$-0.138$&$-0.100$&$-0.054$&$-0.032$\\
&3\arcsec&$-0.484$&$-0.355$&$-0.253$&$-0.202$&$-0.154$&$-0.119$&$-0.069$&$-0.038$\\

\hline
  \end{tabular}
  
\end{table*}

\begin{figure*}
\includegraphics[scale=0.5]{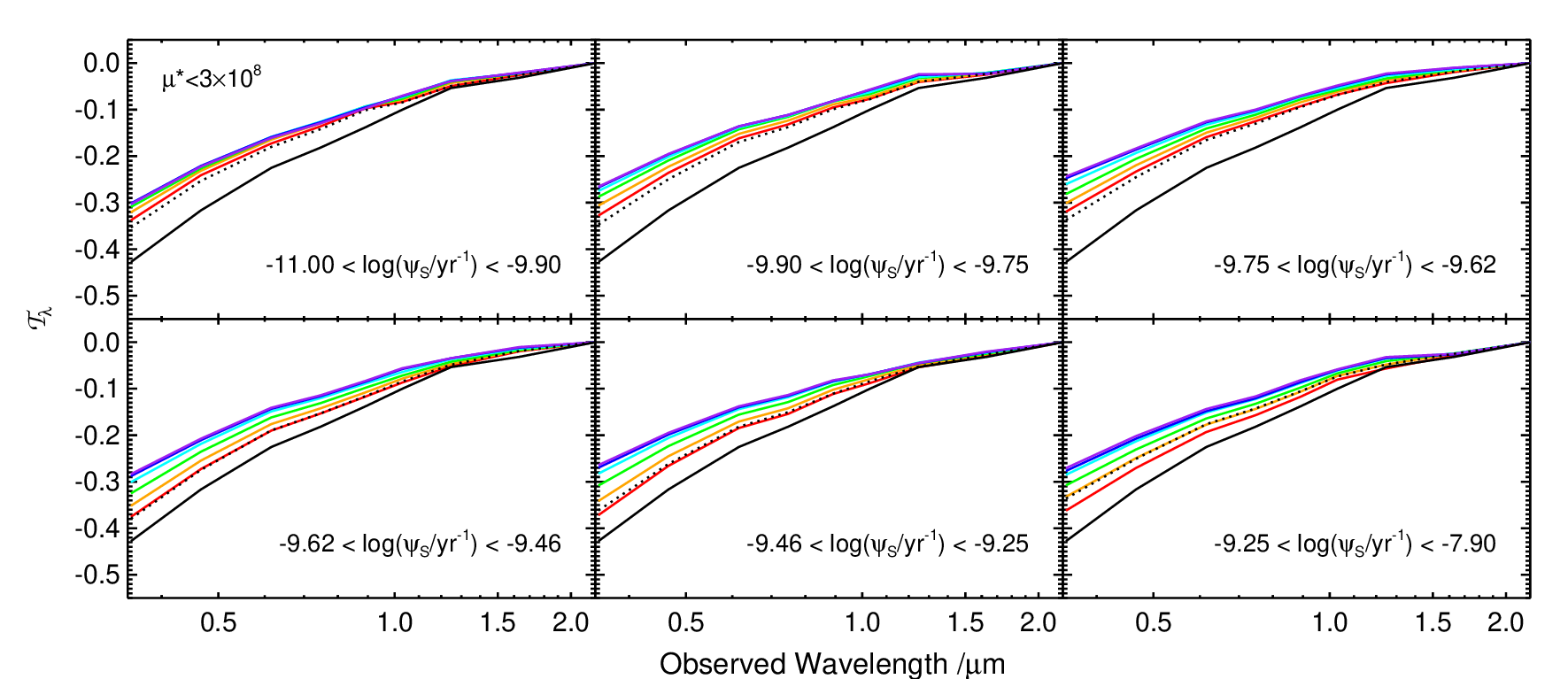}
\includegraphics[scale=0.5]{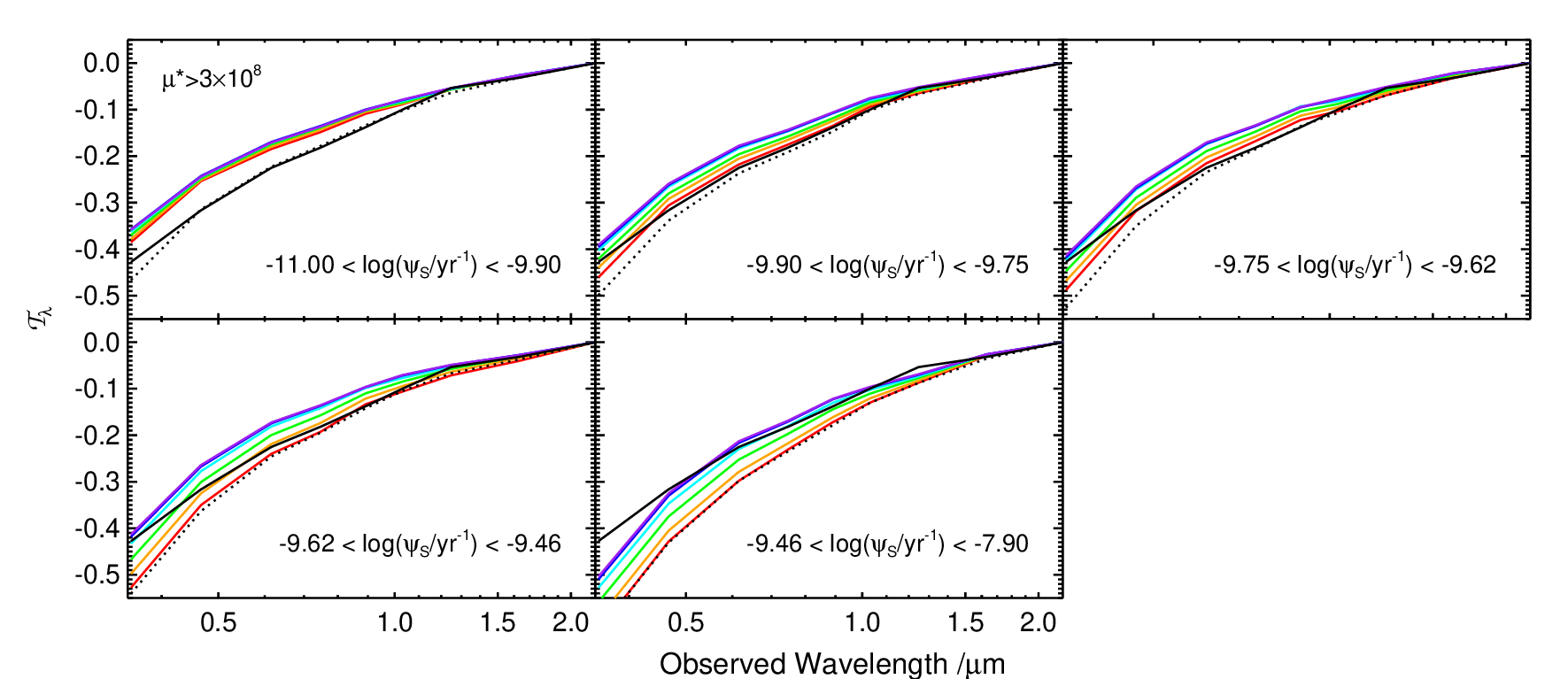}
\caption{Averaged dust attenuation curves, normalised to zero at K
  (Eqn. \ref{eqn:Fobs}). Each panel shows galaxies in bins of \ssfr,
  \emph{top:} with $\log\mu*<8.3$, and \emph{bottom:} with
  $\log\mu*>8.3$. Note that the overall normalisation of the curves in
  each panel depends on the range in dust content of the galaxies that
  make up each sample, and is thus somewhat arbitrary (see text of
  Section \ref{sec:sample}).  Different colour lines indicate
  different physical aperture sizes, from red to purple: 25, 35, 50,
  70, 90 and 100\% of the Petrosian r-band radius. In some panels the
  curves show a progressively lower normalisation with decreasing
  aperture size. This is caused by larger dust contents in the central
  regions of the galaxies.  The dotted black line shows the
  attenuation curve when a 1.76\arcsec\ radius angular aperture is
  used, close to the aperture through which the SDSS spectra are
  taken. To aid comparison between panels, the full black line is the
  attenuation curve measured at 90\%\Rpet\ using all galaxies, and is
  the same in each panel.   }\label{fig:curves_ssfr}
\end{figure*}

\begin{figure*}
\includegraphics[scale=0.5]{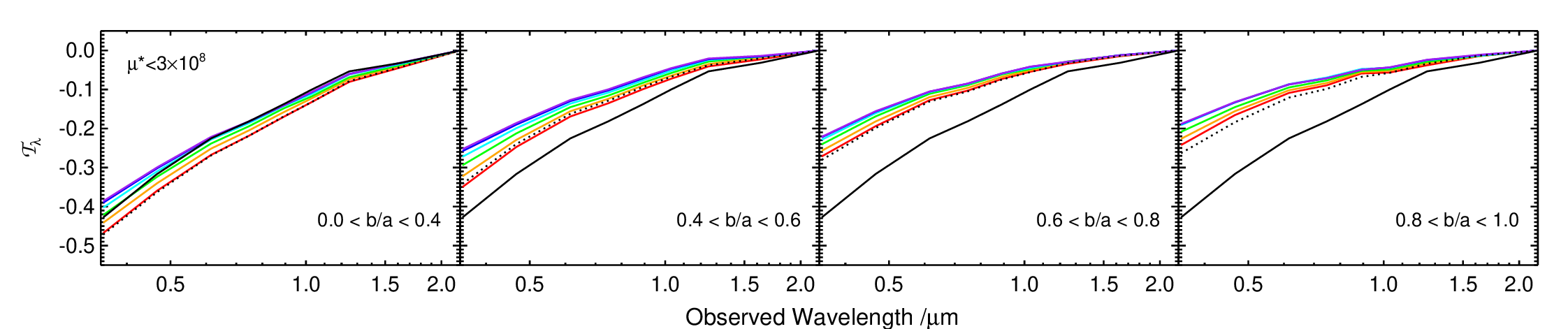}
\includegraphics[scale=0.5]{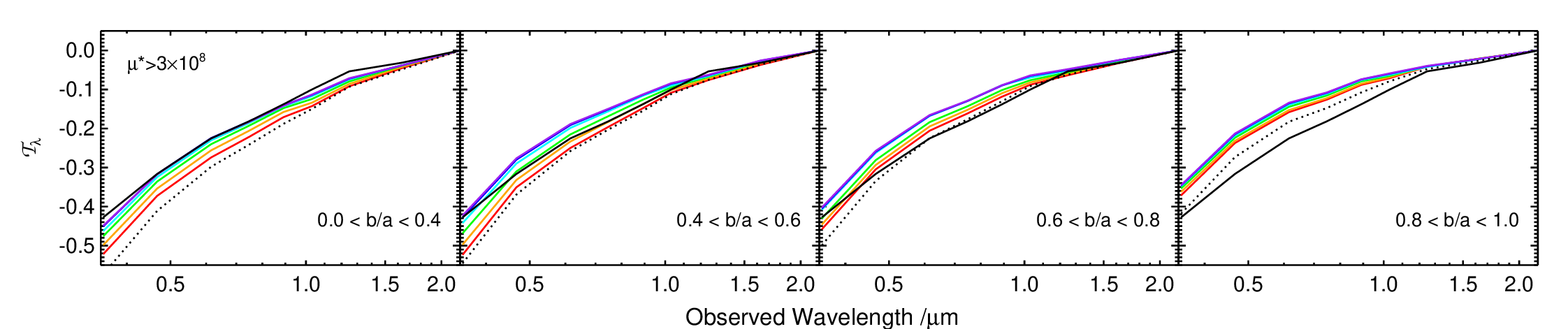}
\caption{Averaged dust attenuation curves, normalised to zero at K
  (see equation \ref{eqn:Fobs}). Each panel shows galaxies in bins
  of axis ratio (a/b), \emph{top:} with $\log\mu*<8.3$, and
  \emph{bottom:} with $\log\mu*>8.3$. Lines as in Figure
  \ref{fig:curves_ssfr}}\label{fig:curves_ab}
\end{figure*}

To begin with we focus on measurements of the optical-NIR attenuation
curves, for which the data is of a uniform and excellent quality, and
the data sample larger. In Section \ref{sec:uv} we study the reduced
UV-NIR sample of galaxies.

We split the galaxies into two independent samples with low and high
stellar surface mass density (\mustar), at the position of the
bimodality observed in the local galaxy population $\mu^*=3\times10^8
{\rm M}_\odot {\rm kpc}^{-2}$ \citep{2003MNRAS.341...54K}. We then
define several sub-samples in bins of $\log_{10}$\ssfr\ and \ab.
Table \ref{tab:sample} presents the details of the samples. The
measured attenuation curves are given in full for each sample and
aperture in an online table, Table \ref{tab:data} shows an extract.

The cut in \mustar\ effectively separates bulge-less galaxies from
galaxies with (even small) bulges. Comparing our \mustar\ cut with
results from the bulge-to-disk decomposition of
\citet{Gadotti:2009p2216} we find that while some pseudo-bulges have
$1\times10^8<\mu^*<3\times10^8$ and therefore fall into our bulge-less
sample, classical bulges and ellipticals all have
$\mu^*>3\times10^8$. Radial gradients may be expected to be more
prevalent in bulge+disk systems, and the presence of a bulge may
significantly affect the attenuation of light from galaxies, compared
to pure disk systems \citep{Pierini:2004p5389}. We therefore
choose to treat these two samples independently.

Figures \ref{fig:curves_ssfr} and \ref{fig:curves_ab} show the
attenuation curves for each bin. Different colour curves indicate
different aperture sizes. To facilitate comparison between panels, the
black line shows the dust curve derived from the whole sample, and is
the same in all panels. Some qualitative comparison between the
different attenuation curves is instructive.  Firstly, we note that in
most panels the overall normalisation of the curves drops with
decreasing aperture size, which is due to radial gradients in dust
content. The inner radii are dustier, and therefore a larger
$\Delta\uptau$ is measured.  We see that radial gradients are more
noticeable in galaxies with higher \ssfr.  We will present a
quantitative study of the radial gradients in Section
\ref{sec:radgrad}.

Secondly, the overall normalisation of the curves in each panel
depends on the range in dust content of the galaxies in the input
sample, simply because of the method used to build the curves. A
larger overall \Dtauvcont\ means a greater range of dust contents in
the input sample. Both samples show a steady decrease in \Dtauvcont\
with increasing \ba\ and increase in \Dtauvcont\ with increasing
\ssfr. This implies that high \ssfr\ and more inclined galaxies
exhibit a larger range of \Dtauvcont. This pure selection effect does
not affect our results, which depend only on the shape of the
attenuation curve, or its amplitude relative to another dust
indicator.

It is noticeable from Table \ref{tab:sample} that the bins of \ssfr\
and axis ratio are not entirely independent: when binning by \ssfr,
bins with lower \ssfr\ have lower median axis ratios, and similarly
when binning by axis ratio. Indeed, when investigating the whole
dataset there is a small but significant positive trend between \ssfr\
and axis ratio, which cannot be a ``fundamental'' property of galaxies
but, as we shall show later in the paper, is likely an effect of
dust. A bias in an observational property such as this one can cause
an intrinsic property to correlate with dust content, and ultimately
bias the shape of the attenuation curve. In this case it shows the
importance of pair-matching galaxies in \ba. 

In the following section, we begin our quantitative study of the
attenuation curves of the binned samples by parametrising them using a
simple analytic form, which allows for easy visualisation of changes
of shape and amplitude with galaxy properties.

\section{Trends of attenuation with galaxy properties}\label{sec:paramfits}

\begin{table*}
  \centering
  \caption{\label{tab:param} Parameters of broken power-law fits to
    attenuation curves for each SDSS+UKIDSS sample (see
    Eqn. \ref{eqn:param}). Measurements of the shape and amplitude of
    the attenuation curves are made from circular
    aperture photometry measured at 90\% of the $r$-band Petrosian
    radius. The line to continuum ratio is measured using a
    1.76\arcsec\ radius fixed angular aperture, to best match the
    emission line measurement through the SDSS fibre.  
    Note that the slopes and $\uptau$ ratios are fundamental
    quantities, whereas the $\Delta\uptau$ values are sample dependent. }
  \vspace{0.2cm}
  \begin{tabular}{cccccccc}\hline\hline
    & \sopt & \snir &
    $\lambda_c/\mu {\rm m}$ & $\Delta\uptau_{\lambda_c}$ &
    $\Delta\uptau_{\mathsf{K}}$ &\Dtauvcont&
    $\frac{\Delta\uptau_{\mathsf{V,line}}}{\Delta\uptau_{\mathsf{V,cont}}}$
    (3\arcsec)\\ \hline 

     $\mu^*<3\times10^8$ & & & & &  & & \\ \hline
All galaxies & $1.02\pm_{0.05}^{0.05}$ & $1.54\pm_{0.50}^{0.33}$ & $0.95\pm_{0.08}^{0.02}$ & $0.10\pm_{0.01}^{0.01}$ & $0.03\pm_{0.01}^{0.01}$ & $0.18\pm_{0.01}^{0.01}$ & $2.56\pm_{0.12}^{0.10}$\\
-11.00$<\log(\psi_S/{\rm yr^{-1}})<$-9.90 & $1.03\pm_{0.09}^{0.09}$ & $1.78\pm_{0.37}^{0.59}$ & $0.95\pm_{0.00}^{0.03}$ & $0.11\pm_{0.03}^{0.03}$ & $0.02\pm_{0.01}^{0.02}$ & $0.19\pm_{0.02}^{0.02}$ & $2.92\pm_{0.35}^{0.26}$\\
-9.90$<\log(\psi_S/{\rm yr^{-1}})<$-9.75 & $1.16\pm_{0.12}^{0.07}$ & $2.61\pm_{1.53}^{0.89}$ & $0.71\pm_{0.00}^{0.03}$ & $0.11\pm_{0.02}^{0.01}$ & $0.00\pm_{0.00}^{0.02}$ & $0.15\pm_{0.01}^{0.02}$ & $2.71\pm_{0.18}^{0.17}$\\
-9.75$<\log(\psi_S/{\rm yr^{-1}})<$-9.62 & $1.14\pm_{0.09}^{0.09}$ & $2.84\pm_{1.22}^{1.25}$ & $0.80\pm_{0.00}^{0.02}$ & $0.09\pm_{0.02}^{0.01}$ & $0.01\pm_{0.01}^{0.02}$ & $0.14\pm_{0.01}^{0.02}$ & $2.78\pm_{0.29}^{0.36}$\\
-9.62$<\log(\psi_S/{\rm yr^{-1}})<$-9.46 & $1.12\pm_{0.09}^{0.07}$ & $1.96\pm_{0.49}^{0.50}$ & $0.94\pm_{0.00}^{0.03}$ & $0.09\pm_{0.01}^{0.01}$ & $0.02\pm_{0.01}^{0.01}$ & $0.17\pm_{0.01}^{0.02}$ & $2.71\pm_{0.15}^{0.18}$\\
-9.46$<\log(\psi_S/{\rm yr^{-1}})<$-9.25 & $0.95\pm_{0.13}^{0.07}$ & $1.29\pm_{0.36}^{0.35}$ & $0.70\pm_{0.21}^{0.07}$ & $0.15\pm_{0.03}^{0.02}$ & $0.04\pm_{0.01}^{0.03}$ & $0.19\pm_{0.02}^{0.03}$ & $2.41\pm_{0.17}^{0.14}$\\
-9.25$<\log(\psi_S/{\rm yr^{-1}})<$-7.90 & $1.04\pm_{0.09}^{0.12}$ & $1.83\pm_{0.40}^{0.41}$ & $0.70\pm_{0.09}^{0.04}$ & $0.14\pm_{0.03}^{0.02}$ & $0.01\pm_{0.02}^{0.03}$ & $0.17\pm_{0.02}^{0.03}$ & $2.08\pm_{0.14}^{0.22}$\\
 
0.0$<b/a<$0.4 & $0.91\pm_{0.04}^{0.06}$ & $1.86\pm_{0.26}^{0.50}$ & $0.96\pm_{0.08}^{0.02}$ & $0.16\pm_{0.01}^{0.01}$ & $0.03\pm_{0.01}^{0.01}$ & $0.27\pm_{0.02}^{0.01}$ & $2.01\pm_{0.11}^{0.15}$\\
0.4$<b/a<$0.6 & $0.97\pm_{0.18}^{0.10}$ & $0.92\pm_{1.35}^{1.38}$ & $0.95\pm_{0.09}^{0.08}$ & $0.11\pm_{0.01}^{0.01}$ & $0.05\pm_{0.01}^{0.04}$ & $0.19\pm_{0.01}^{0.03}$ & $3.15\pm_{0.17}^{0.23}$\\
0.6$<b/a<$0.8 & $1.12\pm_{0.10}^{0.12}$ & $1.29\pm_{0.38}^{0.50}$ & $0.70\pm_{0.08}^{0.07}$ & $0.11\pm_{0.01}^{0.01}$ & $0.02\pm_{0.01}^{0.02}$ & $0.14\pm_{0.01}^{0.02}$ & $3.45\pm_{0.30}^{0.29}$\\
0.8$<b/a<$1.0 & $1.25\pm_{0.10}^{0.08}$ & $2.14\pm_{0.40}^{0.87}$ & $0.72\pm_{0.02}^{0.00}$ & $0.07\pm_{0.01}^{0.01}$ & $0.01\pm_{0.01}^{0.01}$ & $0.10\pm_{0.01}^{0.01}$ & $3.74\pm_{0.34}^{0.30}$\\
 
\hline $\mu^*>3\times10^8$ & & & & &  & &\\ \hline
All galaxies & $1.28\pm_{0.01}^{0.01}$ & $1.57\pm_{0.07}^{0.08}$ & $0.82\pm_{0.01}^{0.01}$ & $0.14\pm_{0.00}^{0.00}$ & $0.03\pm_{0.00}^{0.00}$ & $0.23\pm_{0.01}^{0.00}$ & $2.39\pm_{0.06}^{0.07}$\\
-11.00$<\log(\psi_S/{\rm yr^{-1}})<$-9.90 & $1.19\pm_{0.08}^{0.06}$ & $1.78\pm_{0.31}^{0.50}$ & $0.96\pm_{0.09}^{0.00}$ & $0.11\pm_{0.03}^{0.01}$ & $0.03\pm_{0.01}^{0.01}$ & $0.21\pm_{0.01}^{0.02}$ & $2.72\pm_{0.17}^{0.17}$\\
-9.90$<\log(\psi_S/{\rm yr^{-1}})<$-9.75 & $1.28\pm_{0.08}^{0.08}$ & $1.64\pm_{0.17}^{0.30}$ & $0.89\pm_{0.00}^{0.13}$ & $0.12\pm_{0.02}^{0.01}$ & $0.02\pm_{0.01}^{0.01}$ & $0.21\pm_{0.02}^{0.01}$ & $2.53\pm_{0.17}^{0.17}$\\
-9.75$<\log(\psi_S/{\rm yr^{-1}})<$-9.62 & $1.42\pm_{0.06}^{0.10}$ & $1.74\pm_{0.27}^{0.37}$ & $0.70\pm_{0.09}^{0.04}$ & $0.15\pm_{0.01}^{0.01}$ & $0.03\pm_{0.01}^{0.01}$ & $0.21\pm_{0.02}^{0.02}$ & $2.62\pm_{0.20}^{0.15}$\\
-9.62$<\log(\psi_S/{\rm yr^{-1}})<$-9.46 & $1.34\pm_{0.04}^{0.12}$ & $1.25\pm_{0.17}^{0.33}$ & $0.81\pm_{0.10}^{0.03}$ & $0.14\pm_{0.01}^{0.00}$ & $0.04\pm_{0.01}^{0.01}$ & $0.23\pm_{0.03}^{0.01}$ & $2.34\pm_{0.15}^{0.14}$\\
-9.46$<\log(\psi_S/{\rm yr^{-1}})<$-7.90 & $1.37\pm_{0.05}^{0.06}$ & $1.72\pm_{0.15}^{0.30}$ & $0.85\pm_{0.03}^{0.09}$ & $0.15\pm_{0.00}^{0.01}$ & $0.03\pm_{0.01}^{0.01}$ & $0.26\pm_{0.01}^{0.01}$ & $1.97\pm_{0.08}^{0.12}$\\
 
0.0$<b/a<$0.4 & $1.10\pm_{0.05}^{0.05}$ & $1.70\pm_{0.20}^{0.29}$ & $1.00\pm_{0.08}^{0.04}$ & $0.14\pm_{0.01}^{0.01}$ & $0.04\pm_{0.01}^{0.01}$ & $0.28\pm_{0.01}^{0.01}$ & $2.06\pm_{0.12}^{0.10}$\\
0.4$<b/a<$0.6 & $1.27\pm_{0.05}^{0.04}$ & $1.51\pm_{0.15}^{0.11}$ & $0.83\pm_{0.02}^{0.02}$ & $0.14\pm_{0.01}^{0.00}$ & $0.03\pm_{0.00}^{0.01}$ & $0.24\pm_{0.01}^{0.01}$ & $2.30\pm_{0.09}^{0.15}$\\
0.6$<b/a<$0.8 & $1.39\pm_{0.06}^{0.06}$ & $1.31\pm_{0.20}^{0.21}$ & $0.85\pm_{0.09}^{0.13}$ & $0.11\pm_{0.01}^{0.03}$ & $0.03\pm_{0.01}^{0.01}$ & $0.21\pm_{0.01}^{0.02}$ & $2.58\pm_{0.20}^{0.19}$\\
0.8$<b/a<$1.0 & $1.59\pm_{0.06}^{0.08}$ & $2.59\pm_{0.21}^{0.32}$ & $1.00\pm_{0.00}^{0.00}$ & $0.06\pm_{0.01}^{0.01}$ & $0.01\pm_{0.00}^{0.01}$ & $0.15\pm_{0.01}^{0.01}$ & $2.79\pm_{0.24}^{0.20}$\\
\hline 

    MW (CCM+OD) & 1.06 & 1.6  & 0.82 & 0.20$^b$ & 0.038$^b$ & 0.3$^b$ & -- \\
    C00         & 0.96 & 1.7  & 0.91 & 0.18$^b$ & 0.034$^b$ & 0.3$^b$ & 2.08$^a$ \\ 
    LMC         & 1.05 & 1.7  & 0.90 & 0.17$^b$ & 0.020$^b$ & 0.3$^b$ & -- \\
    SMC-bar$^c$ & 1.2  & --   & --   & 0.16$^b$ & --        & 0.3$^b$ & -- \\
\hline

  \end{tabular}
  \begin{minipage}{\textwidth}
    a: \EBV$_{\rm stars}$ = 0.44\EBV$_{\rm gas}$
    \citep{Calzetti:1997p5726, 2001PASP..113.1449C} converted into a
    ratio of optical depths assuming the emission line attenuation curve from
    \citet{Wild:2011p6538}. \\
    b: In the case of the MW, C00, LMC and SMC-bar curves, the $\uptau$
    values are absolute and given for $\uptau_{\mathsf{V}}=0.3$ and
    $\uptau_{\mathsf{K}}$ is given at rest-frame assuming $z=0.7$. \\
    c: We are unable to reliably measure the NIR slope of the SMC-bar extinction curve (see Section \ref{sec:nir}).  
  \end{minipage}

\end{table*}

As discussed in Section \ref{sec:intro} the shape of the dust
attenuation curve, and strength relative to the ratio of \ha\ to \hb\
luminosity, can tell us about the dust properties and dust-star
geometry within galaxies. In order to visualise trends in the
attenuation curves presented in Figures \ref{fig:curves_ssfr} and
\ref{fig:curves_ab} we must first parametrise the curves. After some
experimentation, a broken power-law was found to provide an adequate
representation of the optical-NIR data, with easily interpretable
fitted parameters. A single power-law was unable to fit the whole
wavelength range, due to a change in the slope between the optical and
NIR wavelength regimes. We fit this function using a least-squares
minimisation routine, weighting each data point equally:
\begin{eqnarray}
  \mathcal{T}_\lambda &=&
  - \Delta\uptau_{\lambda_c}\left(\frac{\lambda}{\lambda_c}\right)^{s_{\mathsf{opt}}}
  +  \Delta\uptau_{\mathsf{K}}
  \hspace{1cm}\lambda < \lambda_c \nonumber\\ 
  &=&
  -\Delta\uptau_{\lambda_c}\left(\frac{\lambda}{\lambda_c}\right)^{s_{\mathsf{nir}}} +  \Delta\uptau_{\mathsf{K}}
    \hspace{1cm}\lambda > \lambda_c  \label{eqn:param}
\end{eqnarray}
where the free parameters describe the wavelength at which the shape
of the attenuation curve changes between the optical and the NIR
($\lambda_c$), the mean amplitude of the attenuation curve at this
wavelength ($\Delta\uptau_{\lambda_c}$), the mean amplitude of the
attenuation curve at the longest wavelength point
($\Delta\uptau_{\mathsf{K}}$), the power-law slope of the attenuation
curve in the optical (\sopt) and in the NIR (\snir).  In this
  section our aim is only to measure the slope of the curves for the
  purpose of visualisation. Therefore, we fit only the $u$, $g$, $r$,
  $i$, Y, J, H and K bands, excluding the $z$-band because of its
  proximity to the break in slopes at $\lambda_c$. For the same reason
  we do not constrain the amplitudes of the two powerlaws
  ($\Delta\uptau_{\lambda_c}$) to be equal. In practice the amplitudes
  of the two functions are similar. Note that by construction the
form of this function is the same as that derived in
Eqn. \ref{eqn:Ftheory}, with $Q_\lambda \propto \lambda^{s}$.

Bootstrap errors on each of these parameters are estimated by
recalculating the geometric mean, using random resampling with
replacement of the galaxy-pairs that contribute. We randomly select
100 samples of the same size as the original dataset, and repeat the
least squares fitting on each sample. The 16th and 84th percentiles on
each parameter are taken to be the $1\sigma$ errors. We verified that
increasing the number of samples to 500 does not change the resulting
estimated errors.

Figures \ref{fig:curves_ssfr} and \ref{fig:curves_ab} show that even
in samples exhibiting strong radial gradients, the attenuation curves
are close to converged between 90\% and 100\% of \Rpet. The fitted
parameters at 90 and 100\% of \Rpet\ are equal within the errors, but
the errors at 90\% are smaller and trends tighter. This is possibly
caused by increasing sky subtraction errors at large radii in the
UKIDSS photometry. Therefore, unless otherwise stated, we measure the
galaxy-wide attenuation curves at 90\% of \Rpet.

The fit is performed at the median rest-frame of the galaxies in the
sample (the median redshift is given in Table \ref{tab:sample}). As
mentioned in Section \ref{sec:procedure}, iteratively K-correcting the
individual flux ratios before combining them into the dust attenuation
curve makes no significant difference to the results and was therefore
discarded from the final analysis.

The resulting fitted parameters and errors are given in Table
\ref{tab:param}, together with comparison numbers for 
curves in the literature measured in exactly the same way as our
data. As expected, the fitted \dtk\ values are close to zero, with
their exact values imposed by the constraint that the NIR slope is a
perfect power-law. In the following subsections we present the trends
of attenuation curve shape, radial gradients and difference between
nebular line and stellar continuum attenuation as a function of \ssfr\
and axis ratio, for both high and low \mustar\ galaxies.

\subsection{The slope of the attenuation curve in the optical}\label{sec:opt}

\begin{figure*}
\includegraphics[scale=0.5]{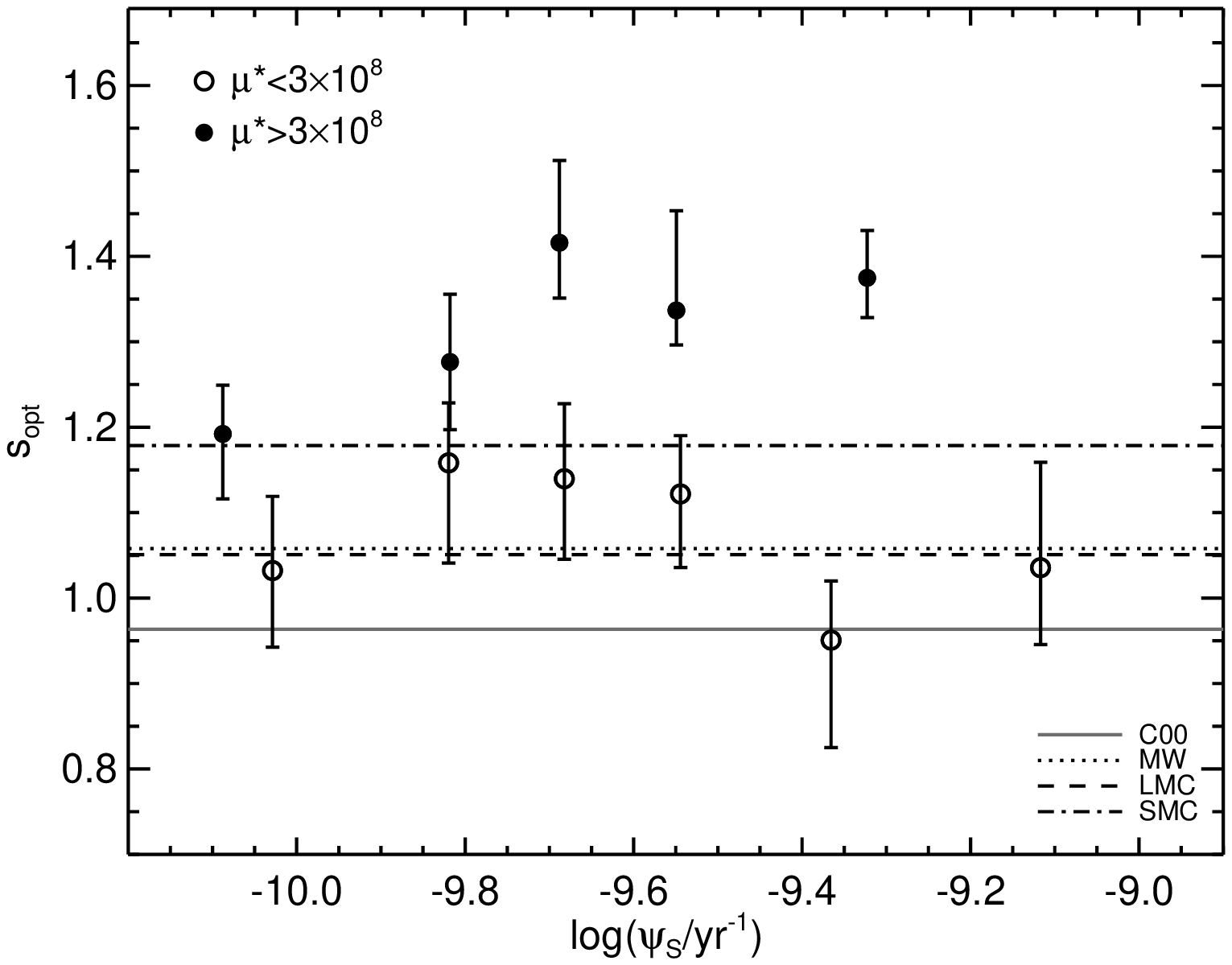}
\includegraphics[scale=0.5]{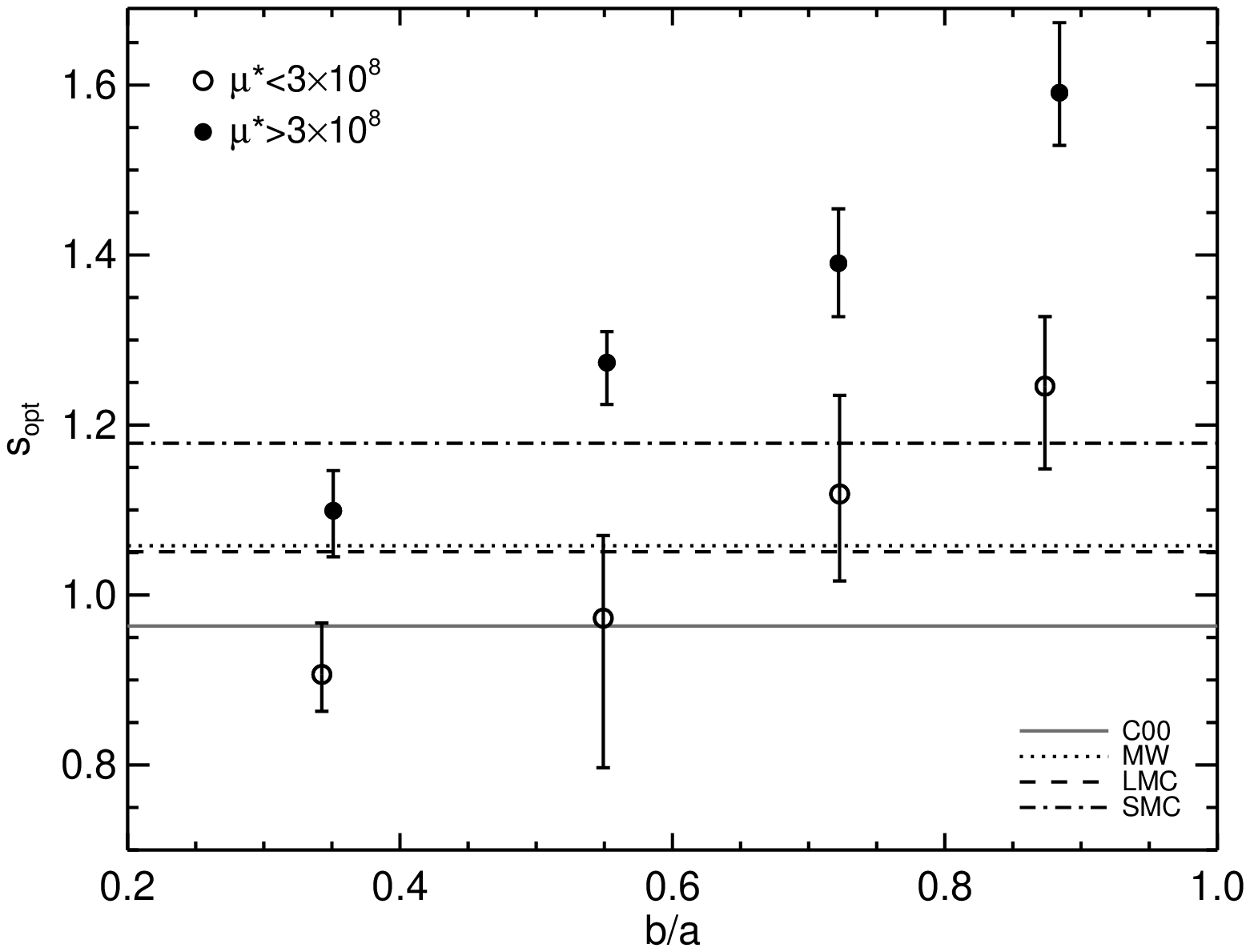}
\caption{Power-law slope of attenuation curve in the optical
  wavelength range, as a function of \ssfr\ (left) and axis ratio
  (right), for low \mustar\ (open circles) and high \mustar\ (filled
  circles) galaxies. The horizontal lines indicate the equivalent
  slopes for the MW (dotted), C00 (gray, full), LMC (dashed) and
  SMC-bar (dot-dash) curves.  }\label{fig:sopt}
\end{figure*}

Figure \ref{fig:sopt} presents the optical slopes of the attenuation
curves (\sopt) as a function of \ssfr\ and \ab. The dotted horizontal
lines indicate the slopes of the MW, C00, LMC and SMC-bar curves
measured using the same method. There are significant trends in
optical slope with galaxy properties. In both high and low \mustar\
samples the curve flattens slowly with decreasing axis ratio, meaning
greyer curves in edge-on galaxies and steeper curves in face-on
galaxies. This is the first clear example of global geometry affecting
the apparent attenuation curves of galaxies. 

The high \mustar\ sample exhibits steeper optical slopes than the low
\mustar\ sample, and significantly steeper even than the Milky Way
extinction curve. This is intriguing as attenuation curves resulting
from mixed stars and dust in galaxies are generally expected to be
flatter than the underlying extinction curves. Steeper optical
extinction curves do exist locally in the SMC and actively
star-forming regions in the MW, possibly caused by the modification of
grain composition. However, it would be surprising if massive, metal
rich, bulge dominated galaxies have grain compositions closer to
active dwarf galaxies than to the MW. Steeper apparent slopes could be
caused by a mismatch in the stellar population of the galaxy pairs.
However, this requires the more-dusty galaxies to have an
intrinsically older and therefore redder stellar population than the
less-dusty galaxies, despite having the same observed \ssfr,
metallicity etc., which is opposite to the slight trend with
inclination that we observe in the data.

It is possible that the steeper slopes are a result of a differential
effect related to the distribution of dust around stars of different
ages. In ordinary star-forming galaxies, younger (older) stars
contribute a larger fraction of the blue (red) optical light. This is
different from low-mass starburst or old elliptical galaxies, in which
the majority of light at all wavelengths comes from young or old
stars.  If young, blue stars suffer greater dust extinction than old,
red stars, then we will observe an apparently steeper attenuation
curve in star-forming galaxies than the underlying extinction
curve. This effect is the common explanation for the difference in
attenuation observed in emission lines and galaxy continua (see
Section \ref{sec:lc}).  We observe that the slope is steeper in
  galaxies with significant bulges. This could occur if the older
  stellar population in the bulge is less obscured by dust than the
  younger stellar population in the disk. Slopes closer
to the extinction curves observed in edge-on galaxies could arise as
the disk obscures a greater fraction of the old stellar population in
the bulge. 

If the steeper slopes are a result of the relative geometry of stars
and dust, we would expect a weaker effect in the smaller apertures where
the light comes mainly from older stars in the bulge, especially in
face-on galaxies: at 0.35\rpet\ we find \sopt$=$1.07 for \ba$=$0.35,
and 1.31 for \ba$=$0.88. This is 3\% and 22\% smaller than the values
for \sopt\ at 0.9\rpet, qualitatively in agreement with the hypothesis
that the steep curves arise due to a differential effect whereby
different stellar populations are attenuated by different dust
columns. 

One final possibility is presented by radiative transfer models of
dust in galaxies. These can also exhibit steeper optical attenuation
curves than the input extinction curves, in the optically thin limit
and with certain geometric configurations
\citep{Tuffs:2004p6123,Pierini:2004p5389}, even before effects for
different stellar populations are included. Ultimately, the cause for
the steep optical slopes must be solved by comparison with such
models. This is beyond the scope of the present paper (see Chevallard
et al., in preparation).

\subsection{The slope of the attenuation curve in the NIR}\label{sec:nir}

\begin{figure*}
\includegraphics[scale=0.5]{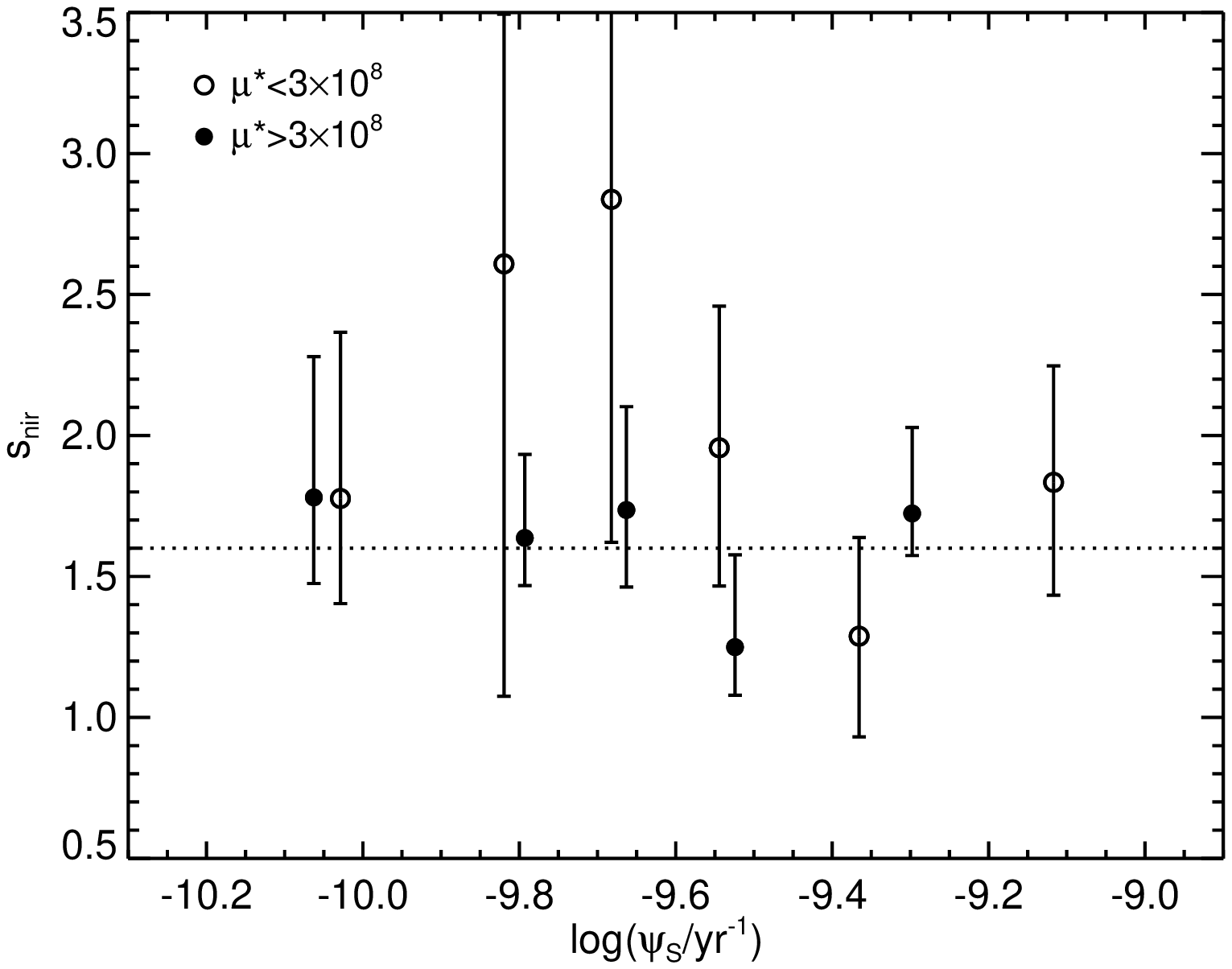}
\includegraphics[scale=0.5]{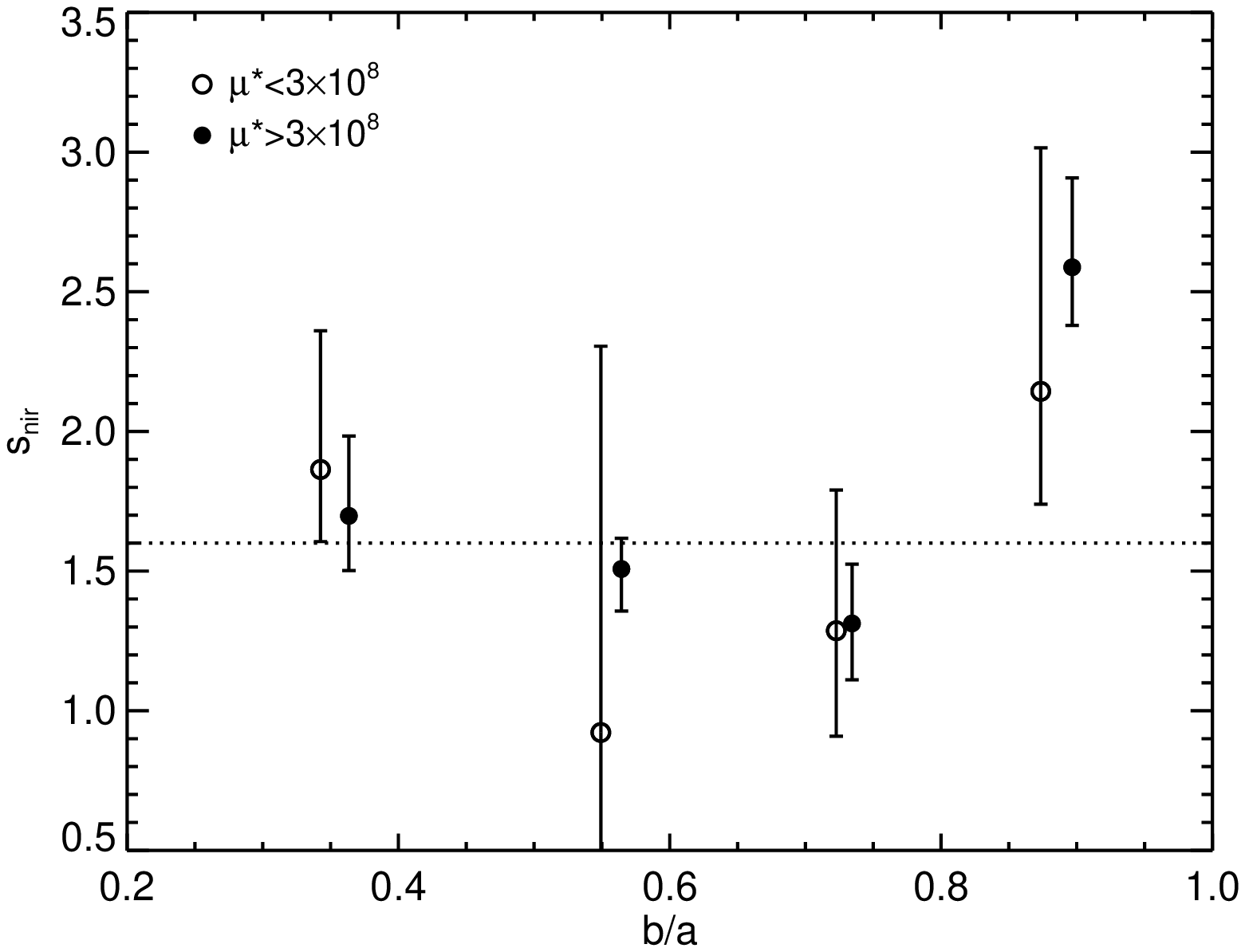}
\caption{Power-law slope of attenuation curve in the NIR wavelength
  range, as a function of \ssfr\ (left) and observed axis ratio
  (right), for low \mustar\ (open circles) and high \mustar\ (filled
  circles) galaxies. For clarity the high \mustar\ symbols have been
  offset slightly to the right. The dotted line indicates the
  equivalent slope for the MW extinction curve. }\label{fig:snir}
\end{figure*}

In Figure \ref{fig:snir} we show the slope of the attenuation curve in
the NIR, as a function of \ssfr\ and axis ratio. Within the errors we
find no significant overall deviation from the MW extinction curve with a
slope of 1.6, and no clear trends with galaxy properties are
evident.  For the LMC and C00 curves we find NIR slopes of 1.7,
  almost identical to the MW curve. The SMC-bar curve of
  \citet{2003ApJ...594..279G} exhibits a strong bump in the NIR that
  prevents us from fitting a robust powerlaw slope.  Inspection of the
  individual curves in \citet{2003ApJ...594..279G} reveals that a
  significant deviation from a power law slope is only evident in one
  out of the four stars that contribute to the average curve. We
  therefore conclude that this bump may result from photometric errors
  and small number statistics, and we therefore do not compare our
  results with the SMC-bar curve in the NIR wavelength regime. 

The constancy of the NIR slope of extinction laws along multiple lines
of sight to stars in the MW is an important feature of the MW
extinction curve, implying a constant size distribution of the largest
particles throughout both diffuse and dense clouds of dust
\citep{Mathis:1990p6269}. The same value found here in external
galaxies leads to the conclusion that the size distribution of the
largest dust particles is universal. It additionally implies that the
factors affecting the attenuation of light in galaxies (dust-star and
global geometry, scattering into the line-of-sight) are unimportant at
long wavelengths.  Recent work by \citet{Stead:2009p6343} suggests
  a slightly steeper MW NIR slope of 2.14 by comparison of UKIDSS data
  with the Besan\c{c}on Galactic model. If this is the case, this
  would instead imply some degree of flattening in the NIR attenuation
  curve of external galaxies. It is clear that further work on
  measuring the shape of extinction curves in the MW and local
  galaxies would be helpful to understand the dust distribution in
  more distant galaxies.

\subsection{Radial Gradients in total dust attenuation}\label{sec:radgrad}

\begin{figure*}
\includegraphics[scale=0.5]{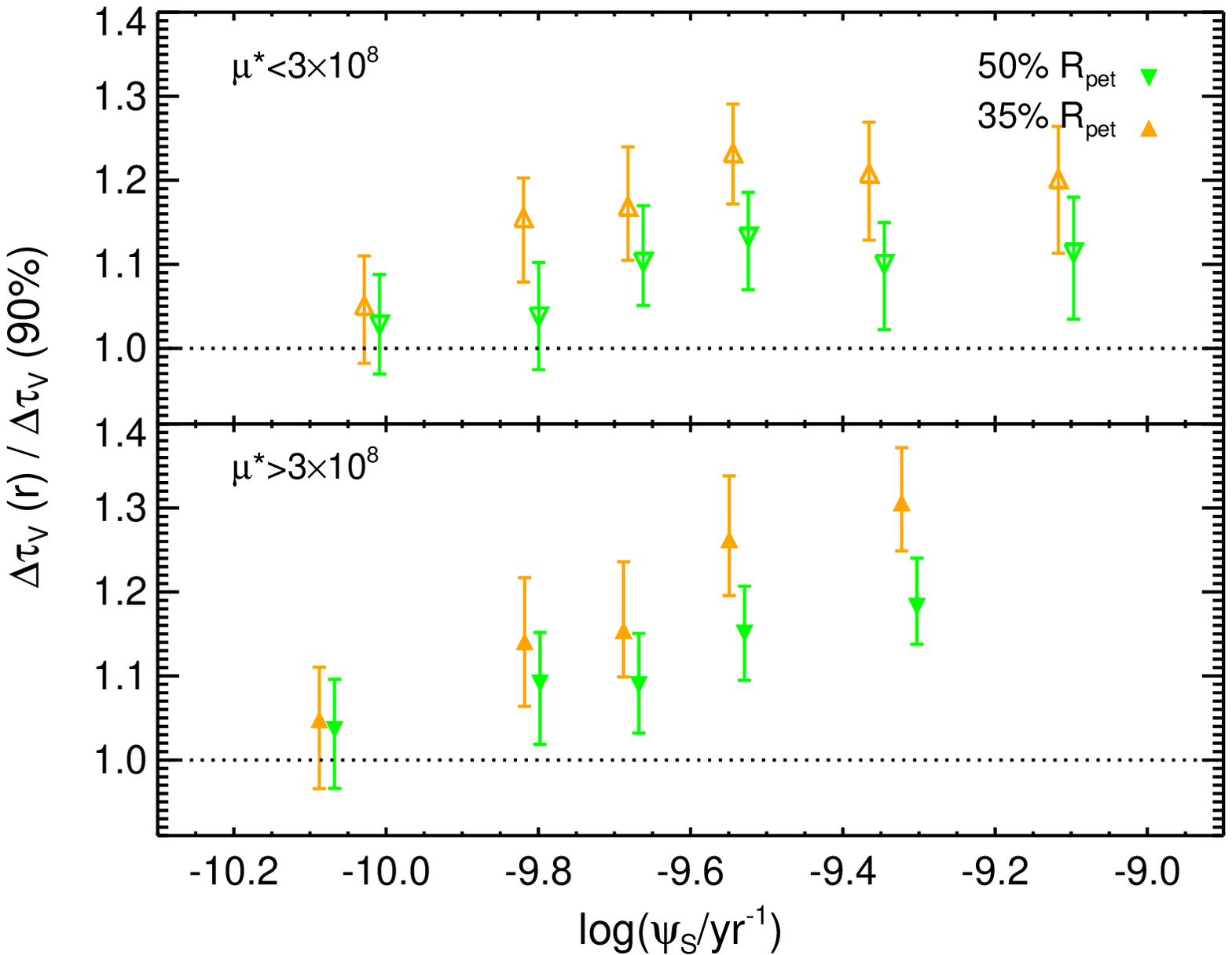}
\includegraphics[scale=0.5]{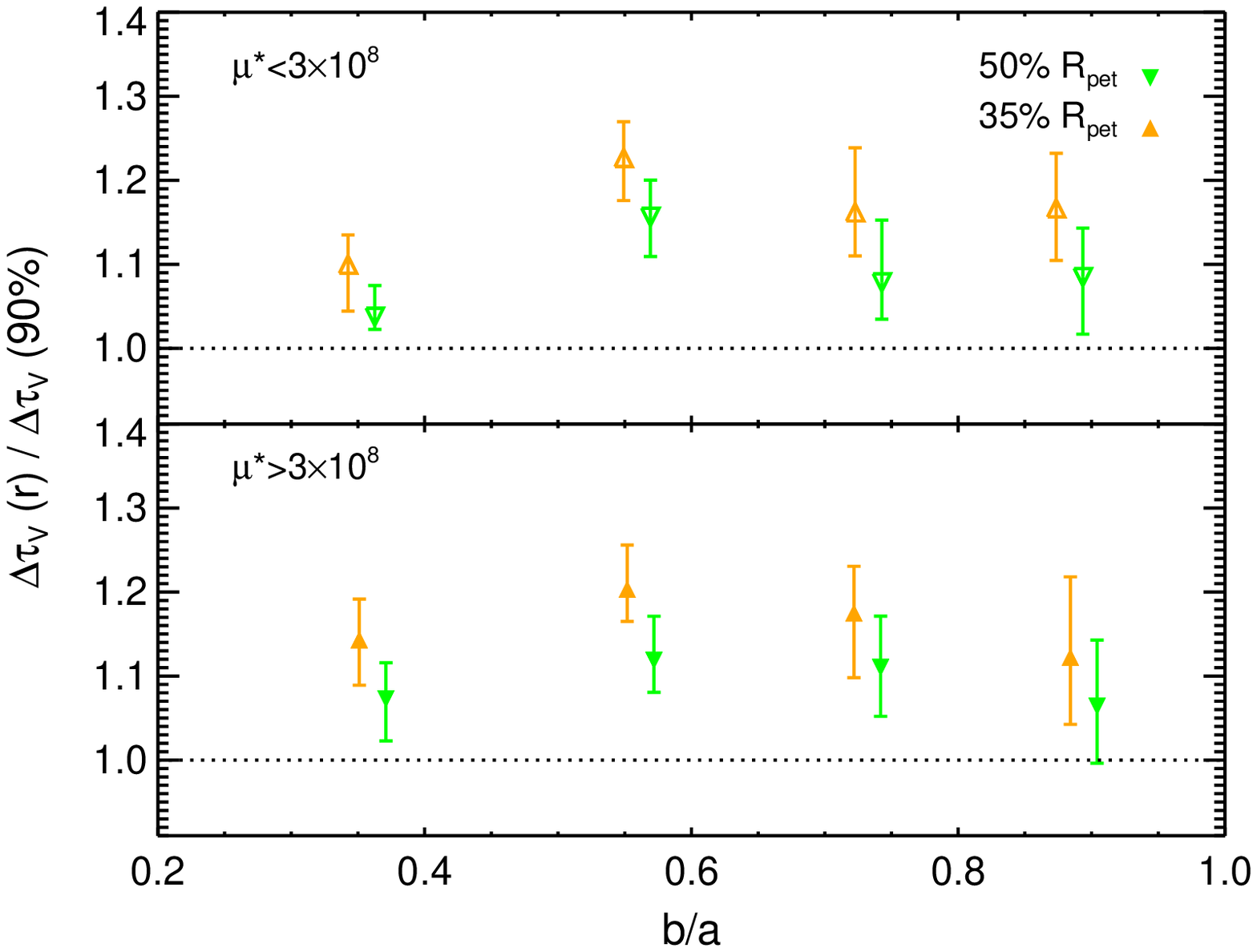}
\caption{Radial gradients in continuum optical depth. Relative optical
  depth at 5500\AA\ towards stars contained within 35\% (yellow,
  upwards triangle) and 50\% (green, downwards triangle) of the
  Petrosian radius of the galaxies, relative to the optical depth
  measured at 90\% of the Petrosian radius, as a function of \ssfr\
  (left) and axis ratio (right). The top (bottom) panel shows the
  results for low- (high-) \mustar galaxies.  }\label{fig:dtauv}
\end{figure*}

In this section we provide a more quantitative analysis of dust radial
gradients than can be achieved by visual inspection of Figures
\ref{fig:curves_ssfr} and \ref{fig:curves_ab}. Because our methodology
only provides a relative difference in dust content between the less-
and more-dusty galaxies, we normalise the measured \dtv\ values at all
radii by the value at 0.9\Rpet. This then allows a direct comparison
between samples. We measure \dtv\ from the independent power-law fits
to the dust curves at each different radius. It can be shown that, in
the case that \tauv($r$)/\tauv($0.9$\rpet) is the same on average for
both the more and less-dusty galaxies that make up the pairs in a
particular sample, i.e. within a particular bin the strength of the
radial gradients is independent of overall dust content, then
\dtv($r$)/\dtv($0.9$\rpet)$=$\tauv($r$)/\tauv($0.9$\rpet). However,
whether this is the case or not does not affect the identification of
trends in which we are primarily interested. 

Figure \ref{fig:dtauv} shows \dtv\ measured within 0.35 and 0.5\Rpet,
relative to \dtv\ measured at 0.9\Rpet. In this section only, we fix
the NIR slope to 1.6 during the fits, because for small apertures the
errors increase substantially on the measured NIR slope.  In the
  left hand panels we see a strong trend in the strength of radial
  gradients with \ssfr: gradients are significantly larger in high
  \ssfr\ galaxies than low \ssfr\ galaxies. In the low \mustar\
  sample, the trend appears to flatten above \ssfr$\sim-9.5$. In the
  right hand panels, we see no strong trends in the strength of the
  radial gradients with axis ratio, although the slightly weaker
  radial gradients in very inclined galaxies is likely a real feature
  caused by large optical depths. It is noticeable that both the high
  and low \mustar\ samples show the same strength of radial
  gradient. Averaging over all bins in \ba, we find that the dust
  optical depth within 0.35 and 0.5\Rpet\ is 1.16 and 1.09 times that
  within 0.9\Rpet, for both samples. This indicates that the presence
  of a bulge does not affect the strength of radial gradients in the
  dust distribution.

Measuring the trends of radial gradient strength with \ssfr\ is
important for understanding the production and destruction of dust
relative to ongoing or past star formation. The dust gradients
reported here, and elsewhere in the literature for spatially resolved
studies of small numbers of local galaxies
\citep[e.g.][]{MunozMateos:2009p5435}, are presumably related to
metallicity gradients in spiral disks. However, because our \ssfr\
measurements relate strictly to the central 3\arcsec\ of the galaxies,
and the scale of star formation episodes is known to depend on the
rate of star formation \citep{1998ARA&A..36..189K}, spatially resolved
information on star formation rates is required before firm
conclusions can be drawn. Suitable spatially resolved spectral studies
of metallicity, dust and star formation gradients in large numbers of
disk galaxies have yet to be undertaken.

\subsection{Dust attenuation of nebular emission lines vs. stellar continua}\label{sec:lc}

\begin{figure*}
\includegraphics[scale=0.5]{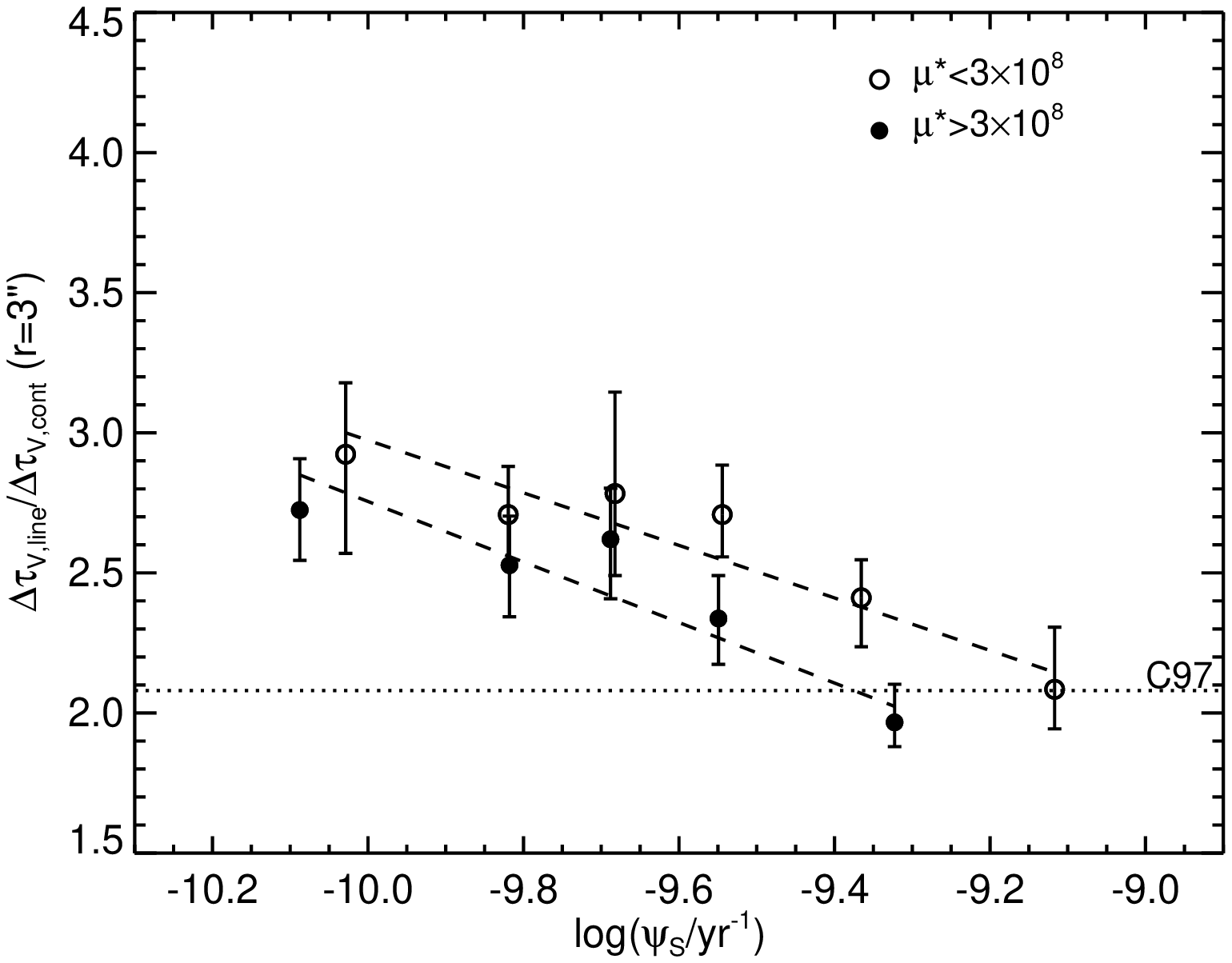}
\includegraphics[scale=0.5]{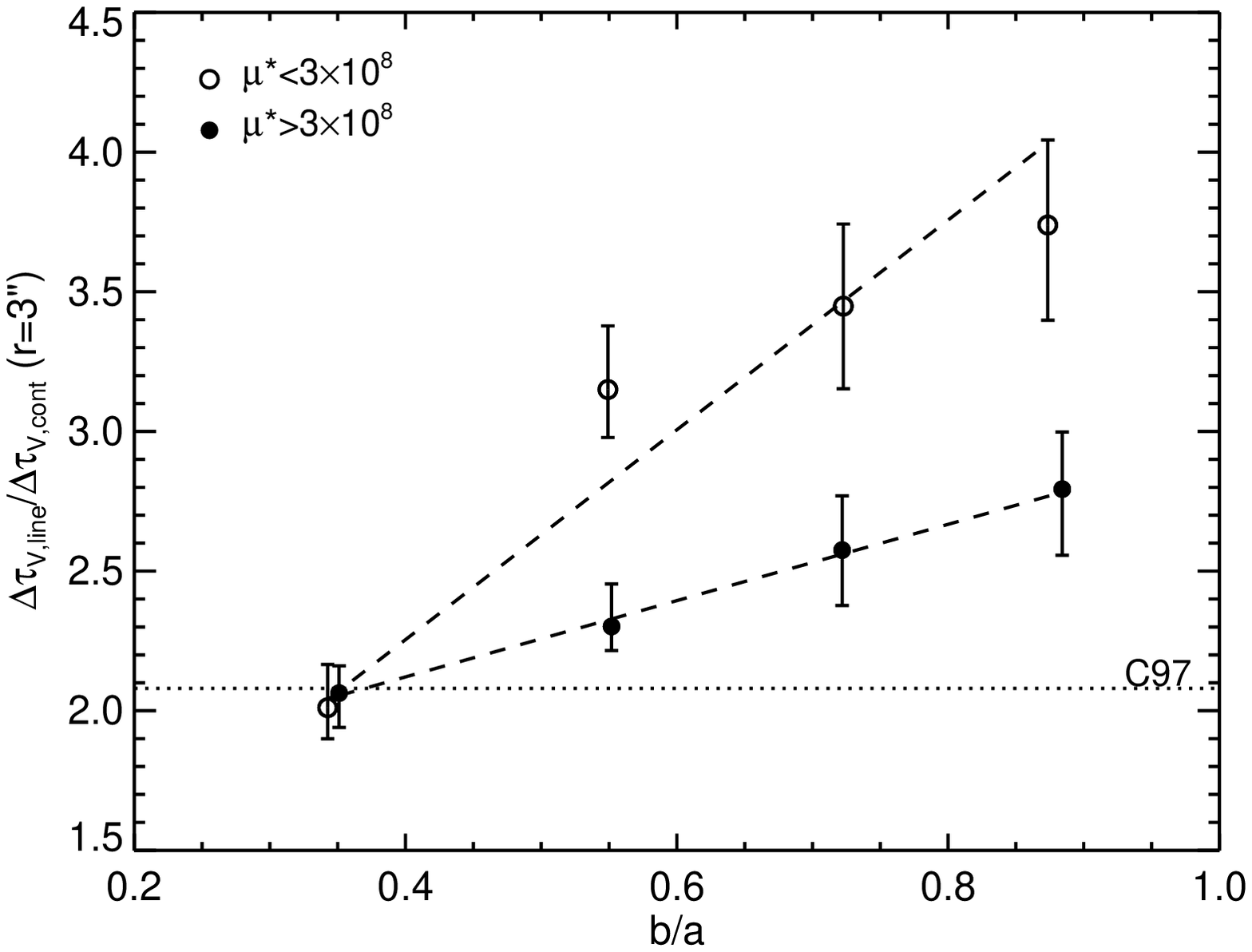}
\caption{Dust optical depth towards the Balmer emission lines,
  relative to dust optical depth in the continuum at 5500\AA, as a
  function of \ssfr\ (left) and axis ratio (right). Unlike previous
  figures, the optical depth in the continuum has been measured using
  a fixed angular aperture of radius 1.76\arcsec, to approximately
  match the emission line measurements which are obtained through an
  optical fibre with 1.5\arcsec radius aperture in generally worse
  seeing conditions. The dotted line indicates the line to continuum
  optical depth ratio from \citet{Calzetti:1997p5726}, assuming the
  same attenuation curve for the emission lines as used in our
  analysis (Eqn. \ref{eqn:emlines}). The dashed lines are fit to the
  data points and are described in the text.  }\label{fig:lc}
\end{figure*}

All of the galaxies in our sample have both \ha\ and \hb\ line
measurements which, until now, we have only used to derive relative
dust content. Here we make use of the line fluxes to compare the dust
attenuation suffered by the nebular emission lines and stellar
continuum. The SDSS spectra are taken using a 3\arcsec\ diameter
fibre, meaning that smaller galaxies, and those at higher redshift,
have a larger fraction of their total light included in the
spectrum. In our sample, the median projected fibre radius is around
0.35\Rpet, but with a wide distribution. In light of the radial
gradients measured in the previous section, this fixed angular
aperture restricts the extent to which we can compare dust attenuation
of lines and continua in our sample. To provide the best comparison,
in this section only, we measure the attenuation curves using a fixed
1.76\arcsec radius angular aperture (the curves are shown as dotted
lines in Figures \ref{fig:curves_ssfr} and \ref{fig:curves_ab}). This
is slightly larger than the 1.5\arcsec radius aperture of the fibre,
however the spectra are taken in worse seeing conditions than the
imaging, and flux measurements at 1.76\arcsec are provided in the SDSS
catalog thus preventing any small errors introduced by interpolation
of the radial flux curves.

Using the same formalism as for dust attenuation of the stellar
continuum (Section \ref{sec:bg}), the amount of dust attenuation of
nebular emission lines in more-dusty relative to less-dusty galaxies
is given by:
\begin{equation}
\Delta\uptau_{\mathsf{\lambda,line}}=\Delta\uptau_{\mathsf{V,line}}Q_{\mathsf{\lambda,line}} 
\end{equation}
where $Q_{\mathsf{\lambda,line}}$ describes the shape of the
attenuation curve applicable for emission lines. \citet{Wild:2011p6538}
compared mid-IR and optical emission line strengths to show that a two
component model \citep{2000ApJ...539..718C}, with diffuse dust
accounting for a fraction ($\mu$) of the optical depth at 5500\AA, and
denser birth-cloud dust accounting for the remainder, is
consistent with the emission line ratios observed in a wide range of
galaxy types:
\begin{equation}\label{eqn:emlines}
Q_{\mathsf{\lambda,line}} =(1-\mu)(\lambda/\lambda_{\mathsf{V}})^{-m} + \mu(\lambda/\lambda_{\mathsf{V}})^{-n}
\end{equation}
 In \citet{Wild:2011p6538} we measured $\mu=0.4$, for the case that
  $m=1.3$ and $n=0.7$. Ultimately the slopes $n$ and $m$ may also be measured
  through comparison between the attenuation
  curves measured in this paper and models. However, for the purposes of this paper
  it is sufficient to know that an accurate \tauvline\ is obtained from the
  Balmer emission line ratios, as shown by \citet{Wild:2011p6538}.  Eqn. \ref{eqn:emlines} leads to the
dust optical depth for emission  lines, as a function of wavelength
and ratio of \ha\ to \hb\ luminosity:
\begin{equation}\label{eqn:dtauline}
\Delta\uptau_{\mathsf{\lambda,line}} = \frac{\ln\left(\frac{f_{\mathsf{H\beta,1}}}{f_{\mathsf{H\beta,2}}}\right) - \ln\left(\frac{f_{\mathsf{H\alpha,1}}}{f_{\mathsf{H\alpha,2}}}\right)}{Q_{\mathsf{H\beta,line}}-Q_{\mathsf{H\alpha,line}}} Q_{\mathsf{\lambda,line}}
\end{equation}
where $Q_{\mathsf{H\beta,line}}-Q_{\mathsf{H\alpha,line}} = 0.310$.

In Figure \ref{fig:lc} we show the ratio of dust optical depth in the
emission lines to optical depth in the stellar continuum as a function
of \ssfr\ and axis ratio.  In both samples we find a strong trend
  with \ssfr\ (left panel), with both low and high \mustar\ galaxies
  having a line-to-continuum attenuation ratio 50\% higher at lower
  \ssfr\ than at high \ssfr. There is also a strong trend with axis
ratio, with high axis ratio galaxies having a higher line-to-continuum
attenuation ratio than low axis ratio galaxies. Similar trends with
axis ratio are observed in both low and high \mustar\ galaxies, but
low \mustar\ galaxies are generally offset to higher overall ratios.

These trends can be interpreted in terms of the two different dust
components in galaxies (diffuse vs. birth-cloud dust as described
above), with young and old stars suffering different attenuation. The
difference between the high and low \mustar\ galaxies may relate to
the lower overall metallicity of the diffuse ISM in the latter sample,
leading to lower diffuse dust contents.  The slightly non-linear
  trend with \ba\ for the low \mustar\ galaxies follows qualitatively
  that expected for a dust disk of finite thickness, with the optical
  depth of the diffuse ISM varying with the ratio of scale-height to
  scale-length. One further point of note is the equal ratio found at
  high \ssfr\ and low \ba. Such a coincidence of ratios is not
  expected a priori, as the former property is intrinsic to the
  galaxies and the latter is observer dependent. This may indicate the
  true limit of optical observations, where significant numbers of
  sight lines are lost due to high column densities of diffuse dust,
  either due to long path lengths or high dust contents.  Such a
  ``skin'' effect would lead to only the stars closest to the observer
  contributing to the observed integrated light. We will return to all
  these points in Section \ref{sec:disc}.

As in the case of the radial gradients presented in the previous
subsection, in the case that \tauvline/\tauvcont\ is independent of
overall dust content, it can be shown that
\Dtauvline/\Dtauvcont$=$\tauvline/\tauvcont. In Appendix
\ref{app:method}, Figure \ref{fig:app1} shows that this is indeed the
case, and therefore we can write (dashed lines in Figure
\ref{fig:lc}):
\begin{eqnarray}\label{eqn:lc_ba}
\frac{\uptau_{\mathsf{V,line}}}{\uptau_{\mathsf{V,cont}}} &=&
0.75(\pm0.2) + 3.8(\pm0.6) b/a \ \ \ \  {\rm (low\ \mu*)} \\
&=&
1.6(\pm0.2) + 1.4(\pm0.4) b/a \ \ \ \ {\rm (high\ \mu*)}
\end{eqnarray}
and 
\begin{eqnarray}\label{eqn:lc_ssfr}
\frac{\uptau_{\mathsf{V,line}}}{\uptau_{\mathsf{V,cont}}} &=&
-6.4(\pm2.6) -0.9(\pm0.3) \psi* \ \ \ \  {\rm (low\ \mu*)} \\
 &=&
-8.0(\pm2.6) -1.1(\pm0.3) \psi* \ \ \ \ {\rm(high\ \mu*)}
\end{eqnarray}

\citet{Calzetti:1997p5726} measured the same quantity for starburst
galaxies, finding \ebv$_{\mathsf{stars}} =
0.44$\ebv$_{\mathsf{line}}$. For the emission line attenuation curve
given in Eqn. \ref{eqn:emlines} this corresponds to
$\uptau_{\mathsf{V,line}} / \uptau_{\mathsf{V,cont}} =
0.44$R$_{\mathsf{V,line}}/$R$_{\mathsf{V,cont}} = 2.08$. This is shown
as a dotted line in Fig. \ref{fig:lc}, which is close to the value
measured for high \ssfr\ galaxies in our sample.

\subsection{The slope of the attenuation curve in the Ultraviolet}\label{sec:uv}

\begin{table}
  \centering
  \caption{\label{tab:uv} Slopes of the UV
    attenuation curves for each SDSS+UKIDSS+GALEX sample (see
    Eqn. \ref{eqn:uv}), and comparison literature curves. }
  \vspace{0.2cm}
  \begin{tabular}{ccc}\hline\hline
    & \sfuvnuv & \sfuvu \\ \hline 

     $\mu^*<3\times10^8$ & &\\ \hline
All galaxies & $0.28\pm_{0.03}^{0.04}$ & $0.54\pm_{0.03}^{0.02}$\\
-11.00$<\log(\psi_S/{\rm yr^{-1}})<$-9.90 & $0.09\pm_{0.12}^{0.14}$ & $0.36\pm_{0.08}^{0.09}$\\
-9.90$<\log(\psi_S/{\rm yr^{-1}})<$-9.75 & $0.26\pm_{0.11}^{0.17}$ & $0.44\pm_{0.08}^{0.08}$\\
-9.75$<\log(\psi_S/{\rm yr^{-1}})<$-9.62 & $0.47\pm_{0.15}^{0.14}$ & $0.68\pm_{0.07}^{0.10}$\\
-9.62$<\log(\psi_S/{\rm yr^{-1}})<$-9.46 & $0.23\pm_{0.08}^{0.09}$ & $0.45\pm_{0.06}^{0.07}$\\
-9.46$<\log(\psi_S/{\rm yr^{-1}})<$-9.25 & $0.26\pm_{0.09}^{0.08}$ & $0.57\pm_{0.06}^{0.07}$\\
-9.25$<\log(\psi_S/{\rm yr^{-1}})<$-7.90 & $0.41\pm_{0.11}^{0.11}$ & $0.65\pm_{0.06}^{0.07}$\\
 
0.0$<b/a<$0.4 & $0.30\pm_{0.08}^{0.06}$ & $0.55\pm_{0.05}^{0.05}$\\
0.4$<b/a<$0.6 & $0.30\pm_{0.08}^{0.06}$ & $0.56\pm_{0.04}^{0.06}$\\
0.6$<b/a<$0.8 & $0.19\pm_{0.08}^{0.12}$ & $0.53\pm_{0.06}^{0.08}$\\
0.8$<b/a<$1.0 & $0.33\pm_{0.12}^{0.10}$ & $0.53\pm_{0.08}^{0.08}$\\
 
\hline $\mu^*>3\times10^8$ & & \\ \hline
All galaxies & $0.32\pm_{0.05}^{0.04}$ & $0.74\pm_{0.02}^{0.03}$\\
-11.00$<\log(\psi_S/{\rm yr^{-1}})<$-9.90 & $0.22\pm_{0.09}^{0.10}$ & $0.88\pm_{0.07}^{0.08}$\\
-9.90$<\log(\psi_S/{\rm yr^{-1}})<$-9.75 & $0.21\pm_{0.13}^{0.15}$ & $0.66\pm_{0.08}^{0.10}$\\
-9.75$<\log(\psi_S/{\rm yr^{-1}})<$-9.62 & $0.30\pm_{0.12}^{0.18}$ & $0.73\pm_{0.10}^{0.09}$\\
-9.62$<\log(\psi_S/{\rm yr^{-1}})<$-9.46 & $0.39\pm_{0.11}^{0.10}$ & $0.72\pm_{0.07}^{0.07}$\\
-9.46$<\log(\psi_S/{\rm yr^{-1}})<$-7.90 & $0.29\pm_{0.09}^{0.10}$ & $0.67\pm_{0.05}^{0.06}$\\
 
0.0$<b/a<$0.4 & $0.07\pm_{0.09}^{0.11}$ & $0.67\pm_{0.08}^{0.07}$\\
0.4$<b/a<$0.6 & $0.23\pm_{0.08}^{0.10}$ & $0.64\pm_{0.06}^{0.06}$\\
0.6$<b/a<$0.8 & $0.48\pm_{0.10}^{0.10}$ & $0.78\pm_{0.05}^{0.06}$\\
0.8$<b/a<$1.0 & $0.56\pm_{0.11}^{0.09}$ & $0.88\pm_{0.05}^{0.08}$\\
\hline 

    MW (CCM+OD) & 0.06 & 0.64 \\
    C00         & 0.58 & 0.62 \\
    LMC         & 0.27 & 0.71 \\
    SMC-bar$^c$ & 1.24 & 1.24 \\
\hline

  \end{tabular}

\end{table}

\begin{figure}
\includegraphics[scale=0.5]{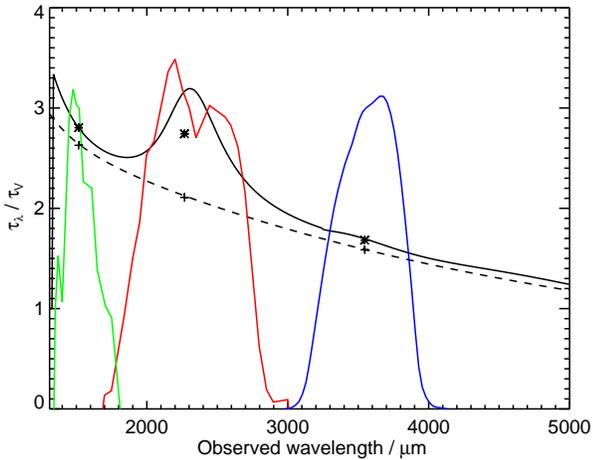}
\caption{The MW (full) and C00 (dashed) attenuation curves in the
  UV-optical wavelength range. Overplotted are the UV filter response
  functions from GALEX and the SDSS u-band. The stars (MW) and crosses
  (C00) indicate the attenuation in each of these bands
  after convolution with the filter functions.  }\label{fig:filter}
\end{figure}

In the previous subsections we used the larger SDSS+UKIDSS galaxy
sample. We now turn to the analysis of the subsample of galaxies with
UV fluxes, repeating the pair-matching procedure from the start. The
results of the optical-NIR fits remain the same as in the full sample,
albeit with larger errors resulting from the smaller sample sizes. In
this section we define the slopes for FUV-NUV
and FUV-$u$ as:
\begin{equation}\label{eqn:uv}
  s' = - \frac{\log(-\mathcal{T}_{\lambda_1}) -
    \log(-\mathcal{T}_{\lambda_2})}{\log\lambda_1 -
    \log\lambda_2}\\
\end{equation}
Note that $s'$ would only be equivalent to the power-law slopes ($s$)
of the previous subsections (Eqn. \ref{eqn:param}), if the attenuation
in the K-band (\dtk) were equal to zero. Due to the smaller dataset
used in this sample, we quote $s'$ rather than $s$ in order to provide
measurements that are independent of the optical-NIR attenuation
curves.  At these short wavelengths, $s$ and $s'$ are not
significantly different. We test this by measuring both $s$ and $s'$
from the literature dust curves, normalising at the rest-frame
wavelength of the K-band at the median redshift of our samples. We
find $s'-s\sim0.03$, well within our measurement errors.

To quantify the influence of the strong bump in the MW extinction
curve on broadband attenuation measurements, we convolve the
comparison dust curves with the filter response functions (see
also the discussion in Section \ref{sec:formalism}). Figure
\ref{fig:filter} shows the MW and C00 dust curves in the UV,
with the FUV, NUV and $u$-band filters overplotted. The stars/crosses
indicate the filter-averaged attenuation of each curve in the three
bands: clearly the filter-averaged attenuation in the GALEX NUV band
for the MW curve is less than the actual attenuation at 2175\AA. This
figure also shows how the slope between the FUV and NUV bands, at
median rest-frame wavelengths of 1430\AA\ and 2138\AA, probes the
depth of the 2175\AA\ dust feature, whereas the slope between the FUV
and $u$ bands measures the shape of the underlying dust continuum.

\begin{figure*}
\includegraphics[scale=0.5]{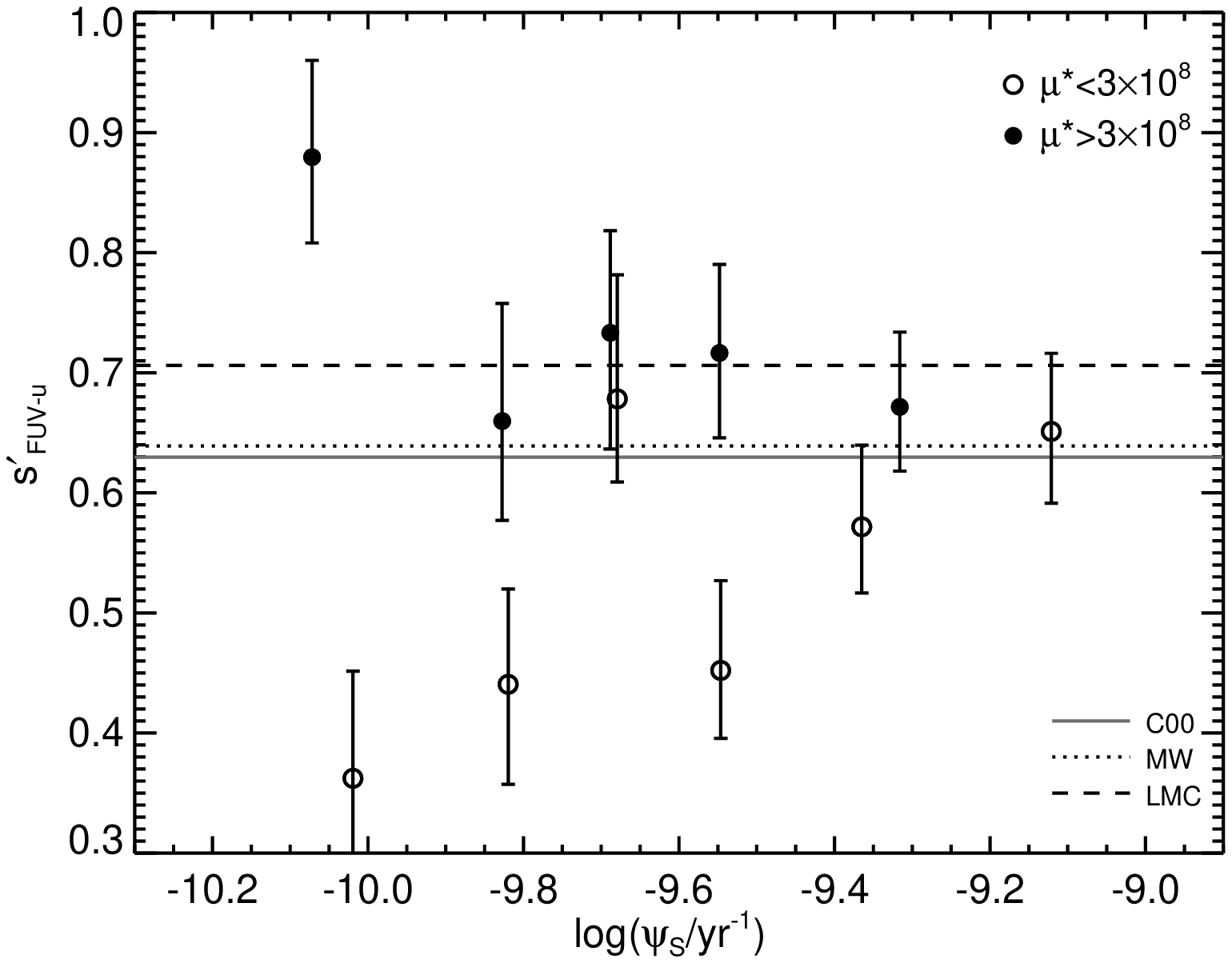}
\includegraphics[scale=0.5]{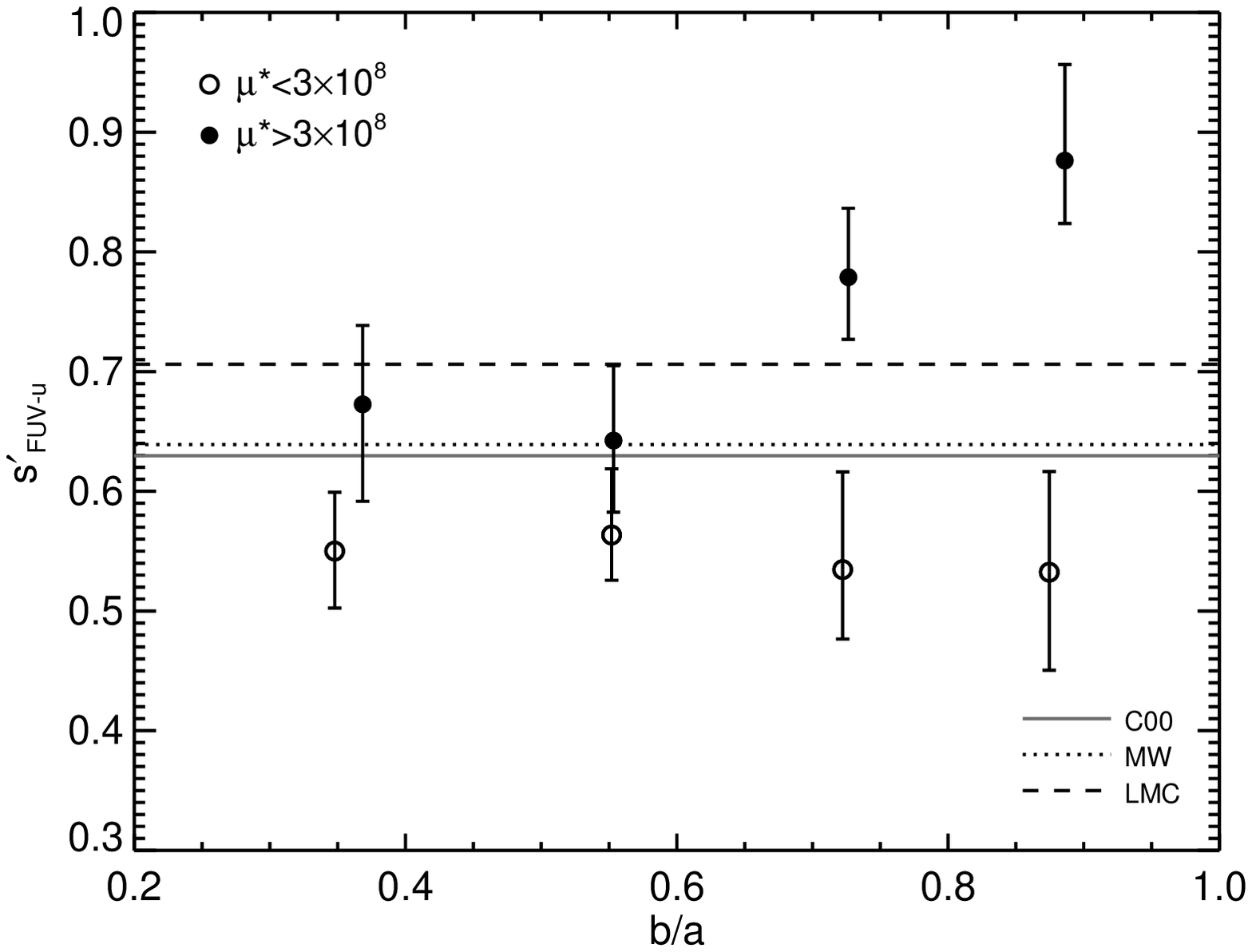}
\includegraphics[scale=0.5]{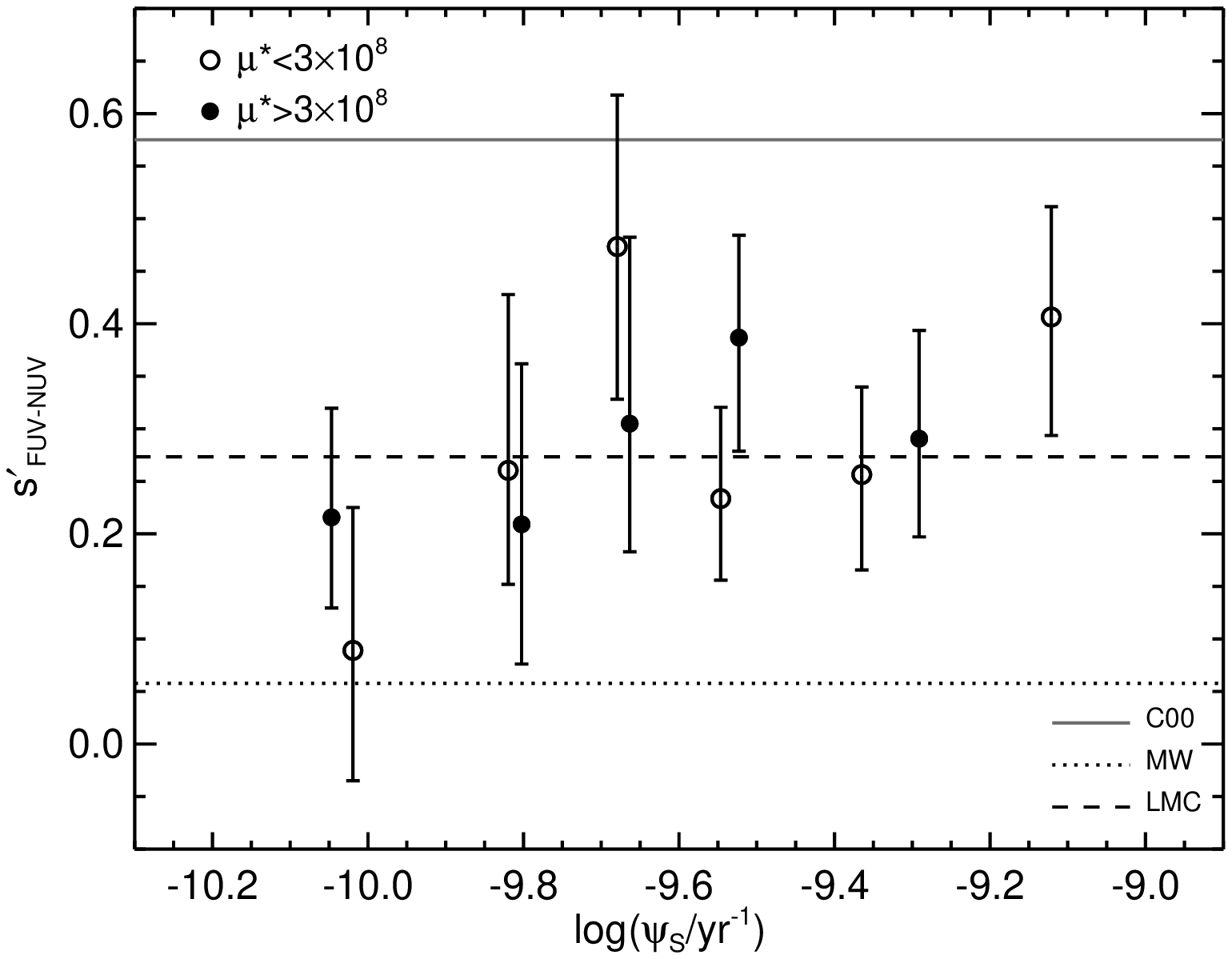}
\includegraphics[scale=0.5]{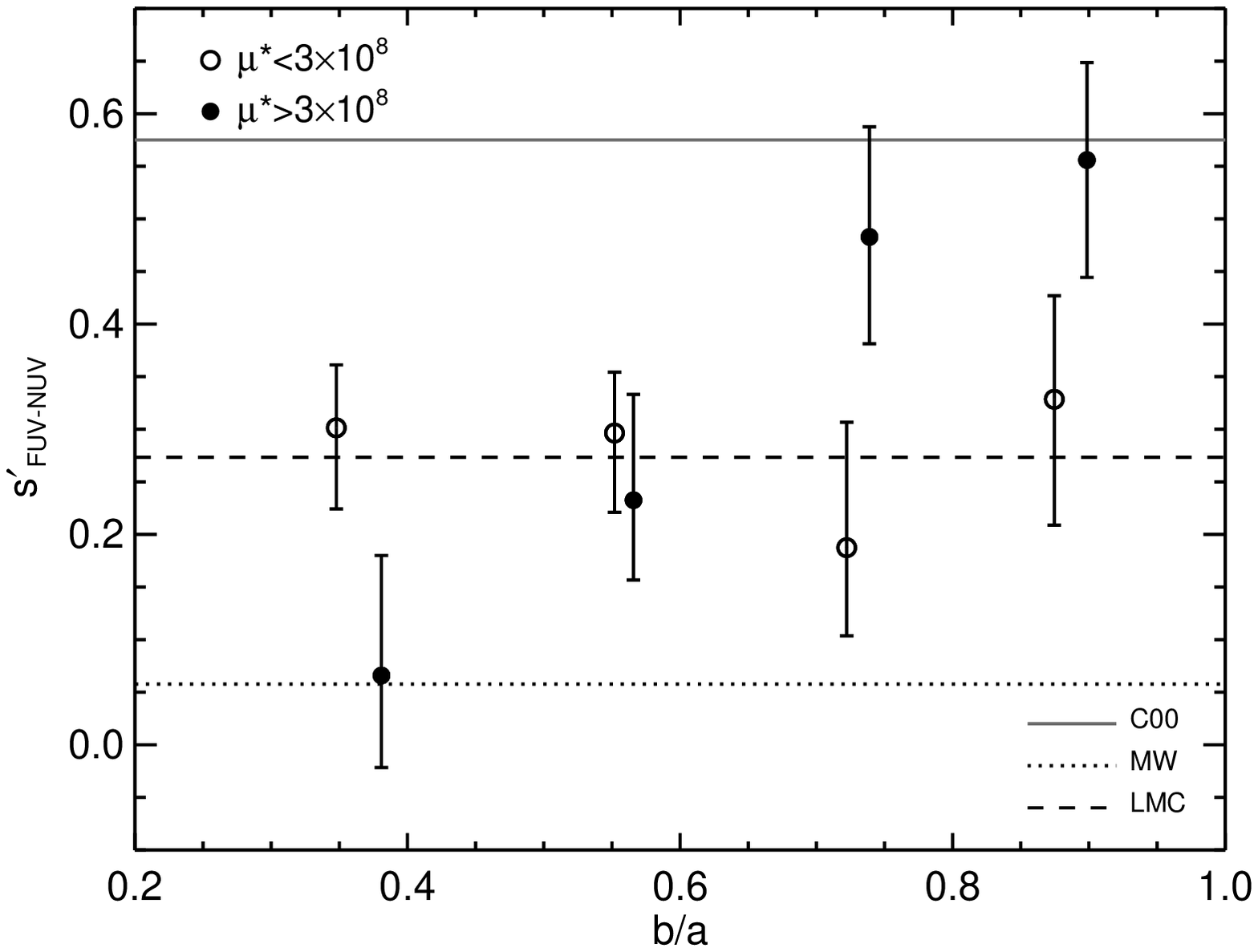}

\caption{Power-law slope of the attenuation curves in the UV
  wavelength range, as a function of \ssfr\ (left) and observed axis
  ratio (right). In the top panels we plot FUV-$u$ band slope. The
  lower panels show the FUV-NUV slope. For clarity in the lower panels
  the high \mustar\ symbols have been offset slightly to the
  right. The horizontal lines mark the corresponding slopes for the MW
  (dotted), C00 (gray, full), LMC (dashed) and SMC-bar (dot-dash)
  curves after convolving with the filter response functions. The
  FUV-$u$ and FUV-NUV slopes of the SMC-bar extinction curve are 1.24,
  which is beyond the axis range. }\label{fig:uv}
\end{figure*}

Table \ref{tab:uv} gives the measured UV slopes for all samples,
together with comparison values for dust curves in the literature. In
the top panels of Fig. \ref{fig:uv} we find that \sfuvu\ is slightly
higher on average in the high \mustar\ samples, than the low \mustar\
samples. The two samples straddle the value measured for the MW and C00
curves (\sfuvu$=0.6$ in both cases). There is some evidence that the
UV attenuation curve steepens with increasing \ba\ in the high
\mustar\ sample, and there is a possible steepening of slope with
\ssfr\ in the low \mustar\ sample.

In the bottom panels of Fig. \ref{fig:uv} we find that \sfuvnuv\
generally lies between the values measured from the MW and C00
curves. The MW extinction curve, with a strong 2175\AA\ dust-bump, has
\sfuvnuv$ = 0.06$, whereas the C00 attenuation curve with no dust bump
has \sfuvnuv$= 0.58$, similar to the global (FUV-$u$) slope of this
curve (\sfuvu$=0.62$).  For the high \mustar\ galaxies we find that
the bump strength increases strongly with increasing \ba, ranging from close to
the C00 curve in face-on galaxies, to close the MW curve in edge-on
galaxies. The same trend is not observed in the low \mustar\
galaxies. A weaker trend of increasing bump strength with decreasing
\ssfr\ may reconcile our results with the weak bump observed in
starburst galaxies by \citet{Calzetti:2000p4473}.

The 2175\AA\ feature is widely believed to be a pure absorption
feature, with its strength in extinction curves varying dependent on
the chemical composition and/or physical environment of the dust
grains \citep{2003ARA&A..41..241D}.  In attenuation curves, the
balance of scattering vs. absorption could affect the strength of the
feature. The similarity in overall strength between the low and
high-\mustar\ galaxies, which have different average metallicities,
could imply that much lower metallicities than are present in our
sample are required to reduce the strength of the feature through
chemical composition effects. The samples studied here are dominated
by luminous and relatively massive galaxies. The C00 curve is well
known for the lack of 2175\AA\ feature, which may be due to dust grain
destruction in the extreme physical environment. Again, although there
is a hint in the data that the strength of the 2175\AA\ feature
weakens with increasing \ssfr, the samples studied here are dominated by
ordinary star-forming galaxies, not extreme starbursts.

The trend with axis ratio in the high \mustar\ galaxies is
interesting, as is the fact that the same trend is not seen in the low
\mustar\ galaxies. The lack of overall difference in strength between
the high and low \mustar\ samples implies that metallicity effects are
not important here, which leaves the possibility of geometric effects.
If dust is distributed uniformly with the stars, then by increasing
the inclination of the galaxy, the amount of interstellar dust that
star light passes through on its way to the observer increases, which
will increase the optical depth of absorption by the grains
responsible for the 2175\AA\ feature.  However, in the case of high
\mustar\ galaxies which have significant bulges, axis ratio is not as
cleanly linked to inclination as in the low \mustar\ galaxies. This
makes our results difficult to interpret without further image
decomposition that is not possible with GALEX data.

We conclude that a 2175\AA\ dust feature is present in most types of
luminous galaxies, except possibly at high \ssfr, but its observed
strength in attenuation curves can be modified by geometric
effects. How these geometric effects act will require better spatially
resolved observations and more detailed evaluation of dust models.

 The implications of a significant 2175\AA\ dust feature extends
  beyond interest in the composition of dust grains, as it impacts
  significantly on the observed UV spectral slope of galaxies and thus
  estimates of the star formation rate density at high redshift
  \citep{Meurer:1999p2873,Bouwens:2009p6639}, and the metallicity and
  dust contents of high redshift galaxies
  \citep{Meurer:1997p6688,Bouwens:2010p6650,Dunlop:2011p6643}. Predictions
  for the exact strength of the effect are difficult given the wide
  variety of bands used to measure the UV spectral slope (commonly
  referred to as $\beta$), and the lack of constraints on the stellar
  population of high redshift galaxies. However, when the reddest band
  used to measure $\beta$ lies close to 2175\AA\ as in
  \citet{Bouwens:2010p6650}, incorrectly assuming a dust curve with no
  2175\AA\ feature would lead to an underestimation in the amount of
  dust, leading to the incorrect conclusion that the stellar
  populations are extremely metal poor. 

\section{Global fit of the dust attenuation curve}\label{sec:empfits}

\begin{table*}
  \centering
  \caption{\label{tab:empfit} Fitted constants for
    Eqn. \ref{eqn:empfit}, which defines the shape of the dust attenuation curve as
    a function of \ssfr\ and axis ratio. }
  \vspace{0.2cm}
  \begin{tabular}{cccccccccc}\hline\hline
    Sample & $c_1$ & $c_2$ & $c_3$ & $c_4$ & $c_5$ & $c_6$ & $c_7$ &
    $c_8$ & $c_9$\\ \hline

       \mustar$< 3\times10^8$&0.15&0.00&0.20&0.70&0.00&0.40&1.10&0.40&-0.10\\

       \mustar$> 3\times10^8$&0.20&0.90&0.10&1.10&0.30&-0.20&1.30&0.60&0.10\\

  \end{tabular}

\end{table*}

\begin{figure*}
\includegraphics[scale=0.5]{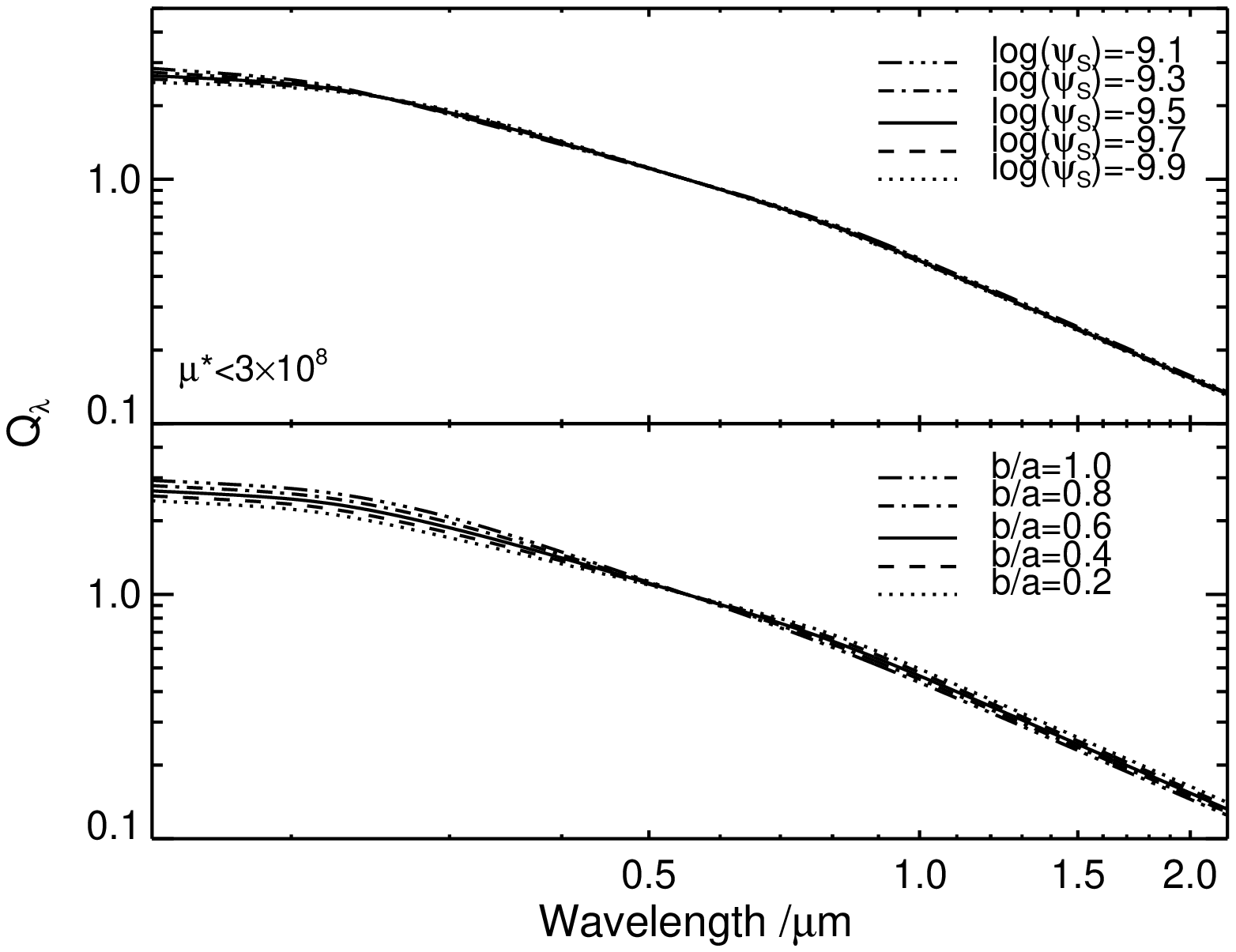}
\includegraphics[scale=0.5]{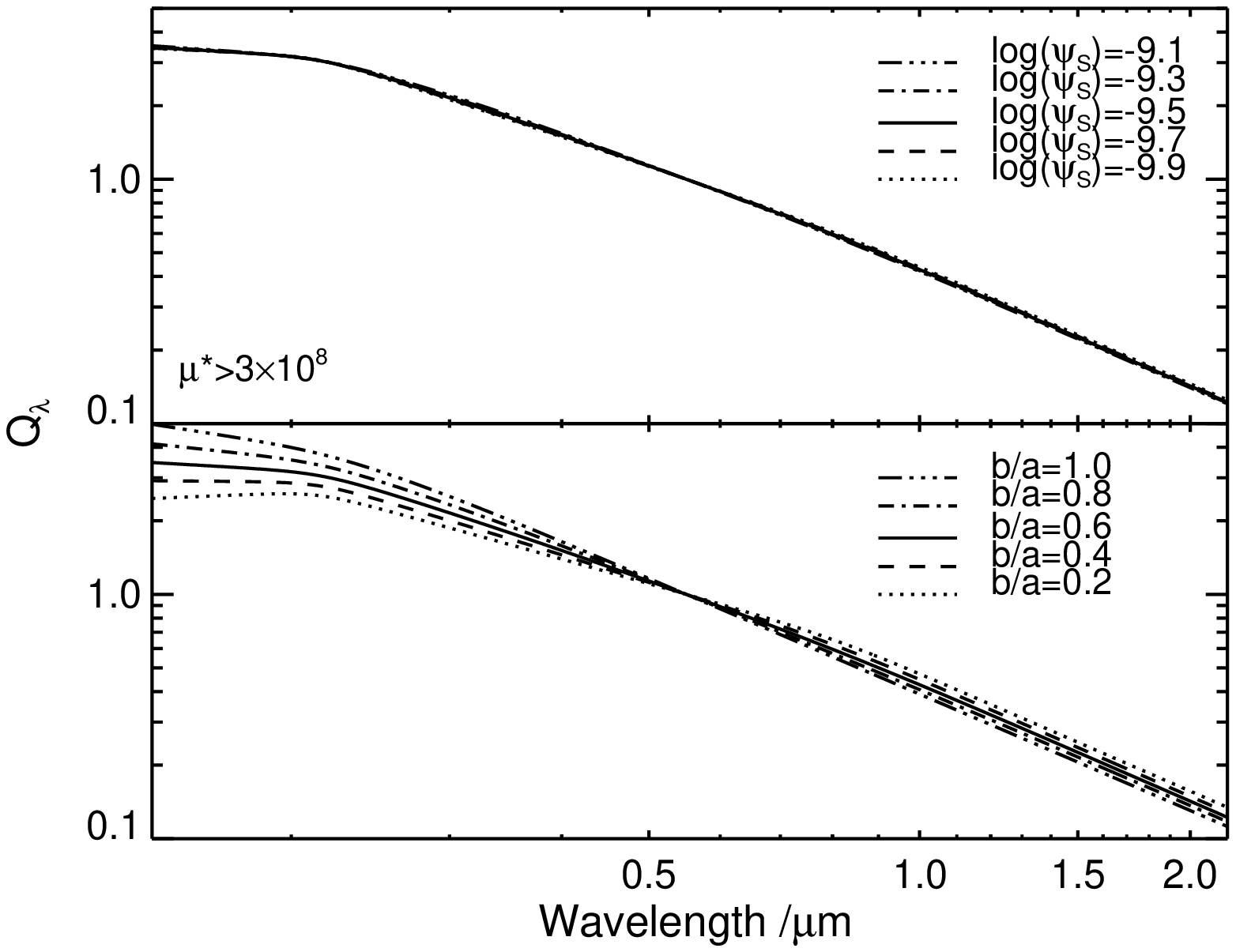}
\caption{Dust attenuation curve shape (\ql),  as given by
    Eqn. \ref{eqn:empfit}, as a function of \ssfr\ (top) and axis
  ratio (bottom) for galaxies with low \mustar\ (left) and high
  \mustar (right). Attenuation curves at values of \logssfr$ =
    [-9.9,-9.7,-9.5,-9.3,-9.1]$ and \ba$=[0.2,0.4,0.6,0.8,1.0]$ are
    shown as different style lines, as indicated in the top right of
    each panel. When variation with \ssfr\ (\ba) is shown, \ba\
  (\logssfr) is held constant at 0.6 (-9.5). }\label{fig:empfit}
\end{figure*}

\begin{figure*}
\includegraphics[scale=0.5]{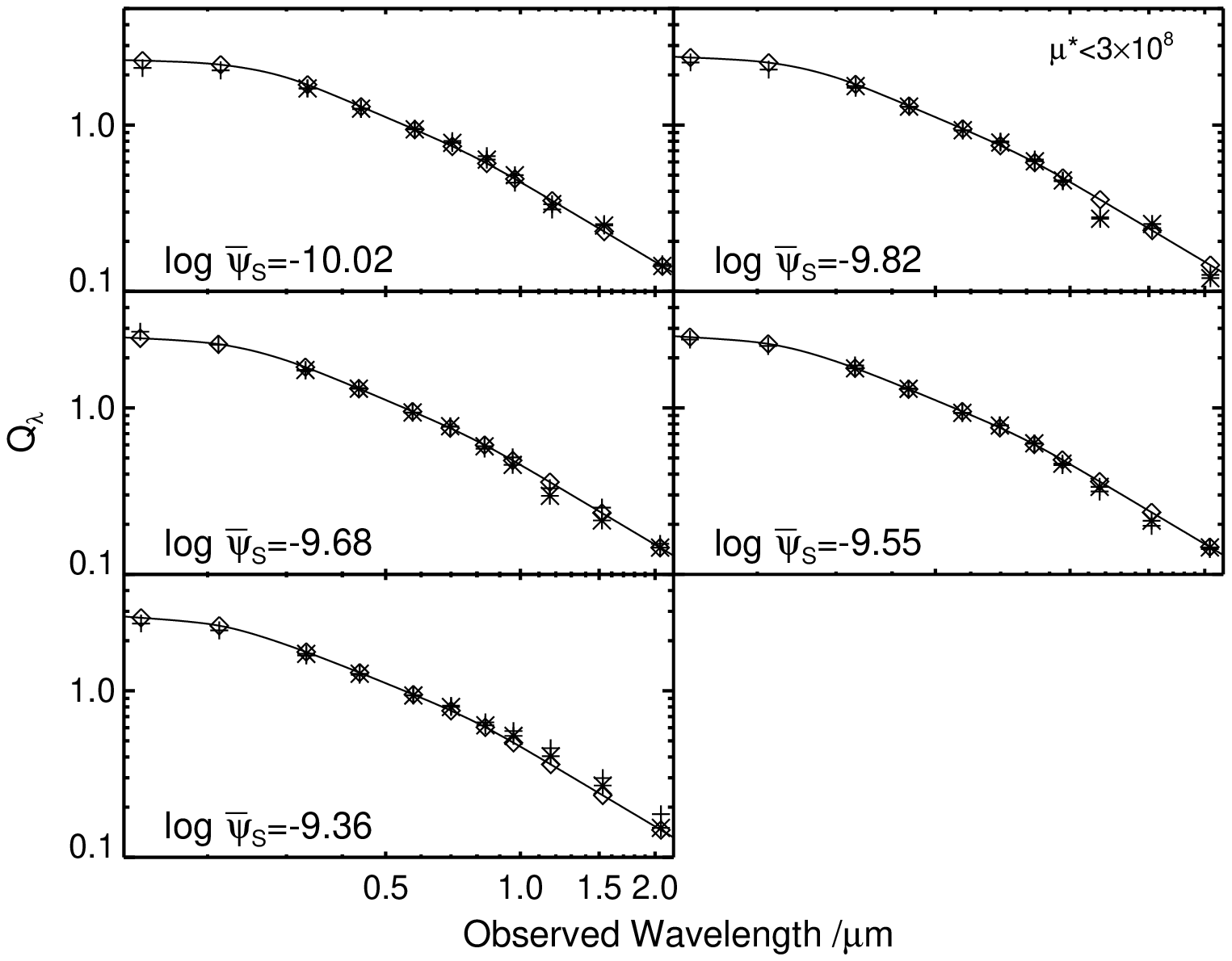}
\includegraphics[scale=0.5]{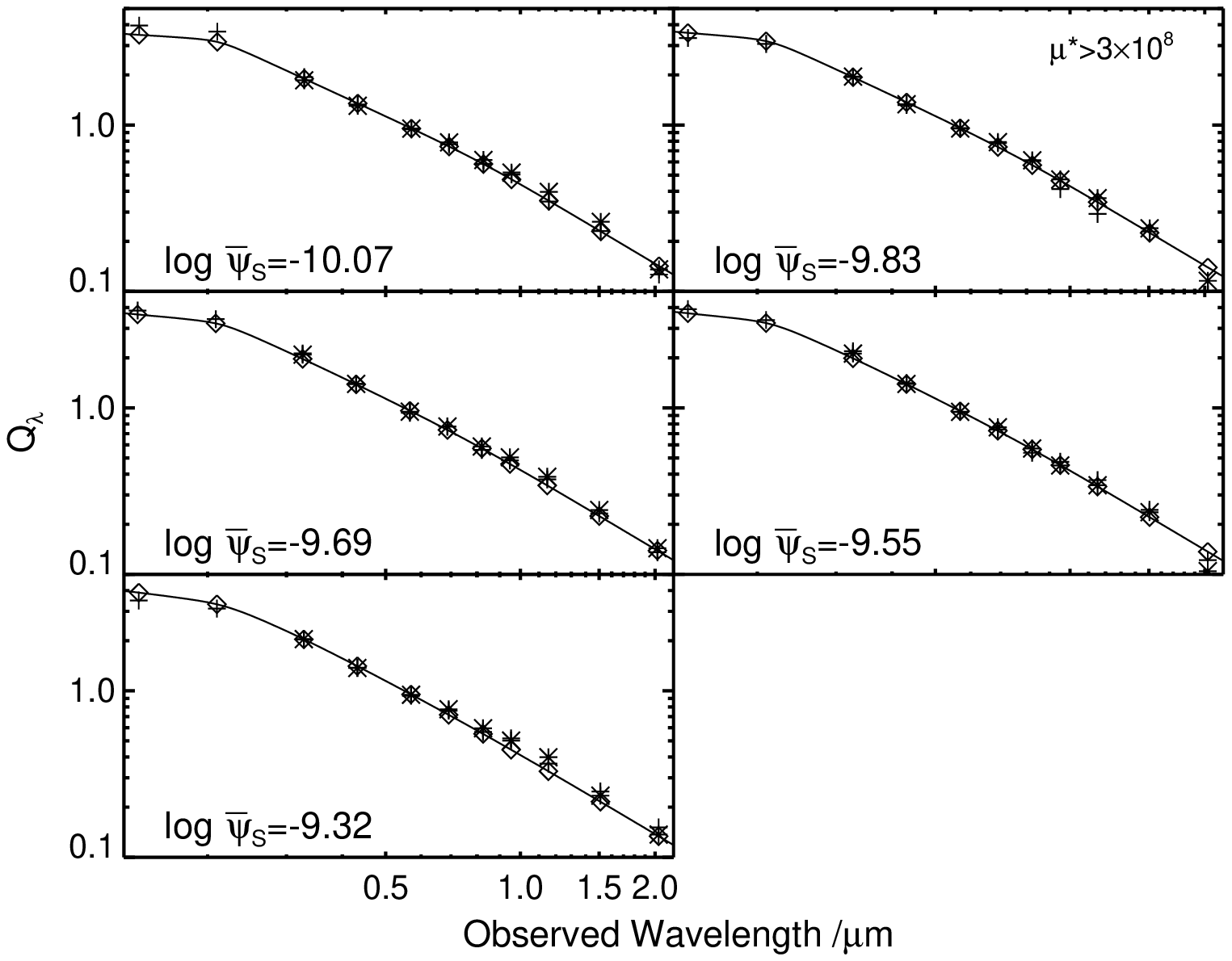}
\includegraphics[scale=0.5]{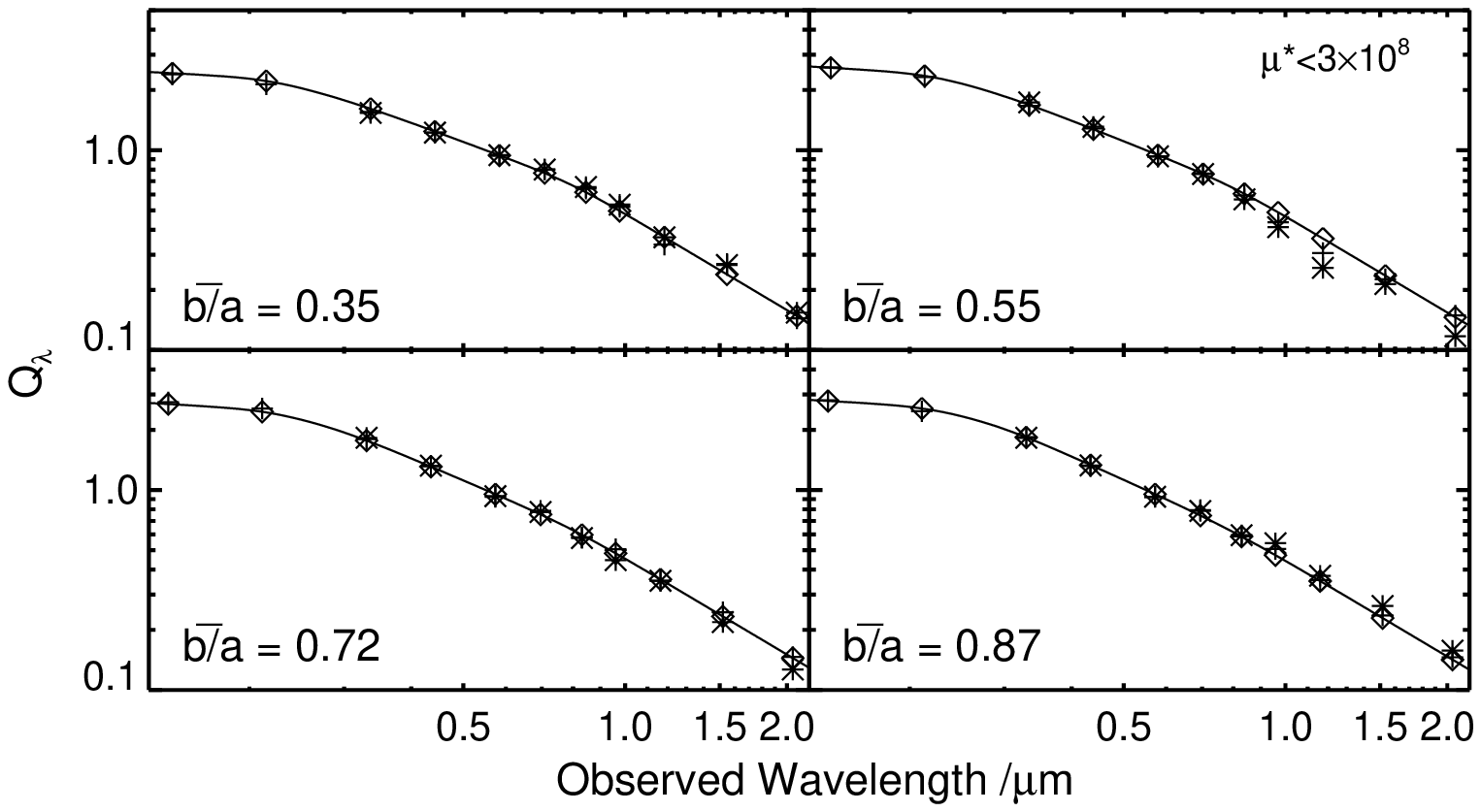}
\includegraphics[scale=0.5]{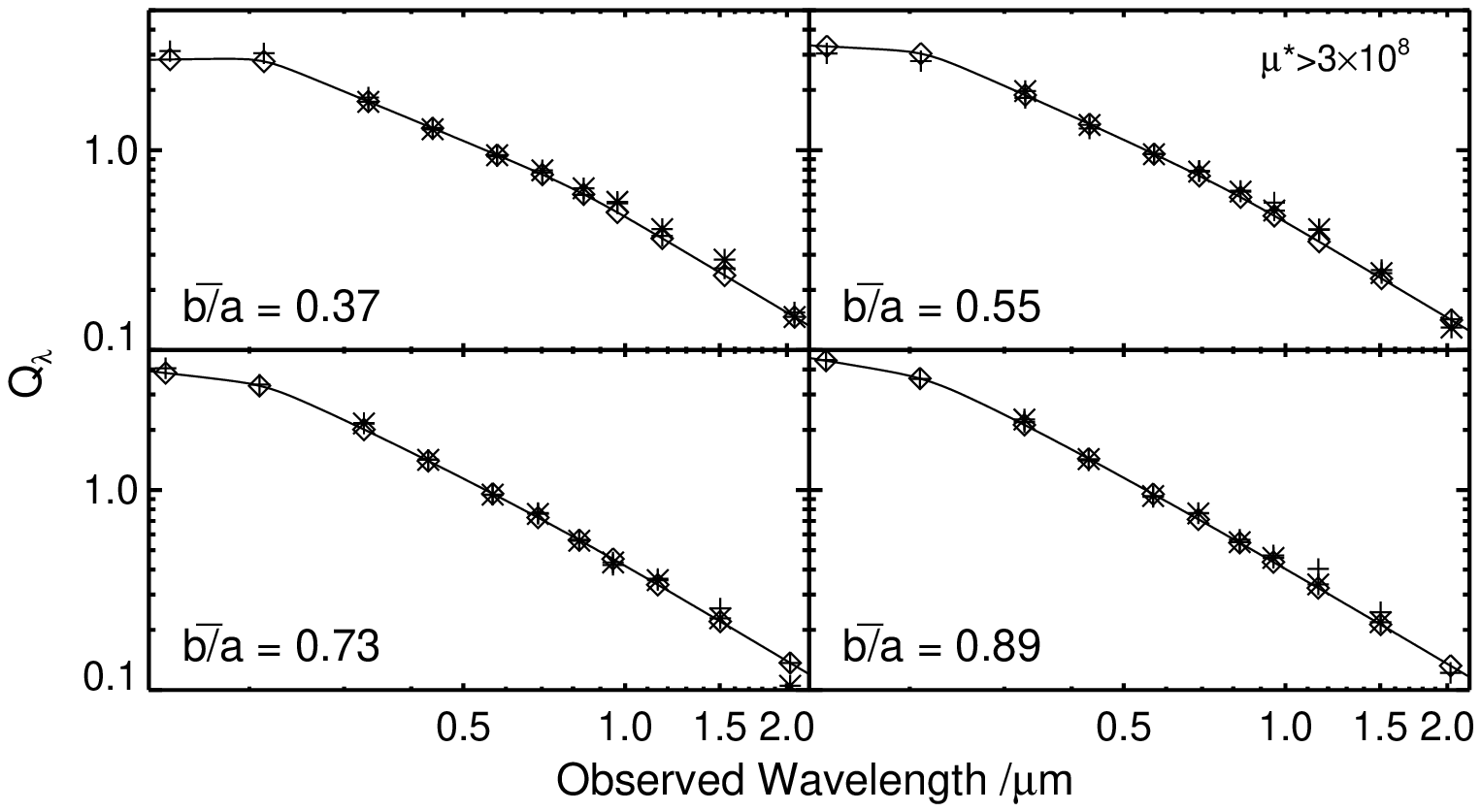}

\caption{The observed shape of the dust attenuation curves in the
  binned samples (with UV data: crosses; without UV data: stars
   [i.e.normalised version of Figs. \ref{fig:curves_ssfr} and
  \ref{fig:curves_ab}]), compared to the empirical dust curve shapes
  derived from the entire (unbinned) samples (line and diamonds). The
  median \ssfr\ and \ba\ of the binned samples have been used to
  extract the appropriate attenuation curve from
  Eqn. \ref{eqn:empfit}, these are given in the bottom left of each
  panel. The left (right) panels show the low (high) \mustar\
  samples. The top (bottom) panels show the samples binned in \ssfr\
  (\ba).  }\label{fig:emppara}
\end{figure*}

In the previous section, we used a combination of a parametrised fit
and direct colour measurements to study the shape of the attenuation
curves for independent galaxy samples binned by \ssfr\ and axis ratio.
Parametrised fits of binned samples are important for robustly
identifying systematic changes, but they do not provide an
``attenuation law'' that can be used at all wavelengths on all
datasets.  In this section, we provide an empirically derived curve to
describe the attenuation by dust at arbitrary wavelength, as a
function of \ssfr\ and axis ratio, suitable for any observational
dataset. Where \ssfr\ and axis ratio are unknown, and no informed
guess can be made, the ``typical'' curve may be defined at the central
points of \ssfr$\sim-9.5$ and axis ratio \ab$\sim0.6$.

In this section the data are not binned in \ssfr\ or \ba, so that each
galaxy pair can be included at its rest-frame wavelength and a smooth
function with \ssfr\ and \ba\ can be derived. We use a multiple
smoothly broken power-law to describe the shape of the curve
\citep{Guidorzi:2009p5965}. 

In order to derive an average attenuation curve shape (\ql) without prior binning of the data,
we require an estimate of the optical depth in the continuum
(\Dtauvcont) for each pair, a quantity not known a priori. We
therefore use the emission line optical depth, taking into account the
variation in line to continuum dust opacity (\Dtauvline/\Dtauvcont)
with \ssfr\ and \ba\ observed in Section \ref{sec:lc}, and the radial
gradients in overall dust content observed in Section
\ref{sec:radgrad}.

With only 2 broad bands in the UV, an accurate description of the
2175\AA\ region of the attenuation curve is not possible. While there
is a definite change in slope of the curves, indicative of the
2175\AA\ feature observed along most lines-of-sight in the MW, we are
unable to constrain the precise shape of this feature. We have
therefore chosen to define a break in the attenuation curve at the
central wavelength of the feature observed in the MW, but
interpolate smoothly over the broad band observations. This will not
cause significant errors in the correction of broad band fluxes for
dust attenuation. However, when using our empirical curve to correct
spectral observations, or narrow band fluxes, readers should be aware
of the limitations of our observations.

The curve is composed of four power-law functions with exponents
$s_{[1-4]}$ smoothly joined with a smoothness parameter $n$. We define
the position of the three break points ($\lambda_{\rm b[1-3],eff}$) at
0.2175, 0.3 and 0.8\mum. The first is motivated by the central
position of the dust bump in the MW curve. The second break point is
positioned to produce a smooth transition between the steeper optical
and shallower NUV slopes observed. The final break position is
motivated by the results of our parametrised fits in Section
\ref{sec:paramfits}. The sampling of the curves is too sparse to fit
for the smoothness parameter $n$, and therefore we fix this to be 20
which results in a sufficiently smooth curve.

Following the formalism set out in Section \ref{sec:formalism}, the
shape of the attenuation curve is given by:
\begin{equation}\label{eqn:empfit}
Q_\lambda = \!\!\frac{1}{N}\left[
\left(\frac{\lambda}{\lambda_{\rm b1}}\!\right)^{n\,s_1}\!\!\!\!+\!\!
\left(\frac{\lambda}{\lambda_{\rm b1}}\!\right)^{n\,s_2}\!\!\!\!+\!\!
\left(\frac{\lambda}{\lambda_{\rm b2}}\!\right)^{n\,s_3}\!\!\!\!+\!\!
\left(\frac{\lambda}{\lambda_{\rm b3}}\!\right)^{n\,s_4}
\right]^{-1/n}
\end{equation}
where $N$ is the normalisation, defined such that the curve is unity
at 5500\AA, $Q_{V}=1$; $n=20$ defines the smoothness of the
breaks. Note that due to the smoothing, the exponents of
Eqn. \ref{eqn:empfit} are not all trivially related to the slopes
measured in the previous section.

The effective position of the break points are converted into the
$\lambda_{\rm b[1-3]}$ values according to:
\begin{eqnarray}
\lambda_{\rm b1} &=& \lambda_{\rm b1,eff}\\
\lambda_{\rm b2} &=& \left(\frac{\lambda_{\rm b1}^{s_2}}{\left(\lambda_{\rm b2,eff}\right)^{s_2-s_3}}\right)^{1/s_3}\\
\lambda_{\rm b3} &=& \left(\frac{\lambda_{\rm b2}^{s_3}}{\left(\lambda_{\rm b3,eff}\right)^{s_3-s_4}}\right)^{1/s_4}
\end{eqnarray}
with the effective positions of the breaks as motivated above.

The power-law slopes are allowed to vary with both axis ratio and
\ssfr\ according to simple linear functions:
\begin{eqnarray}
s_1 &=& c_1 + c_2 (b/a)_c + c_3 (\psi_c)\\
s_2 &=& c_4 + c_5 (b/a)_c + c_6 (\psi_c)\\
s_3 &=& c_7 + c_8 (b/a)_c + c_9 (\psi_c)\\
s_4 &=& 1.6
\end{eqnarray}
where $(b/a)_c = (b/a)-0.6$ and $\psi_c = \psi_S+9.5$. The functions
are valid in the primary parameter ranges covered by the binned
samples studied in this paper: $0.3<b/a<0.9$ and $-10.2<\psi_S<-9.3$
for high \mustar\ and $-10.0<\psi_S<-9.1$ for low \mustar, and should
not be extrapolated beyond these values. For galaxies outside of these
ranges the functions could be calculated at the minimal or maximal
extent of the range probed in our samples. The free parameters for
high and low \mustar\ galaxies are provided in Table \ref{tab:empfit}.

Figure \ref{fig:empfit} shows the shape of the attenuation curves
(\ql) for different values of \ssfr\ and axis ratio.  Figure
\ref{fig:emppara} compares the observed attenuation in our binned
galaxy samples to the appropriate empirical curves for the median
\ssfr\ and axis ratio of each sample. Overall the agreement is
excellent, with no trend in residuals with either parameter. 

\subsection{Comparison with other curves}

\begin{figure}
\includegraphics[scale=0.5]{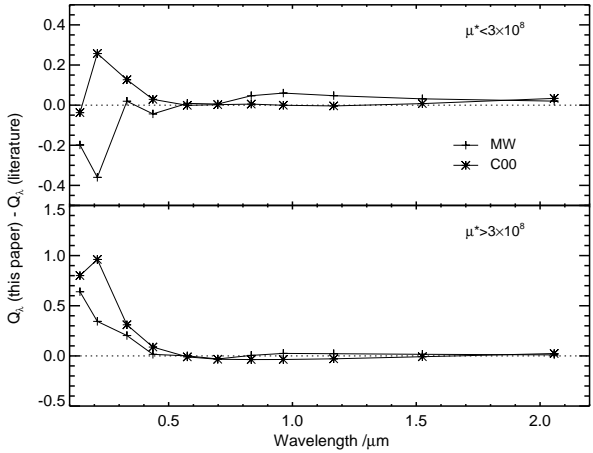}
\caption{The difference between the shape of the attenuation curve derived in this
  paper for galaxies with \logssfr$=-9.5$ and \ba$=0.6$, and the C00
  (stars) and MW (crosses) curves. The top (bottom) panel shows the
  comparison with the attenuation curve for low (high) \mustar\
  galaxies. The NUV point on the MW curve has been convolved with the
  GALEX NUV filter for this comparison, this convolution is not
  necessary for other wavelengths or for the C00 curve (see Figure
  \ref{fig:filter}). }\label{fig:resid} 
\end{figure}

In Fig. \ref{fig:resid} we compare the dust curves derived here, to
the MW and C00 dust curves, for galaxies with \ssfr$=-9.5$ and
\ba$=0.6$. For a galaxy with attenuation in the stellar continua
  of $A_V=1$, the $y$-axis represents the difference in attenuation
in magnitudes. The very steep attenuation in the blue found for the
high \mustar\ galaxies is evident here, as is the strong 2175\AA\
feature compared to the C00 curve. For the low \mustar\ galaxies, the
attenuation is almost indistinguishable to C00 redwards of 5500\AA,
although the curves differ substantially in the blue, and the lack of
a 2175\AA\ feature in the C00 curve is again evident. However, it is
clear that for the low \mustar\ galaxies, the feature is not as strong
as in the MW.

A complete dust attenuation curve is built from the shape (\ql), and
the normalisation i.e. the amount of dust attenuation suffered by the
\emph{continuum} at 5500\AA\ ($\tau_V$)\footnote{Note that this is
  different from Calzetti et al. who define the normalisation by the
  amount of dust in the Balmer emission lines, and include the
  conversion between the amount of dust in the lines and continuum in
  the formula for the attenuation curve itself.}. The most appropriate
estimator for $\tau_V$ will depend on the particular dataset. If the
Balmer emission lines are used, Eqns. \ref{eqn:lc_ba} and
\ref{eqn:lc_ssfr} provide the conversion between $\tau_{V,line}$ and
$\tau_{V,cont}$ when the Balmer emission lines and continuum
measurements are made within the same aperture. This is the best
estimate possible with the SDSS data, but it assumes that the ratio
measured within a 3\arcsec diameter fibre is representative of the
entire galaxy. Other possible estimations of $\tau_V$ could come from
the measurement of the dust emission in the FIR
\citep{2008MNRAS.388.1595D}, or depth of the mid-IR silicate
absorption feature \citep{Wild:2011p6538}.

\section{Discussion and Conclusions}\label{sec:disc}

\begin{figure*}
\includegraphics[scale=0.5]{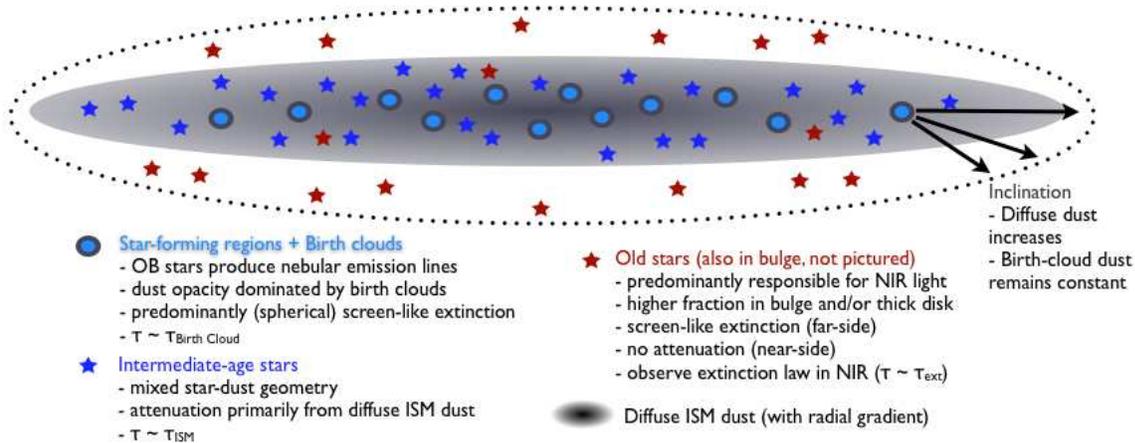}
\caption{Pictorial representation of the distribution of dust in
  galaxies, which qualitatively can account for most results found in
  this paper. The bulge has been omitted for clarity, but is expected
  to contain predominantly old stars with little diffuse dust. }\label{fig:cartoon} 
\end{figure*}

Our main results are:
\begin{description}
\item[{\bf NIR slope: }] The slope of the attenuation curve in the NIR
  is consistent with that measured for the MW extinction curve
  ($\sim1.6$) for all types of galaxies, irrespective of \ssfr, inclination
  angle or the presence or absence of a bulge. 
\item[{\bf Optical/UV slope: }] We find significant trends in the
  optical/UV slope of the attenuation curve with galaxy
  properties. The high \mustar\ galaxies show a steeper blue/UV slope
  than the low \mustar\ galaxies and the average MW extinction curve. A
  flattening of the slope with increasing inclination is evident in
  both the high and low \mustar\ galaxies. There is no trend in
  optical slope with total optical depth.
\item[{\bf 2175\AA\ feature: }] Most samples of galaxies exhibit a
  discontinuity in the UV attenuation curve that is suggestive of a
  2175\AA\ dust feature at a level that is slightly weaker than seen
  in the MW. In the high \mustar\ galaxies the strength of the feature
  varies with axis ratio, from as strong as that observed in the MW
  extinction curve at low axis ratios, to undetected at high axis
  ratio.  There is some evidence that the bump weakens with increasing
  \ssfr.
\item[{\bf UV dust attenuation:}] For low (high) \mustar\ galaxies
  with $A_V=1$, we find typically 0.3 (1.0) magnitudes more
  attenuation in the NUV compared to the C00 curve for starburst
  galaxies.
\item[{\bf Emission lines: }] We find that Balmer emission lines
  experience 2 - 4 times more attenuation than the continuum at
  5500\AA\ within the 3\arcsec\ SDSS fibre aperture.  The ratio of
  emission line to continuum dust optical depths (\tauvline/\tauvcont)
  varies strongly with galaxy properties. It decreases with increasing
  \ssfr, increases with increasing axis ratio (decreasing inclination) and is lower
  in high-\mustar\ than in low-\mustar\ galaxies.
\item[{\bf Radial gradients: }] Radial gradients are significant in all
  samples, and increase with increasing \ssfr. The strength of the
  gradients is not modified by the presence of a bulge. 
\end{description}
 
\subsection{Comparison to the two component dust model}
The primary aim of this paper is to measure the shape of the dust
attenuation curve of star-forming galaxies of different types. We now provide
simple qualitative arguments to interpret the observed trends in the
context of the distribution of dust and stars in
galaxies. Fig. \ref{fig:cartoon} shows a pictorial representation of
one possible configuration for the global distribution of dust in
star-forming galaxies, which can account for the majority of our
results. This model for the distribution of dust relative to stars is
based upon the popular two component dust model, in which star-forming
regions are surrounded by dense birth clouds, which evaporate on
timescales similar to the main sequence lifetimes of B stars
($\sim10^7$ years). The remainder of the interstellar medium contains
diffuse dust.

This two component dust distribution has been included in several
models for dust in galaxies \citep{Silva:1998p4065,
  2000ApJ...539..718C, Popescu:2000p6154, Tuffs:2004p6123}. However,
many important details of the model are unconstrained by
observations. For example, how is the diffuse dust distributed
relative to stars in the galaxy disks, how does dust content change
with star formation rates, and what is the covering factor of the
stellar birth clouds?  Here, we attempt to address these questions in
light of our new results combined with results in the literature.

\subsubsection{Covering factor of birth clouds}
As discussed in detail in Section \ref{sec:intro}, one popular method
for studying dust attenuation in galaxies, is by comparing the SEDs of
inclined and face-on galaxies. Through such a study,
\citet{Yip:2010p4536} found that the distribution of the ratio of \ha\
to \hb\ luminosity does not vary with galaxy inclination. There are
two plausible explanations for this result: (i) Emission lines are
attenuated predominantly by the birth cloud from which they
originate. The increase in dust opacity caused by an increase in path
length through the diffuse ISM dust is insignificant in comparison to
the difference in dust opacity between the birth clouds and ISM. (ii)
Sightlines to star-forming regions that lie deeper within a galaxy are
entirely opaque, causing us to measure the dust opacity only towards
star-forming regions in an outer ``skin''.

These different scenarios can alternatively be expressed in terms of
the covering factor of stellar birth clouds in galaxies. If the
covering factor is high, optical radiation from stars and gas produced
in the inner layers of the galaxy will have near-unit probability of
being absorbed by a dense cloud before escaping from the galaxy (skin
model). Such an effect is not supported in general by the fact that
SFR estimates from optical emission lines correlate well with those
estimated from mid-IR and far-IR radiation; nor by the weak trend
between \ssfr\ and \ab\ mentioned in Section \ref{sec:sample}; nor by
the weak trend between \ha\ equivalent width and inclination measured
by \citet{Yip:2010p4536}. Additional evidence against a
  significant skin effect comes from measurements of \ha\ rotation
  curves, which find less than unity optical depths to \ha\ at all but
  the steepest inclination angles \citep{Goad:1981p6624,
    Giovanelli:2002p6625, Misiriotis:2005p6630}. 

These observational results suggest that optically thick birth
  clouds have a low covering factor in ordinary star-forming
  galaxies. Only in highly inclined galaxies do we see some evidence
  for a skin effect, as we observe a slight decrease in observed
  radial gradient strength (Figure \ref{fig:dtauv}). It is well known
  that optical data alone is insufficient to study some unusual
  galaxies, due to large dust optical depths. Does this limitation of
  optical data apply to the general galaxy population? Our results
  suggest that only in highly inclined galaxies and galaxies with very
  high \ssfr, where the diffuse dust optical depths are high due to
  long pathlengths or large diffuse dust contents, is optical data
  limited by dust.

\subsubsection{Dust scale height and mixing with stars }
If stellar birth clouds have a small covering factor, most of the
light dominating the UV-NIR emission from a star-forming galaxy has
been attenuated by diffuse ISM dust alone. The geometrical arrangement
of stars mixed with a dusty ISM can lead to an attenuation curve that
is flatter (i.e. greyer) than the underlying extinction curve. This is
because, at any wavelength, the emergent radiation is dominated by
photons that are emitted by sources lying behind dust columns with
less than unity optical depth. Thus, emergent red light originates
from deeper within the galaxy than blue light, and attenuation appears
more constant with wavelength. In Section \ref{sec:opt} and Figure
\ref{fig:sopt} we found that, in contrast, galaxies with high \mustar\
have an attenuation curve that is \emph{steeper} than expected at
UV-optical wavelengths. This could occur if older/redder stars suffer
less attenuation on average than younger/bluer stars, for example if
both the younger stars and diffuse dust have a smaller scale height
than older stars. In the case of high \mustar\ galaxies, there could
be an additional effect from the presence of a bulge: galaxy bulges
tend to have older stellar populations than galaxy disks, and
sightlines not obscured by the disk may have lower dust contents. 

The observed universality of the slope of the NIR attenuation curve
(Fig. \ref{fig:snir}), with a value equal to that of the MW extinction
curve, implies a universal size distribution for large dust grains,
and additionally sets further constraints on the relative geometry of
old stars and dust in galaxies. For the old stars to suffer
screen-like dust extinction, they must predominantly lie outside of
the dust disk. This could occur through heating/migration effects of
stars in galaxy disks, or through the thick disk being built through
minor mergers of satellite galaxies which cease star formation as soon
as they loose their gas supplies. 

\citet{Xilouris:1999p6157} and \citet{Dalcanton:2004p6401} have also
suggested that diffuse interstellar dust is distributed throughout the
disk with a smaller scaleheight than that of the stars, following
observations of a small number of local galaxies. A variable scale
height for dust is also included in the models of
e.g. \citet{Tuffs:2004p6123}. Matching the trends presented in this
paper to detailed models of the dust and star distribution in galaxies
will reveal quantitative values such as the relative scale heights of
stars and dust and the total opacity in galactic disks, as a function
of specific star formation rate.

\subsubsection{Diffuse and birth cloud dust}
In Section \ref{sec:lc} and Figure \ref{fig:lc} we observe strong
trends in the ratio of line-to-continuum dust opacity as a function of
axis ratio, \ssfr\ and \mustar. As discussed above, within the
framework of the two component dust model the greater attenuation
observed in the nebular emission lines compared to the stellar
continuum is attributed to the additional attenuation caused by the
stellar birth clouds surrounding the hot OB stars that produce the
emission lines. The covering factor of these birth clouds is unknown,
although we argued above that it is likely to be small. If this is the
case then we can, to first approximation, assume that the attenuation
of the emission lines arises primarily from dust in the birth clouds,
and the attenuation of the stellar continuum arises from dust in the
diffuse ISM. 

In more inclined galaxies, the pathlength through the diffuse dust is
greater: the observed opacity towards all sources lying within
the dust disk increases with increasing inclination. This can have a
significant effect on the optical depth measured in the continuum. By
virtue of the significantly larger opacities of the birth clouds
compared to the diffuse ISM dust, inclination will have less impact on
the total optical depth measured in the Balmer emission lines so long
as the opacity of the ISM is low enough. Thus, the observed
correlation between the ratio of line-to-continuum dust opacity and
axis ratio, and the invariance of ratio of \ha\ to \hb\ luminosity
with inclination found by \citet{Yip:2010p4536}, is consistent with a
two-component dust model, in which the birth clouds have significantly
greater dust optical depths than the diffuse ISM. The increase in
strength of the 2175\AA\ absorption feature with inclination in the
bulge-dominated galaxies may also be related to an increase in diffuse dust
opacity. However, the lack of a similar trend in the low \mustar\
galaxies is puzzling. The answer to this question may provide clues to
the nature of the grain responsible for the 2175\AA\ absorption feature.

At least two factors could cause the trend found between the ratio of
line-to-continuum dust opacity and \ssfr. Firstly, the greater
contribution of young stars to the V-band continuum light in high
\ssfr\ galaxies may cause the measured \tauvcont\ to approach
\tauvline\ at high \ssfr\ if these younger stars suffer greater dust
attenuation relative to older stars.  Secondly, a decrease in diffuse
dust content from star-forming to quiescent galaxies would reproduce
the observed trend. Given the significant amplitude of the observed
trend, and the small quantities of diffuse dust found in elliptical
galaxies, we believe this to be the most plausible explanation. Such
an evolution would have implications for theories of dust grain
formation during star formation, and dust grain destruction in the
diffuse ISM.  The larger ratio of line-to-continuum dust opacity in
low \mustar\ galaxies may relate to a decreased diffuse dust content
in these lower metallicity galaxies. The trend is in qualitative
agreement with observations of diffuse FIR emission, which show that
later type spirals having a lower fraction of diffuse dust than early
type spirals \citep{Sauvage:1992p6767}.

Finally, the strong radial gradients found in the attenuation of the
stellar continua (Fig. \ref{fig:dtauv}) imply that the central regions
of galaxies are dustier than the outskirts, a result presumably
related to the metallicity gradients observed in galaxy disks. The
link between \ssfr\ and dust content is intriguing, but
observationally we are limited by the 3\arcsec\ SDSS fibre; spatially
resolved spectroscopic studies of spiral galaxies will allow us to
explore this area further.

\subsubsection{Summary}
Taken together, our observations are broadly consistent with the
popular two-component model for the distribution of dust in galaxies,
although several points remain unexplained. In Fig. \ref{fig:cartoon}
we have suggested some improvements to the basic picture, based upon
results in this paper. In detail, we suggest a low covering factor for
the dense birth clouds surrounding the youngest stars, we infer a
scale height for the dust that is smaller than that of the old stars,
and we suggest a radial gradient in the amount of diffuse ISM dust
with more dust found towards the center of the galaxy, particularly in
galaxies with high \ssfr. Clearly, a more quantitative comparison
between our results and detailed models will lead to a deeper
understanding of the production and destruction of dust in the
Universe.

\subsection{Applying the dust attenuation curve}

The general properties of the galaxies studied in this paper are
summarised in Figure \ref{fig:distbn}, and the dust curves presented
here should be applied with caution to galaxies outside of these
ranges. In order to apply the dust attenuation curves provided in this
paper to the SED of a star-forming galaxy:
\begin{itemize}
\item Does the galaxy have a significant bulge? If so, use the coefficients for
  the high \mustar\ galaxies. Otherwise use the coefficients for the
  low \mustar\ galaxies.
\item Determine the inclination and \ssfr\ of the galaxy, if either
  are unknown set \ba$=0.6$ and $\log$\ssfr$=-9.5$.
\item Determine the \tauvcont\ of the galaxy. If using the Balmer
  emission lines:
  \begin{itemize}
  \item Account for strong dependence of \tauvline/\tauvcont\ on
    inclination and \ssfr. 
  \item Ensure that the emission lines are measured using the same
    aperture as the continuum data, or correct for radial gradients
    using Figure \ref{fig:dtauv}.
  \end{itemize}
\item Determine the shape of the appropriate attenuation curve from
  Eqn. \ref{eqn:empfit}, or using the IDL code provided in Appendix
  \ref{app:code}. 
\end{itemize}

The attenuation curves presented in this work, like that of C00,
account for the effects of mixed dust-star geometry and global
geometry in galaxies. Unlike the MW, or Magellanic extinction curves,
our attenuation curve does not assume a screen-like dust
geometry. Additionally, including the dependence on \ssfr\ accounts
for the different forms of attenuation that different types of stars
suffer, and the changing balance of these types of star to the
integrated light of the galaxy depending on the SFR of the galaxy.


\section*{Acknowledgements}
The authors would like to thank the referee, M. Boquien, for his
thorough reading of the text and helpful comments that improved the
clarity of the text. We gratefully acknowledge David Hill and Simon
Driver for their help assessing photometric errors using GAMA
comparison data; Nicholas Cross, Nigel Hambly, Paul Hewett, Simon
Hodgkin, Mike Irwin and Steve Warren for their help with using the
UKIDSS dataset; Masataka Fukugita for useful discussions; Christy
Tremonti for her initial work on the project; Sundar Srinivasan and
Gustavo Bruzual for sharing their AGB star expertise. 

VW acknowledges support from a Marie Curie Intra-European fellowship,
the visitor programme of the Institut d'astrophysique de Paris, and
European Research Council Grant (P.I. J. Dunlop).  OV is supported by
the Ministry of Science and Technological Development of the Republic
of Serbia through the project no. 176021.

\bibliographystyle{mn2e}


\begin{appendix}

\section{Tests of  the pair-matching methodology}\label{app:method}

\begin{figure*}
\includegraphics[scale=0.5]{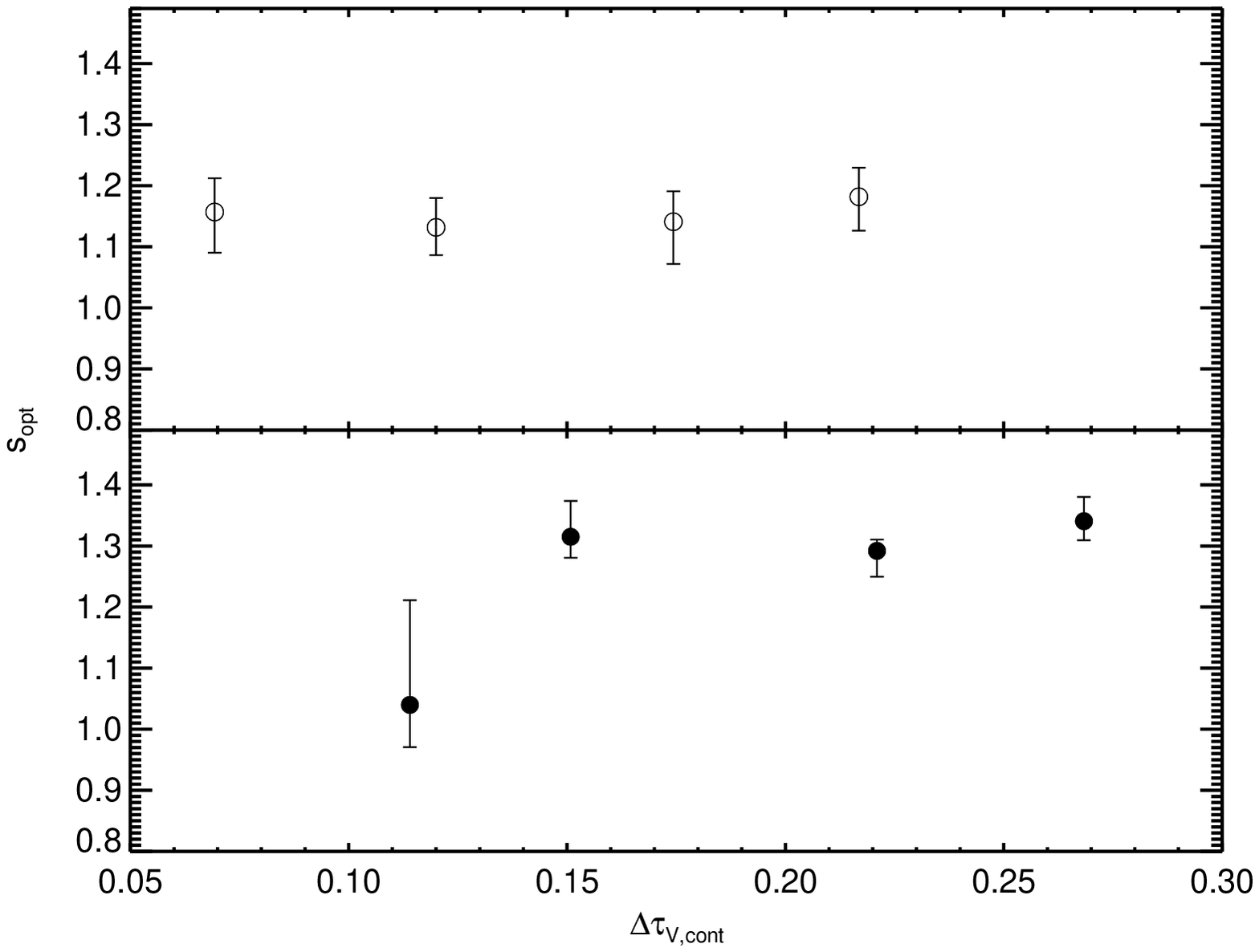}
\includegraphics[scale=0.5]{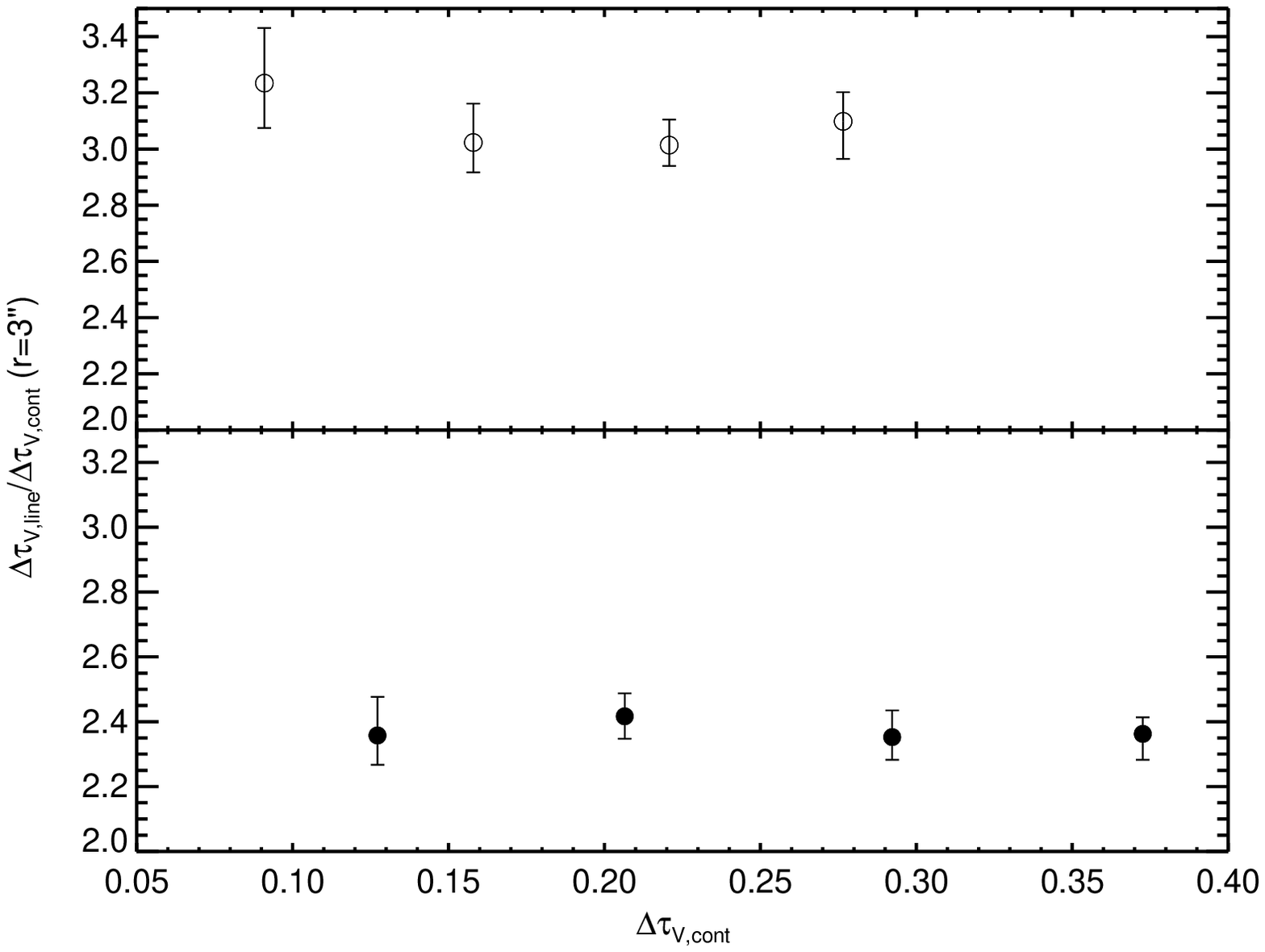}
\caption{The optical slope (left) and \Dtauvline/\Dtauvcont\ ratio
  measured from samples of low (top panel) and high (bottom panel)
  \mustar\ galaxies, where the pairs of galaxies are required to have
  \Dtauvline$ = [0.3,0.5,0.7,0.9]\pm0.2$ (points from left to
  right). This shows that there is no variation in
  the shape of the dust curves or in the line-to-continuum optical
  depth ratio, with continuum optical depth.  }\label{fig:app1}
\end{figure*}

\begin{figure*}
\includegraphics[scale=0.35]{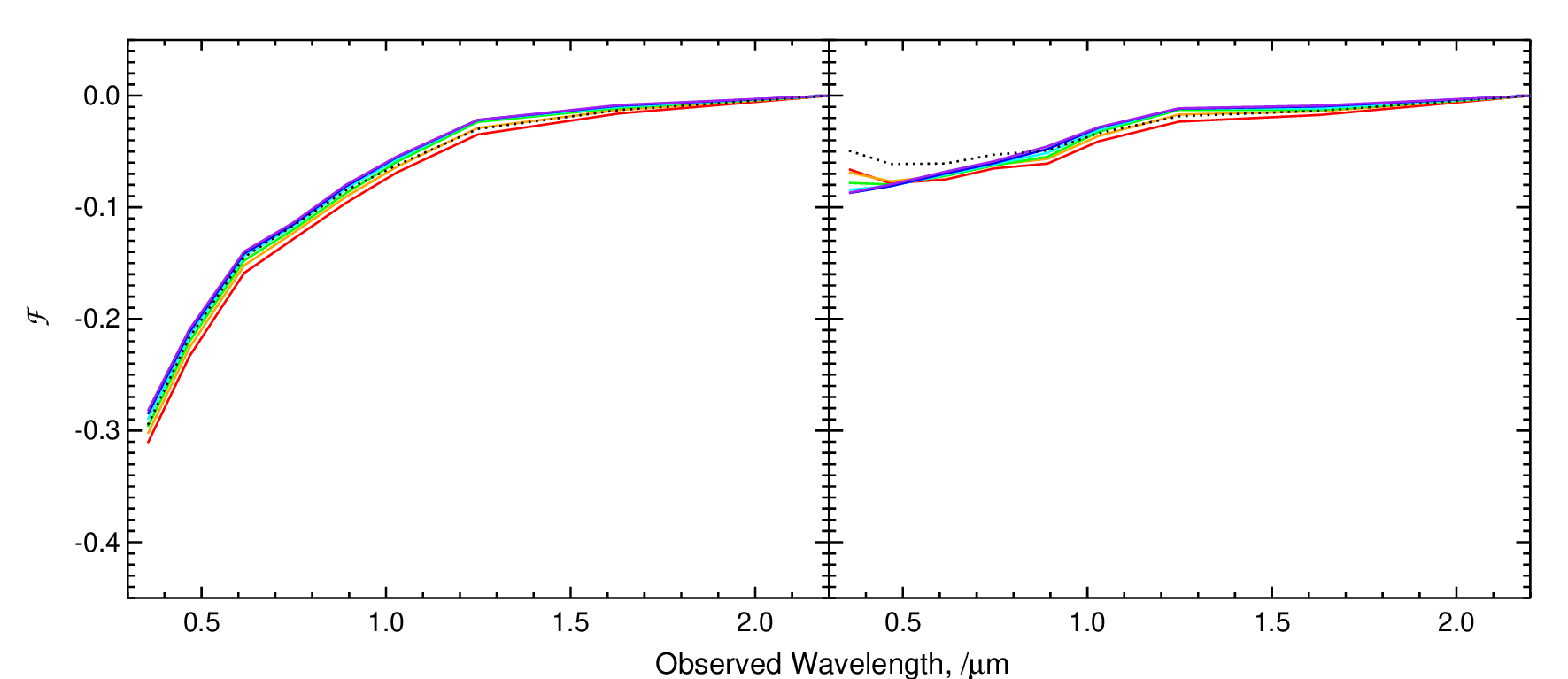}
\includegraphics[scale=0.4]{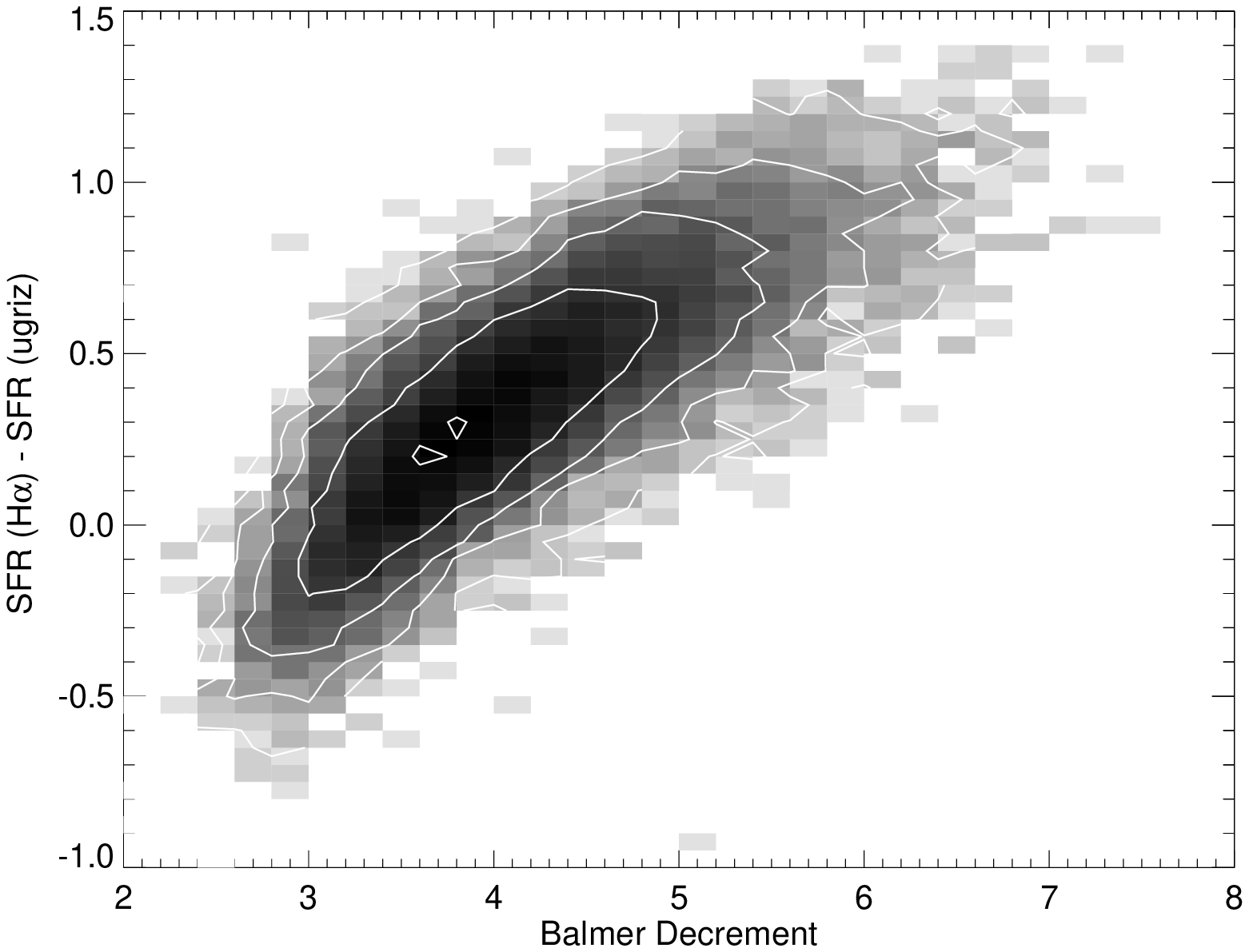}
\caption{On the left we compare attenuation curves derived from galaxy
  pairs matched on \ssfr\ estimated from \ha\ within the 3\arcsec SDSS
  fibre divided by stellar mass estimated from the optical photometry
  (left), and a pure photometric estimate of \ssfr (right). We find
  that the observed attenuation curves are substantially weaker when
  using the pair matching based upon photometric estimates of
  \ssfr. On the right we investigate the reason for this, comparing
  the two SFR estimates as a function of ratio of \ha\ to \hb\ luminosity (2D
  histogram in log number of galaxies). We find that the photometric
  SFR$\equiv M^*$\ssfr is systematically lower in galaxies with a high
  ratio of \ha\ to \hb\ luminosity. This is not unexpected as it is difficult to
  distinguish dust reddening from an older stellar population using
  optical photometry alone. The resulting correlation between stellar
  population and dust content causes the attenuation curve to appear
  considerably flatter.  }\label{fig:app2}
\end{figure*}

As described in Section \ref{sec:formalism} there are some important
features of the new pair-matching method to be aware of if the same
experiment is to be repeated on different datasets. In Figures
\ref{fig:app1} and \ref{fig:app2} we present some simple tests to show
that our method is valid for the dataset in this paper.

\pagebreak
\section{SQL queries}\label{app:sql}

The queries below are run on the SDSS Catalog Archive Server
(http://casjobs.sdss.org/casjobs/), WFCAM Science Archive
(http://surveys.roe.ac.uk/wsa/) and MAST Catalog Archive Server
(http://mastweb.stsci.edu/gcasjobs/) respectively. The resulting data
tables are joined manually on SDSS specobjid, using a purpose designed
IDL code. Here we provide an extract of each query, the full SQL code
is available online.

\begin{code*}
{\bf /* To obtain parameters for spectroscopic galaxies in the SDSS */}
\begin{verbatim}
/* Basic parameters */
SELECT  s.bestobjid,s.specobjid, mjd,plate,fiberid,s.z,sn_0,sn_1,
petror50_r,petror90_r,petror50_z,rowindex, p.extinction_g, p.petrorad_r,

/* Petrosian magnitudes */
p.petromag_u, p.petromag_g, p.petromag_r, p.petromag_i, p.petromag_z,

/* The number of annuli for which there is a measurable signal */
p.nprof_u, p.nprof_g, p.nprof_r, p.nprof_i, p.nprof_z,

/* The u-band AperFlux */
up0.profMean as uAperFlux0, up1.profMean as uAperFlux1, 
up2.profMean as uAperFlux2, up3.profMean as uAperFlux3, 
up4.profMean as uAperFlux4, up5.profMean as uAperFlux5, 
up6.profMean as uAperFlux6, up7.profMean as uAperFlux7, 

/* The g-band AperFlux */
gp0.profMean as gAperFlux0, gp1.profMean as gAperFlux1, [....]

/* mydb.sdssmpa_rowind contains the SDSS-MPA rowindex and specobjid for all galaxies */
FROM specobj as s, photoobjall as p, mydb.sdssmpa_rowind as m, 
PhotoProfile as up0, PhotoProfile as gp0, PhotoProfile as rp0, PhotoProfile as ip0,
PhotoProfile as zp0, PhotoProfile as up1, [....]
 

/* Join Specobj, PhotoobjAll, SDSSMPA_rowind and select only spectroscopic galaxies */
WHERE s.specobjid=p.specobjid AND s.specobjid=m.specobjid AND s.specclass=2 AND 

/* Join u-band PhotoProfiles */
s.bestobjid = up0.objid AND up0.band = 0 AND up0.bin = 0 AND
s.bestobjid = up1.objid AND up1.band = 0 AND up1.bin = 1 AND 
s.bestobjid = up2.objid AND up2.band = 0 AND up2.bin = 2 AND 
s.bestobjid = up3.objid AND up3.band = 0 AND up3.bin = 3 AND 
s.bestobjid = up4.objid AND up4.band = 0 AND up4.bin = 4 AND
s.bestobjid = up5.objid AND up5.band = 0 AND up5.bin = 5 AND
s.bestobjid = up6.objid AND up6.band = 0 AND up6.bin = 6 AND
s.bestobjid = up7.objid AND up7.band = 0 AND up7.bin = 7 AND

/* Join g-band PhotoProfiles */
s.bestobjid = gp0.objid AND gp0.band = 1 AND gp0.bin = 0 AND
[....]

\end{verbatim}
\end{code*}

\clearpage

\begin{code*}
{\bf /*To obtain parameters for UKIDSS sources which are spectroscopic galaxies
in the SDSS. */}
\begin{verbatim}
SELECT  ss.specobjid, ls.sourceid, x.distancemins, 

/* Petrosian magnitudes */
ls.yPetroMag, ls.j_1PetroMag, ls.hPetroMag, ls.kPetroMag, 

/* dust attenuation */
ls.aY, ls.aH, ls.aJ, ls.aK,

/* Petrosian radii */
yd.PetroRad as yPetroRad, jd.PetroRad as j_1PetroRad, hd.PetroRad as hPetroRad, 
kd.PetroRad as kPetroRad, 

/* UKIDSS Aper mags */
yd.Aperflux1 as yAperFlux1, jd.AperFlux1 as j_1AperFlux1, hd.AperFlux1 as hAperFlux1, 
kd.AperFlux1 as kAperFlux1,
yd.Aperflux2 as yAperFlux2, [....]

/* J band exposure time */
mf.expTime

FROM lasYJHKSource AS ls,  bestdr7..specobj as ss, lasSourceXDR7PhotoObj as x, 
lasYJHKMergeLog AS ml, lasDetection as jd, lasdetection as hd, lasdetection as yd, 
lasdetection as kd, multiframe as mf

WHERE x.masterobjid=ls.sourceid AND ss.bestobjid=x.slaveobjid AND 
ls.framesetID = ml.framesetID AND ml.j_1mfID = mf.multiframeID AND

/* Join merge log to detection for the J frame: */
ml.j_1mfID = jd.multiframeID AND ml.j_1eNum = jd.extNum AND
/* Join source to detection to look up the detection required: */
ls.j_1SeqNum = jd.seqNum AND 

/* Join merge log to detection for the Y frame: */
ml.ymfID = yd.multiframeID AND ml.yeNum = yd.extNum AND
/* Join source to detection to look up the detection required: */
ls.ySeqNum = yd.seqNum AND 

/* Join merge log to detection for the H frame: */
ml.hmfID = hd.multiframeID AND ml.heNum = hd.extNum AND
/* Join source to detection to look up the detection required: */
ls.hSeqNum = hd.seqNum AND 

/* Join merge log to detection for the K frame: */
ml.kmfID = kd.multiframeID AND ml.keNum = kd.extNum AND
/* Join source to detection to look up the detection required: */
ls.kSeqNum = kd.seqNum AND 

/* remove blended objects */
ls.yppErrBits&0x00000010 = 0 AND ls.j_1ppErrBits&0x00000010 = 0 AND 
ls.hppErrBits&0x00000010 = 0 AND ls.kppErrBits&0x00000010 = 0 AND

/* now get spectroscopic galaxies and select the nearest match */
ss.specclass=2 AND distanceMins<0.033333 AND 
distanceMins IN ( SELECT MIN(distanceMins) FROM lasSourceXDR7PhotoObj 
WHERE masterObjID=x.masterObjID AND sdssPrimary=1)
\end{verbatim}
\end{code*}

\clearpage

\begin{code*}
{\bf /*To obtain parameters for GALEX sources which are spectroscopic galaxies
in the SDSS. */}
\begin{verbatim}

/* Basic parameters */
SELECT ss.bestobjid, ss.specobjid, gg.objid, gg.distance, gg.reversemultiplematchcount,

/* GALEX Fluxes */  
ga.nuv_flux, ga.nuv_fluxerr,ga.fuv_flux, ga.fuv_fluxerr, ga.e_bv

/* myDB.dr7SFgal contains an uploaded list of all the specobjids we want to match */
FROM myDB.dr7SFgal as ss, galexgr4plus5..xsdssdr7 as gg, galexgr4plus5..photoobj as ga

WHERE ss.bestobjid=gg.SDSSobjid and gg.objid=ga.objid and gg.reversedistancerank=1

\end{verbatim}
\end{code*}

\clearpage

\section{IDL code to compute dust curve}\label{app:code}
\begin{code*}
\begin{verbatim}

FUNCTION VW_DUSTCURVE, wave, flux, tau_v, flag_mustar, ssfr=ssfr,ba=ba, 
                       silent=silent, magnitudes=magnitudes

;+
; NAME: VW_DUSTCURVE
;
; AUTHORS: 
;    Vivienne Wild <vw@roe.ac.uk>
;
; PURPOSE: 
;    Calculate the unattenuted galaxy spectral energy distribution
;    followingt the prescription of Wild et al. 2011     
;
; CALLING SEQUENCE:
;    unattenuated_flux = VW_DUSTCURVE(wave,flux,tau_v, flag_mustar, [ssfr=, ba=])
;
; INPUTS:
;    wave        = wavelength array in microns, not to extend beyond [0.14,2.05]
;    flux        = flux array (or magnitudes, see /magnitudes keyword) 
;    tau_v       = optical depth in the stellar continuum at 5500Angstrom
;    flag_mustar = -1 for high stellar surface mass density, or bulge-dominated
;                =  1 for low stellar surface mass density, or disk-dominated
;
; OPTIONAL INPUTS:
;    ssfr        = logarithm of the specific star formation rate of
;                  the galaxy in years (log(psi_s/yr^-1). If none is
;                  supplied, -9.5 is assumed  
;    ba          = minor/major axis ratio of galaxy. If none is
;                  supplied then 0.6 is assumed. 
;
; KEYWORD PARAMETERS:
;    silent      = set to prevent some messages
;    magnitudes  = set to correct an array of magnitudes, rather than fluxes
;         
; OUTPUTS:
;    unattenuated_flux    = unattenuated flux array
;
; REFERENCES: [1] Wild, Charlot, Brinchmann et al., 2011, MNRAS, submitted
; 
; NOTES: 
;
; MODIFICATION HISTORY:
;   2011 First implementation in IDL V. Wild
;-
;****************************************************************************************;
;  Copyright (c) 2011, Vivienne Wild 
;                                                                                        ;
;  Permission to use, copy, modify, and/or distribute this software for any              ;
;  purpose with or without fee is hereby granted, provided that the above                ;
;  copyright notice and this permission notice appear in all copies.                     ;
;                                                                                        ;
;  THE SOFTWARE IS PROVIDED "AS IS" AND THE AUTHOR DISCLAIMS ALL WARRANTIES              ;
;  WITH REGARD TO THIS SOFTWARE INCLUDING ALL IMPLIED WARRANTIES OF                      ;
;  MERCHANTABILITY AND FITNESS. IN NO EVENT SHALL THE AUTHOR BE LIABLE FOR               ;
;  ANY SPECIAL, DIRECT, INDIRECT, OR CONSEQUENTIAL DAMAGES OR ANY DAMAGES                ;
;  WHATSOEVER RESULTING FROM LOSS OF USE, DATA OR PROFITS, WHETHER IN AN                 ;
;  ACTION OF CONTRACT, NEGLIGENCE OR OTHER TORTIOUS ACTION, ARISING OUT OF               ;
;  OR IN CONNECTION WITH THE USE OR PERFORMANCE OF THIS SOFTWARE.                        ;
;****************************************************************************************;
\end{verbatim}
\end{code*}

\begin{code*}
\begin{verbatim}
; Return to caller on error.
On_Error, 2

n = 20d                         ;smoothing parameter

lc_1_eff = 0.2175d              ;rest-frame break points
lc_2_eff = 0.3d
lc_3_eff = 0.8d

V = 0.55                        ;normalised here

;;-- verify inputs
if N_params() LT 4 then begin
   print, 'syntax: VW_DUSTCURVE(wave,flux,tau_v, flag_mustar, [ssfr=, ba=])'
   return, -1
endif

;;-- check for b/a and ssfr
if n_elements(ssfr) eq 0 then begin
   ssfr=-9.5
   if not(keyword_set(silent)) then print, 'Assuming log(psi_s)=-9.5'
endif
if n_elements(ba) eq 0 then begin
   ba = 0.6
   if not(keyword_set(silent)) then print, 'Assuming b/a=0.6'
endif

;;-- define parameters for high / low mustar cases
if flag_mustar eq -1 then P = [1.3, 0.6, 0.1, 1.1, 0.3, -0.2, 0.2, 0.9,0.1] $
else if flag_mustar eq 1 then P = [1.1, 0.4, -0.1, 0.7, 0., 0.4, 0.15, 0.,0.2] $
else begin
   print, 'Please specify: flag_mustar = -1 for high stellar surface mass density'
   print, '                            =  1 for low stellar surface mass density'
   print, 'syntax: VW_DUSTCURVE(wave,flux,tau_v, flag_mustar, [ssfr=, ba=])'
   return, -1
endelse

;;-- check parameters are within specified limits
ind = where(wave lt 0.14 or wave gt 2.05)
if ind[0] ne -1 then begin
   print, 'Dust curve only valid in wavelength range 0.14-2.05microns'
   return, -1
endif

if ba lt 0.3 or ba lt 0.9 then begin
   print, 'Dust curve only valid in axis ratio range 0.3-0.9, see Section 6 of [1]'
   return, -1
endif

if flag_mustar eq -1 then begin ;high mustar
   if ssfr lt -10.2 or ssfr lt -9.3 then begin
      print, 'For bulge-dominated galaxies dust curve only valid in ssfr range -10.2 -> -9.3'
      return, -1
   endif
endif

if flag_mustar eq 1 then begin ;low mustar
   if ssfr lt -10.0 or ssfr lt -9.1 then begin
      print, 'For disk-dominated galaxies dust curve only valid in ssfr range -10.0 -> -9.1'
      return, -1
   endif
endif
   
\end{verbatim}
\end{code*}

\begin{code*}
\begin{verbatim}
;;------------------------------------------------------------------
;;-- define relations with SSFR and b/a

ba_c = ba - 0.6
ssfr_c = ssfr + 9.5

;; sopt
sopt = P[0]+ba_c*P[1]+ssfr_c*P[2]

;; snir
snir = 1.6

;; snuv 
snuv = P[3]+ba_c*P[4]+ssfr_c*P[5]

;; sfuv
sfuv = P[6]+ba_c*P[7]+ssfr_c*P[8]


lc_1 = lc_1_eff
lc_2 = ((lc_1^snuv)/(lc_2_eff^(snuv-sopt)))^(1/sopt)
lc_3 = ((lc_2^sopt)/(lc_3_eff^(sopt-snir)))^(1/snir)
   
norm = (((V/lc_1)^(sfuv*n)+(V/lc_1)^(snuv*n)+(V/lc_2)^(sopt*n)+$
         (V/lc_3)^(snir*n))^(-1/n)) ;Qlambda=1 at V

Q_lambda = (((wave/lc_1)^(sfuv*n)+(wave/lc_1)^(snuv*n)+(wave/lc_2)^(sopt*n)+$
             (wave/lc_3)^(snir*n))^(-1/n)) / norm ;eq. 18

tau_lambda = Q_lambda*tau_V 

if not(keyword_set(magnitudes)) then return, flux/exp(-tau_lambda) $ ;e.g. eq 5
else return, flux - 1.086*tau_lambda ;eqn. 3

END

\end{verbatim}
\end{code*}

\end{appendix}

\end{document}